\documentclass[10pt,twocolumn]{IEEEtran}
\usepackage{authblk}
\providecommand{\keywords}[1]{\textbf{\textit{Index terms---}} #1}

\usepackage{cite}
\usepackage{amsmath,amssymb,amsfonts}
\usepackage{algorithmic}
\usepackage[ruled,vlined]{algorithm2e}
\usepackage{graphicx}
\usepackage{textcomp}
\usepackage[numbers]{natbib}
\usepackage[labelformat=simple]{subcaption}

\usepackage{booktabs}
\usepackage{adjustbox}
\usepackage{url}
\usepackage{hyperref}
\usepackage{comment}

\def\BibTeX{{\rm B\kern-.05em{\sc i\kern-.025em b}\kern-.08em
    T\kern-.1667em\lower.7ex\hbox{E}\kern-.125emX}}

\usepackage{array}
\usepackage{fontawesome5}
\usepackage{ccicons}
\usepackage{pifont}
\newcommand{\cmark}{\ding{51}}%
\newcommand{\xmark}{\ding{55}}%

\usepackage{longtable}
\usepackage{tabularx}
\usepackage{ltxtable}
\usepackage{ragged2e}
\usepackage{xltabular}
\usepackage{cellspace}
\newcommand\cincludegraphics[2][]{\raisebox{-0.2\height}{\includegraphics[#1]{#2}}}

\newcolumntype{Y}{>{\RaggedRight\arraybackslash}X}
\newcolumntype{P}[1]{>{\centering\arraybackslash}p{#1}}

\begin{document}

\title{Systematic Literature Review of\\EM-SCA Attacks on Encryption}

\author[1]{Muhammad Rusyaidi Zunaidi}%
\author[2]{Asanka Sayakkara}%
\author[1]{Mark Scanlon}%
\affil[1]{School of Computer Science, University College Dublin, Ireland \authorcr (e-mail: muhammad.zunaidi@ucdconnect.ie, mark.scanlon@ucd.ie)}%
\affil[2]{School of Computing, University of Colombo, Colombo, Sri Lanka \authorcr (e-mail: asa@ucsc.cmb.ac.lk)}%

\maketitle

\begin{abstract}
Cryptography has become an essential tool in information security, preserving data confidentiality, integrity, and availability. However, despite rigorous analysis, cryptographic algorithms may still be susceptible to attack when used on real-world devices. Side-channel attacks (SCAs) are physical attacks that target cryptographic equipment through quantifiable phenomena such as power consumption, operational times, and EM radiation. These attacks are considered to be a significant threat to cryptography since they compromise the integrity of the algorithm by obtaining the internal cryptographic key of a device by seeing its physical implementation. The literature on SCAs has focused on real-world devices, yet with the growing popularity of sophisticated devices like smartphones, fresh approaches to SCAs are necessary. One such approach is electromagnetic side-channel analysis (EM-SCA), which gathers information by listening to electromagnetic (EM) radiation. EM-SCA has been demonstrated to recover sensitive data like encryption keys and has the potential to identify malicious software, retrieve data, and identify program activity. This study aims to evaluate how well EM-SCA compromises encryption under various application scenarios, as well as examine the role of EM-SCA in digital forensics and law enforcement. Regarding this, addressing the susceptibility of encryption algorithms to EM-SCA approaches can provide digital forensic investigators with the tools they desire to overcome the challenges posed by strong encryption, allowing them to continue playing a crucial role in law enforcement and the justice system. Furthermore, this paper seeks to define the current state of EM-SCA in terms of attacking encryption, the encryption algorithms and encrypted devices that are most vulnerable and resistant to EM-SCA, and the most promising EM-SCA on encryption approaches. This study will provide a comprehensive analysis of EM-SCA in the context of law enforcement and digital forensics and point towards potential directions for further research.

\end{abstract}


\section{Introduction}
\label{intro}

In recent years, cryptography has emerged as a key tool in information security, preserving the availability, confidentiality, and integrity of data. By convention, cryptographic algorithms are rigorously analyzed in terms of computational complexity to assure their security. However, when used on real-world devices, these algorithms may still be susceptible to attack. The focus has switched to side-channel attacks (SCA) since cryptographic algorithms are made to resist conventional methods of breaking secrecy~\cite{cryptoeprint:2005/388}. SCA are physical attacks that are not invasive and are seen as a severe threat to cryptographic equipment. To extract a device's internal cryptographic key, SCAs take advantage of quantifiable phenomena present in the device, such as its power consumption, EM radiation, and operational times~\cite{SAYAKKARA201943}. SCAs can be carried out with low-cost, generic tools, but invasive attacks call for a more expensive setup that physically contacts the target device to harvest internal signals. In side-channel attacks, attackers gain knowledge of the algorithm's internal state by observing its physical implementation, which can undermine the algorithm's integrity. The academic literature on side-channel attacks focuses on real-world devices, yet these devices could be simple in terms of architecture or electronic design. The growing popularity of sophisticated devices with extensive functionality, such as smartphones, emphasizes the need for fresh approaches to side-channel attacks. When compared to other techniques like power analysis, such as electromagnetic (EM) side-channel leakage is a desirable alternative because it is less invasive and versatile~\cite{Das2020c}. On a System-on-Chip (SoC), it also enables the targeting of particular regions or components.

Electronic equipment that is in operation emits EM noise at different frequencies. This is a result of the electrical currents being used, which vary over time. Running computers and mobile devices also produce significant amounts of EM noise~\cite{Sayakkara2019a}. Due to the quick clock pulses that are employed on them, CPUs are regarded as one of the strongest EM noise producers in computing. The software instructions being followed and the data being handled determine the pattern of electrical pulses sent through the CPU of a computer. Consequently, it has been demonstrated that CPU EM emissions leak data as well as information about software processes~\cite{GetzRMoeckel}. Furthermore, the power usage of a smart card running an unprotected implementation of an algorithm is one illustration of a side-channel attack that can be used to guess the key and break the cryptosystem~\cite{quisquater2001electromagnetic}. While EM attacks are a more covert alternative that uses EM emissions that coincide with specific computations, measuring power usage requires direct physical access. Although the feeble emissions of low-power devices restrict the range of these attacks, EM attacks use magnetic-field antennas within millimeters of the target chip.

The discipline of information security known as EM side-channel attack (EM-SCA) analysis, collects data by capturing EM signals. A single such capture is called an EM trace, which contains three signal characteristics; amplitude, phase, and frequency~\cite{Du2020}. Once a sufficient number of traces are captured, these can be fed into an EM-SCA analysis algorithm to gain insights into the patterns of data operations captured. The approach can be applied to tasks like identifying malicious software, retrieving data, and detecting software behavior. Cryptographic keys are particularly valuable for evidence collection. There are several methods for doing this, such as Simple EM Analysis (SEMA)~\cite{de2005electromagnetic}, Differential EM Analysis (DEMA)~\cite{Gandolfi2001}, and Correlation EM Analysis (CEMA)~\cite{Danial2020a}. EM side-channel attacks can offer an easy access point for the attacker without the target device running any specialized software or allowing the attacker to access its internal hardware. Recent developments in the field demonstrate the capability of such attacks to recover sensitive data, such as encryption keys~\cite{Sayakkara2018a}. 


In the context of digital forensics and law enforcement, EM-SCA can be a valuable tool for accessing encrypted data in legally authorized situations~\cite{Sayakkara2019a}. Law enforcement authorities have a legitimate need to break encryption under warrant in certain situations. For instance, when criminal and illegal activities involve electronic and computing devices, such as child exploitation, terrorism, drug trafficking, and other serious crimes, access to encrypted data may be essential to collect potentially pertinent evidence~\cite{CASEY2011129}. In such cases, digital forensics plays a crucial role in investigating and prosecuting these crimes. Additionally, law enforcement authorities may require access to encrypted data to recover important information lost due to hardware failure, accidental deletion, or other reasons. Another situation where breaking encryption may be necessary is in cases of ransomware attacks. Ransomware is a type of malware that encrypts the victim's data, making it inaccessible, and then demands payment in exchange for the decryption key. In such cases, law enforcement authorities may need to break the encryption to recover the victim's data and track down the perpetrators.  However, it is important to note that any such access to encrypted data must be done under a proper warrant and following due process. Digital forensics specialists use specialized tools and procedures to acquire and analyze digital evidence in a forensically-sound manner, ensuring that the evidence is admissible in a court of law~\cite{David2016}. Any new methods of acquiring digital evidence must be thoroughly scrutinized to ensure reliability and admissibility in a court of law.

The main focus of these systematic literature reviews is to analyze the use of EM-SCA to break cryptographic algorithms, while the paper by \citet{10.1145/3456629} provides a wider survey of microarchitectural side channels with a focus on attacks and defenses in cryptographic applications, including power consumption and acoustic emission; however the authors did not cover physical side-channel attacks including EM-SCA, which was stated as out of scope. Although both papers discuss side-channel attacks in cryptography, these papers provide a more specialized and detailed analysis of the EM-SCA, which is particularly significant for law enforcement and digital forensics purposes. In contrast, \citet{10.1145/3456629} covers a broader range of side-channel attacks and defenses, which results in a more comprehensive, but less targeted analysis. While some topics overlap between the two papers, this research analysis is more practical and specific to the use of EM-SCA. Therefore, this study endeavors to assess the effectiveness of EM-SCA in compromising encryption under a variety of application conditions. The primary contributions of this research are as follows:

\begin{itemize}
\item A thorough and systematic review of the existing literature on EM side channel analysis as it relates to breaking encryption is provided, with a focus on recent advancements.
\item The encryption algorithms and encrypted devices that are most vulnerable and resilient to EM-SCA are identified.
\item The practical application of EM side channel analysis is demonstrated through the examination of multiple digital device investigation scenarios.
\item The most promising approaches to EM side channel analysis on encryption are demonstrated, highlighting avenues for future research.

\end{itemize}

\subsection{Review Questions}

This systematic literature review's objective is to synthesize and retrieve information from a substantial body of EM-SCA on encryption. This is achieved by identifying the characteristics of empirical studies in data encryption operations – to better understand how EM-SCA is represented in user studies, the value of these publications and any relevant documentation for any research findings they represent. Three review questions were defined to focus the analysis performed as part of this paper:

\begin{enumerate}
    \item[\textbf{RQ1}] What is the state of the art of EM-SCA with respect to attacking encryption?
    \item[\textbf{RQ2}] Which encryption algorithms and encrypted devices are most susceptible to, and resilient against, EM-SCA attacks?
    \item[\textbf{RQ3}] Which approaches to EM-SCA on encryption prove most fruitful, or demonstrates the most promise for the future?
\end{enumerate}

\section{Methodology}

This section provides a thorough overview of the study's systematic literature review, including the article selection method and research process. The standard guidelines for systematic literature reviews by \citet{kitchenham2004procedures} were adopted for this review in order to identify and select the articles that would be included. For this paper, the strategy and implementation steps comprised of:

\begin{enumerate}
    \item Specifying the review questions and focus.
    \item Defining the data sources and search strategy.
    \item Identifying the articles/studies to be included or excluded.
    \item Data extraction.
    \item Data synthesis.
\end{enumerate}



Each of these steps is outlined in greater detail in the following subsections:


\subsection{Defining the Review Questions}

A fundamental component of the systematic review process is defining the research questions, as it serves as the foundation for all other tasks. The review questions will inform the data that should be gathered from the reviewed literature, as well as which relevant articles should be included or removed from the review. The systematic literature review's conclusions should include the answers to the identified research questions.\\

\textbf{RQ1: What is the state of the art of EM-SCA with respect to attacking encryption?}\\

EM-SCA is a highly effective approach for attacking encryption systems. Cryptographic algorithms are executed in software or hardware on physical devices that connect with and are affected by their contexts. On the other hand, side-channel attacks are predicated on the fact that when cryptosystems exist, they generate unintentional physical effects, and the information from these effects can reveal evidence about the system, e.g., the EM radiation emitted by an operation. EM-SCA is one approach that could potentially overcome the encryption challenge facing legitimate and lawful investigation by recovering cryptographic keys and other types of critical information. The effectiveness of EM-SCA is dependent on a variety of factors, such as the number of plaintext-ciphertext pairs available for analysis, the complexity of the encryption algorithm, and the quality of the equipment used for data collection. Investigating the state of the art of EM-SCA with regard to combating encryption is therefore crucial in order to pinpoint current limitations and future development areas. The evaluation needs to take into account both new trends in hardware architecture and their possible effects on EM-SCA performance, as well as recent advancements in EM-SCA and how they have affected the strategy's efficacy. A comprehensive overview of the state of the art of EM-SCA will be presented by addressing these important questions. \\

\textbf{RQ2: Which encryption algorithms and encrypted devices are most susceptible to, and resilient against, EM-SCA attacks?}\\

In order to conduct a thorough and informative systematic literature review on EM-SCA in attacking encryption, it is crucial to not only evaluate the state of the art of EM-SCA but also to investigate which encryption algorithms and encrypted devices are most vulnerable and resistant to these attacks. RQ2 aims to address this issue by exploring the current knowledge on the susceptibility of various encryption algorithms and encrypted devices to EM-SCA attacks, as well as the effectiveness of countermeasures proposed to mitigate these attacks. To provide a comprehensive answer to RQ2, the systematic literature review will consider a range of factors that may affect the susceptibility of encryption algorithms and devices to EM-SCA. These factors could include the design and implementation of the algorithm, the hardware and software used to run the algorithm, the environment in which the algorithm is executed, and the characteristics of the attacker. Additionally, this review question will evaluate the quality and reliability of the studies that investigate the susceptibility of encryption algorithms and devices to EM-SCA, and the effectiveness of the proposed countermeasures. The RQ2 is to critically evaluate the existing literature on the susceptibility of encryption algorithms and encrypted devices to EM-SCA, and to identify any gaps in the knowledge or inconsistencies in the findings. In order to help digital forensics experts and law enforcement authorities make informed choices about which encryption algorithms and encrypted devices are more susceptible/resilient against EM-SCA in various scenarios, this RQ2 can address this question and offer insightful information about the strengths and weaknesses of various encryption approaches. \\

\textbf{RQ3: Which approaches to EM-SCA on encryption prove most fruitful or demonstrates the most promise for the future?}\\




In the field of electronic device security, approaches to EM-SCA have demonstrated success in retrieving encryption keys from specific devices. However, it is important to determine which approach holds the most potential for future development in this area. In particular, it is crucial to evaluate and compare various EM-SCA methods in terms of their practicality and usability under real-world conditions, where attackers may not have access to ideal laboratory conditions or cherry-picked victim devices. Thus, investigating how EM-SCA attacks can be conducted in real-world contexts is necessary to determine the most useful and practical approaches. This investigation becomes even more relevant given the ongoing trend toward hardware architectures featuring multiple-core processors and dynamic voltage-frequency scaling (DVFS), which is an energy-saving technique that adjusts the CPU core's frequency to meet processing demand while minimizing energy usage. As processor design advances, the number of cores per chip increases, and each core operates at its frequency, which could significantly increase power consumption. The usefulness of existing EM-SCA approaches on hardware with multiple core processors, and dynamic voltage-frequency scaling must therefore be investigated and evaluated, to determine whether these techniques could still extract meaningful signals from the EM emanations of a device, given the changing and unpredictable nature of the EM environment. In particular, it is critical to investigate the effectiveness of these approaches under emerging hardware trends, as the techniques may become less effective or even obsolete as hardware evolves. In addition, dedicated hardware components for cryptographic operations are being increasingly incorporated into CPUs for embedded systems. In contrast to software solutions, these hardware components produce EM radiation in a unique manner and are built to conduct cryptographic operations more quickly and efficiently. Therefore, the evaluation should explore how specialized hardware affects EM-SCA performance and decide whether new strategies are required to address the problems these components pose. This RQ3 finding can contribute to the development of novel and potent EM-SCA methods for breaking encryption by assessing the impact of specialized cryptographic hardware on EM-SCA performance.

\begin{figure*}[htp]
    \centering
    \includegraphics[width=\textwidth]{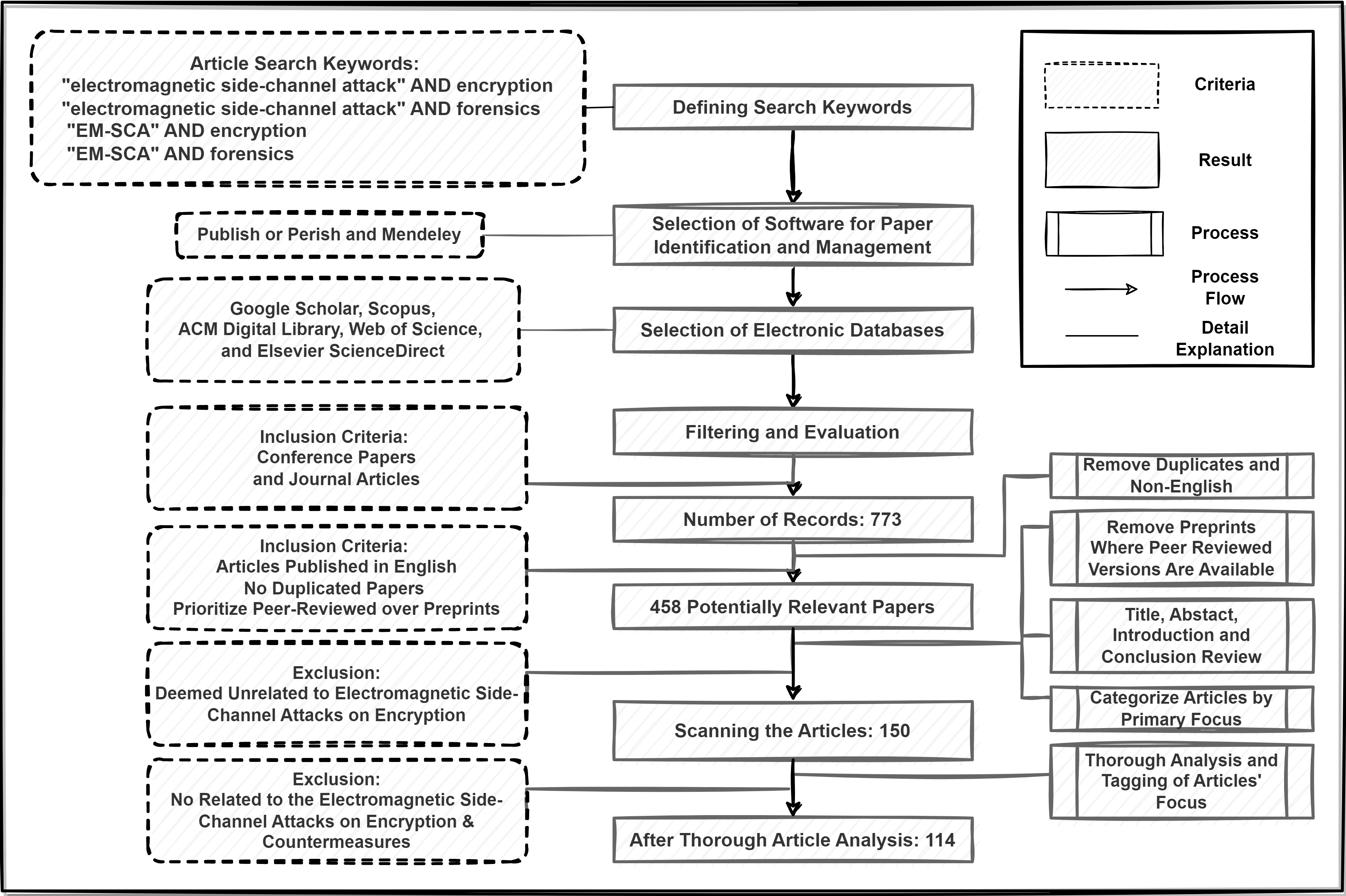}
    \caption{The process followed as part of this systematic review}
    \label{processfig}
\end{figure*}

\subsection{Data Source Selection and Conducting Searches}

The first stage in the planning of the review involves a keyword list for the search and identifying the appropriate search engines. Figure~\ref{processfig} depicts how the systematic review was carried out. Defining search keywords is crucial to ensure that the search keywords used are specific and relevant to this topic. The primary goal of this systematic literature review is to provide insights into how law enforcement and digital forensics experts can break encryption using EM-SCA methods, and to identify the vulnerabilities and strengths of various encryption algorithms and devices. To achieve this, a keyword list was compiled to cover the primary focus of the paper. The search phrases chosen were selected based on their ability to provide a comprehensive overview of EM-SCA in the context of encryption and digital forensics. It was necessary to avoid keywords that were either too general, which could lead to irrelevant publications, or too specific, which could result in relevant papers being overlooked. The following search phrase combinations were selected, which covers the focus of this paper and was used to search across several electronic academic article databases:


\begin{itemize}
\item ``EM side-channel attack'' AND encryption
\item ``EM side-channel attack'' AND forensics
\item ``EM-SCA'' AND encryption 
\item ``EM-SCA'' AND forensics
\end{itemize}

The selection of appropriate search keywords is a critical factor for the success of this systematic literature review. It is imperative that the chosen search phrases effectively capture the focus of the paper and yield relevant and targeted search results. The chosen search phrases were carefully selected to align with the review questions and to yield comprehensive and relevant results. The systematic review strives to offer critical analysis and evaluation of the retrieved studies, ultimately leading to an informative and insightful assessment of the state-of-the-art EM-SCA techniques in attacking encryption.

Furthermore, the software \textit{Publish or Perish} was used to retrieve and evaluate academic citations. It leverages a range of information sources to gather the raw citations and generates a variety of citation metrics, i.e., the quantity of papers and total citations, and the h-index~\cite{HarzingPOP}. \textit{Publish and Perish} facilitated the literature review's search strategy -- focusing on gathering peer-reviewed articles and papers in journals and conference proceedings that matched the chosen keywords anywhere in the resulting publications and electronic databases. These included Google Scholar, Scopus, the ACM Digital Library, Web of Science, and Elsevier ScienceDirect. The search terms were then applied to selected databases' titles, abstracts, and keyword fields. The search was conducted on 14 July 2022.

The one-by-one keyword search began by conducting a keyword search in each electronic database to ensure that the paper was not duplicated, and then folders were created based on the databases examined. The total number of articles found before eliminating duplicates was 773. Duplicate papers and non-English articles were deleted from each folder during the first round of paper elimination. In the second stage, all collected data were merged into a single folder, and duplicate papers were removed. This is required because most of the papers in ACM Digital Library, Science Direct, Scopus, Google Scholar and Web of Science were also found elsewhere. Executing the above deduplication on the five databases resulted in 458 articles to proceed to the next analysis step.

\subsection{Selecting Studies}

The next phase of a systematic review is to determine the papers that will represent its basis. Mendeley, a bibliography management tool, was used to collaboratively manage the sorting and organization of each paper. There were 4 steps in the article selection process: Step 1 -- involved sorting the articles based on title and abstract analysis, Step 2 -- addition of the conclusion to the analysis, Step 3 -- scanning the full article, and Step 4 -- final sorting of the articles based on thorough article analysis and classification of each article's primary contribution, i.e., attack approach, data gathering technique, devices, targeted algorithms and topic eliminated. While each article was assigned to its corresponding primary focus, Mendeley's \textit{tags} feature was employed to make it efficient to find articles later and to identify which folder the article belongs to. Any articles that did not clearly address the use of EM-SCA to break encryption or countermeasures to such attacks were deemed irrelevant and categorized as ``topic eliminated''. Papers that were not full conference papers have also been eliminated at this stage, e.g., workshop articles, extended abstracts, or where they explicitly marked as ``work in progress''. This selection of articles resulted in 150 papers remaining for consideration.

\subsection{Data Extraction}

The information gathered at this point regarding each article consisted of creating brief summaries of the main contributions of each article as they related to the subjects under consideration. At this phase, each of this paper's authors annotated and classified the articles in Mendeley. After further reading of each article, several were reclassified based on the article's primary focus, while other papers were dropped from the review due to an out-of-scope focus. To respond to the above-stated review questions, the following data was taken from the source studies:

\begin{itemize}
    \item The source of publication
    \item The accessibility of the full text
    \item Relevance of paper on EM-SCA on encryption
    \item The quality of the paper reviewed in journals and conferences.
\end{itemize}

Executing this data extraction phase resulted in 114 articles remaining for inclusion in the final systematic literature review.

\subsection{Data Synthesis}

\begin{figure*}[htp]
    \centering
    \includegraphics[width=\textwidth]{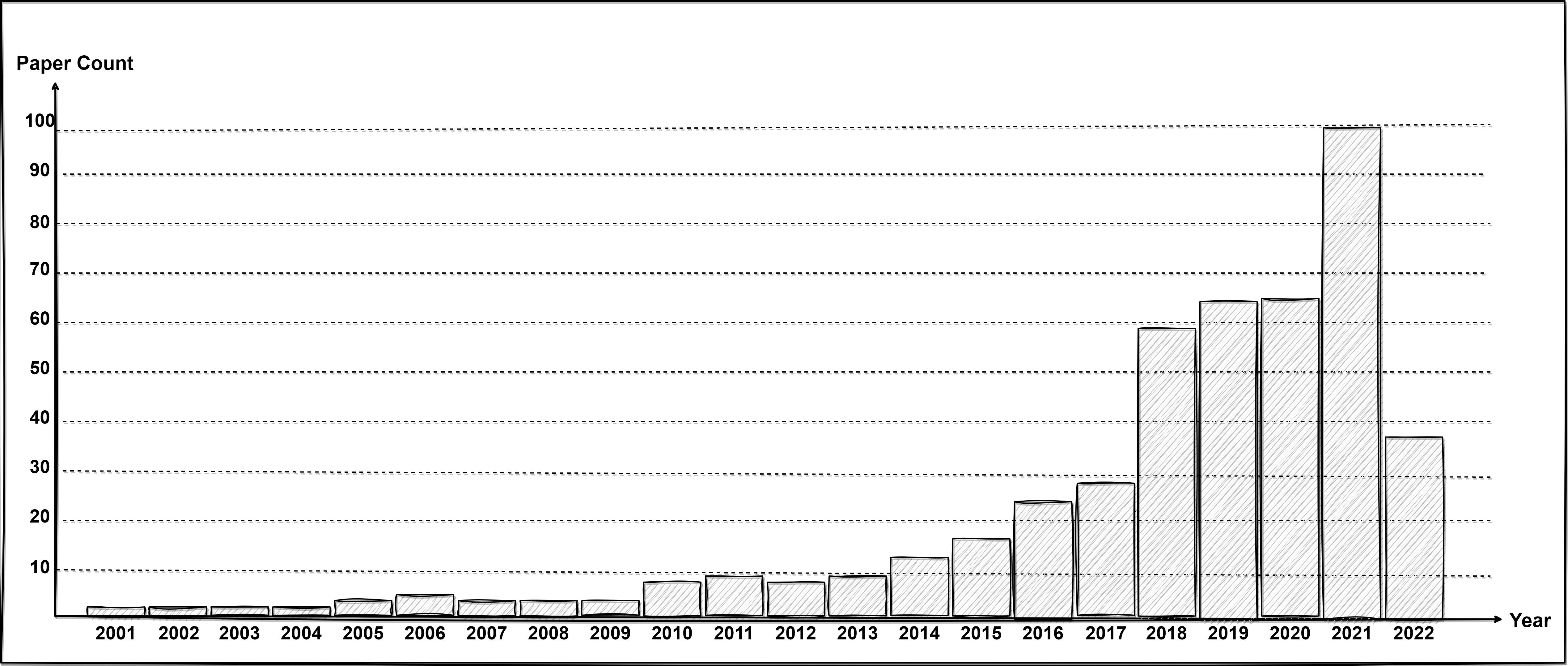}
    \caption{EM-SCA Publications by Year deemed ``Potentially Relevant'' from Initial Search}
    \label{fig:PaperFoundPoP}
\end{figure*}

\begin{figure*}[htp]
    \centering
    \includegraphics[width=\textwidth]{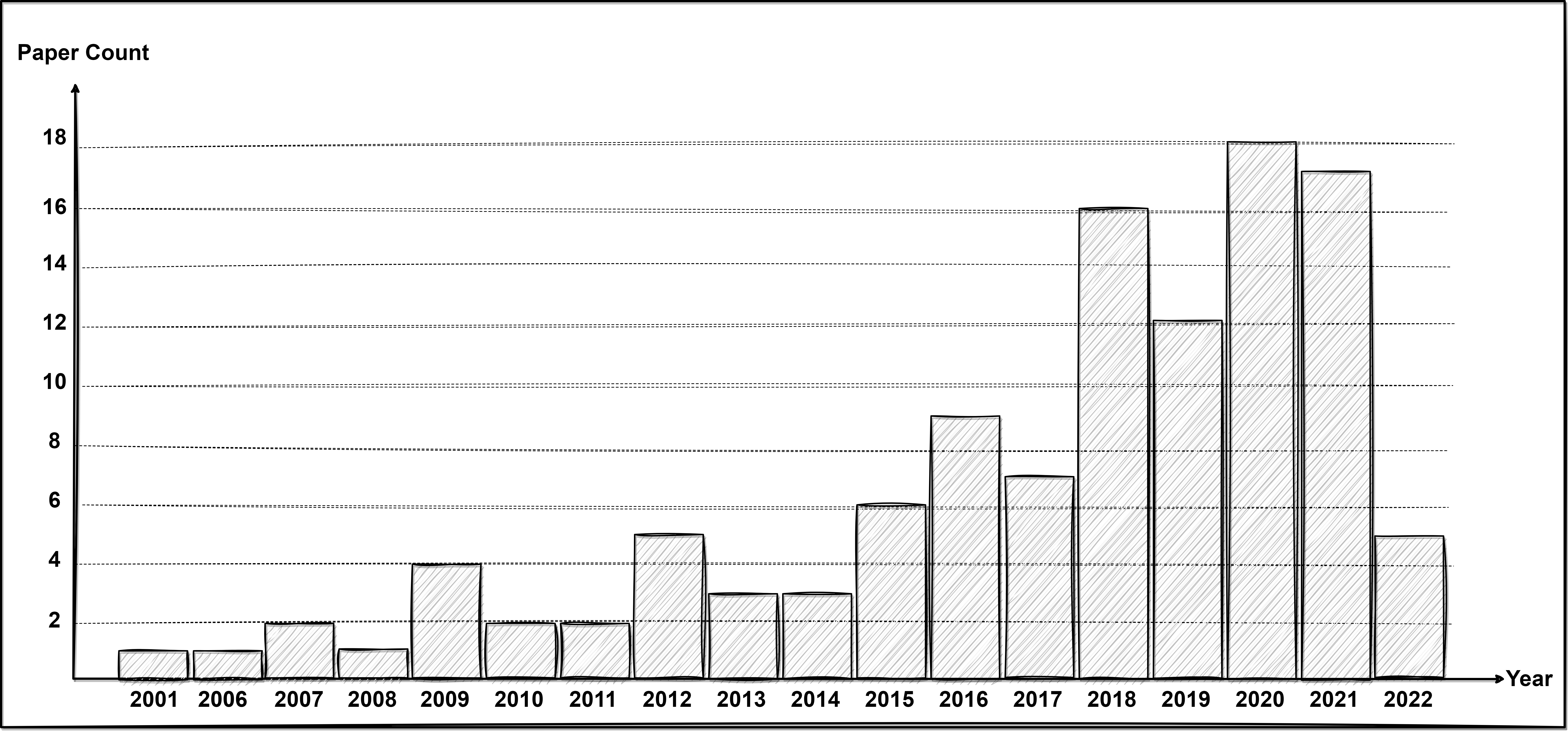}
    \caption{EM-SCA Publications Determined to be Focused on Encryption by Year}
    \label{fig:PaperFoundEncrypt}
\end{figure*}

As mentioned in Kitchenham and Charters' Systematic Literature Review guidelines~\cite{kitchenham2004procedures}, there are many data synthesis strategies. The majority of the data in this review is based on qualitative research, and the use of Mendeley to keep all the data has been maximized. The strategy for combining data involves categorizing the papers based on the type of encryption and device being analyzed, the analytic technique, and the findings. The initial step was to perform a search on Publish or Perish and extracted the data to the Mendeley, which, after removing duplicate results, revealed a total of 458 research papers on EM-SCA. Figure~\ref{fig:PaperFoundPoP} depicts papers that were recognized as having an upward trend in the EM-SCA study. Then, each document was carefully examined as a part of the filtering procedure. Figure~\ref{fig:PaperFoundEncrypt} illustrates the 114 publications that are exclusively those that concentrate on EM-SCA on encryption. The second histogram likewise showed a trend toward more EM-SCA research being done on encryption. These results imply that EM-SCA usage in the field of encryption is becoming more popular. This might be a result of the growing demand for secure data storage and transfer, the development of technology, and the accessibility of cutting-edge equipment for conducting EM-SCA.

\begin{figure*}[htp]
    \centering
    \includegraphics[width=0.8\textwidth]{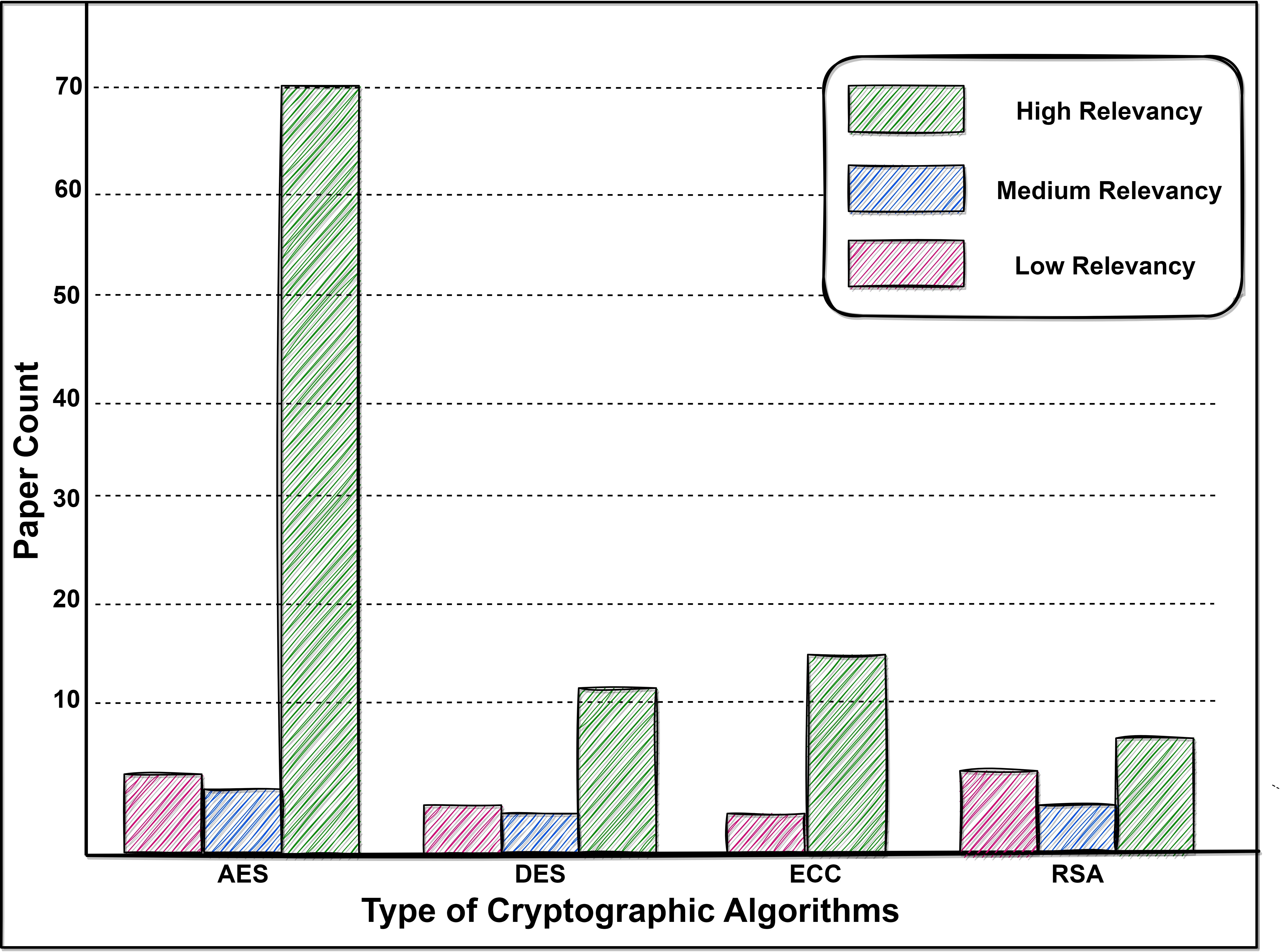}
    \caption{Frequency Distribution of EM Side Channel Analysis Papers by Relevance Level and Encryption Type}
    \label{fig:EncryptionTable}
\end{figure*}

Furthermore, the review analyzed 114 research papers on EM-SCA in attacking encryption, categorized by encryption type and relevance level, as illustrated in Figure~\ref{fig:EncryptionTable}. The term low, medium, and high relevancy are commonly used in systematic literature reviews to indicate the level of relevance of a paper to the research topic. Categorizing collected papers into these group streamlines will make this analysis of the literature more efficient. It allows focusing on the most relevant papers that provide useful insight into the research question. This study classifies papers into low, medium, and high relevancy categories based on the specific focus on particular encryption algorithms. Papers with a strong focus on the algorithms are considered highly relevant, while papers with a weaker focus are rated as low relevancy. This approach helps to identify the most valuable papers for the systematic review and ensure that the analysis is focused on the research questions.

As demonstrated in Figure~\ref{fig:EncryptionTable}, A significant number of studies focused on AES encryption, emphasizing the utilization of EM-SCA to compromise encryption. The majority of the papers focused on AES encryption with high relevance, which suggests that AES encryption is a popular target for EM-SCA due to its widespread use. In contrast, DES encryption had a lower frequency of papers, which may indicate that it is less vulnerable to EM-SCA or less popular as a target for such attacks. Interestingly, the number of publications focused on ECC encryption was relatively small, which may suggest a lack of research in this area. Two papers on EEA encryption with low relevance may indicate that it is less commonly used or less vulnerable to EM-SCA. Finally, the data shows that RSA encryption had a relatively low frequency of papers with high relevance, which suggests that EM-SCA less commonly targets it.  Results indicated that AES encryption is vulnerable to EM side channel analysis under various application conditions. These data syntheses allow for conclusions to be drawn regarding the effectiveness of EM-SCA in compromising encryption and for areas for future research to be identified.

Additionally, Table~\ref{tab:legend} provides a comprehensive overview of each publication analyzed in the review. The table includes the reference, encryption type, targeting device, Instrumented/Uninstrumented, the device that captures EM, machine learning, deep learning, key recovery and the relevance level assigned to each paper. This level of detail in the table allows for a more thorough understanding of the focus of each publication, aiding in the identification of the most valuable papers for the systematic review. Figure~\ref{fig:EncryptionTable} provides an overview of the relevance levels for each encryption type, while Table~\ref{tab:legend} offers a more in-depth analysis for each individual publication.

\section{Background}

\subsection{EM radiation}

\begin{figure*}[htp]
    \centering
    \includegraphics[width=0.8\textwidth,trim={0 0 0 5cm},clip]{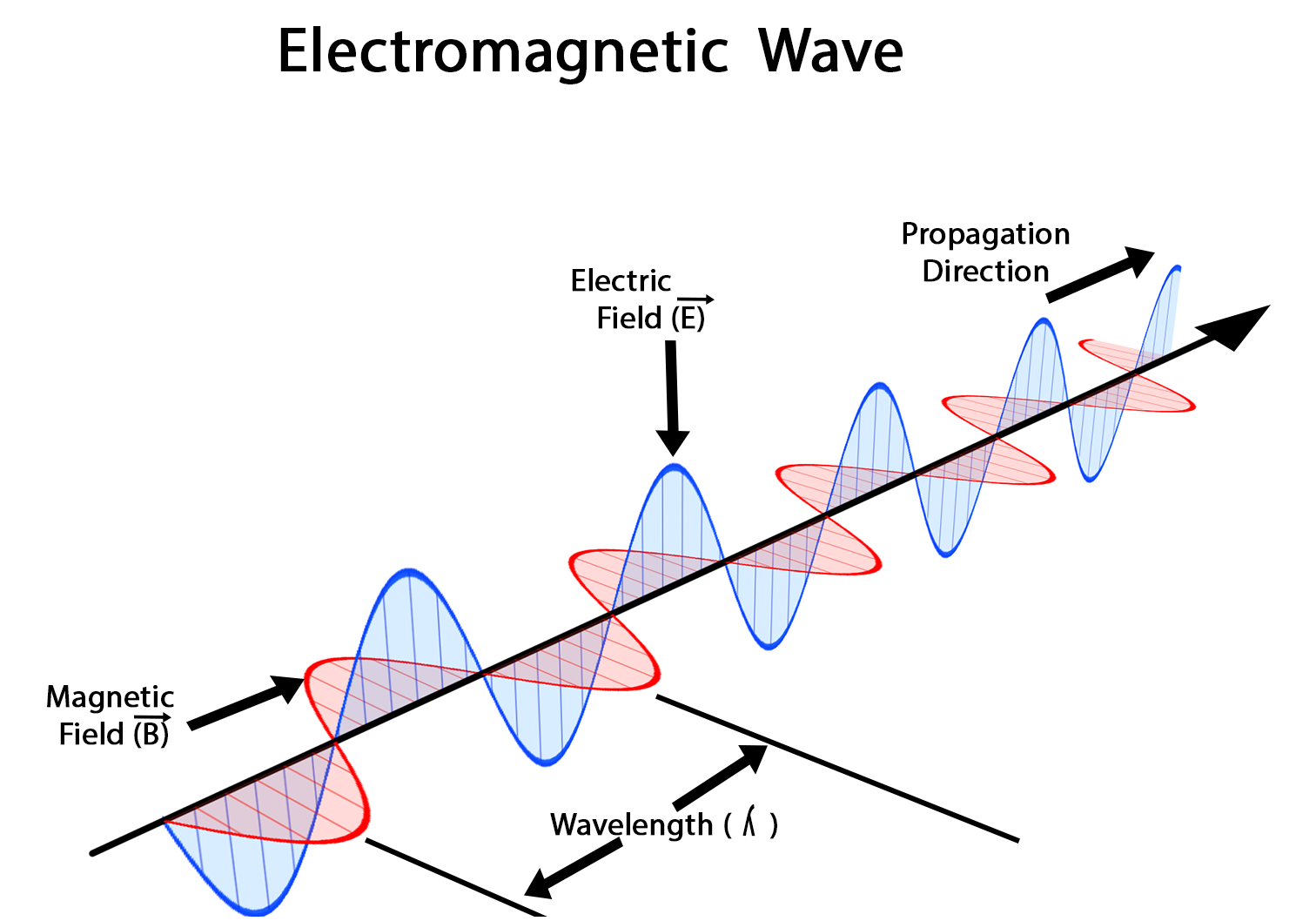}
    \caption{EM Radiation \ccbysa~\cite{EMDiagram}}
    \label{EMradfig}
\end{figure*}

EM radiation is strongly tied to human existence and generated by various electrical systems, including mobile phones, the Internet of Things (IoT), wearable devices, communication base stations, electronic devices, and other EM technology~\cite{9016183}. EM radiation is a type of energy that is produced when electricity is transmitted through conductive material. The fundamentals of wave theory show that all EM radiation has common characteristics and acts in predictable ways. 
An EM radiation consists of electrical (E) and magnetic (B) fields that are oscillating in two perpendicular planes.
The resulting EM wave travels in a direction perpendicular to both E and B at the speed of light in vacuum~\cite{ishimaru1144/3294442}. Figure~\ref{EMradfig} depicts this phenomenon. Due to the electric and magnetic fields' right-angle intersection, which results in the disturbance, the combination wave moves perpendicular to the oscillating electric and magnetic field. EM radiation is characterized by a magnetic field (B) directed at a right angle to the electrical field and an electrical field (E) whose magnitude varies in a direction perpendicular to the direction of the radiation.

Each electronic device, e.g., IoT devices, can gather data, transmit it to other devices using wired or wireless networks, and receive data from other devices utilizing EM waves~\cite{rizvi2022network}. Several technologies may be used, depending on the needs of wireless communication range and data transfer rate, e.g., WLAN (Wi-Fi), satellite communication, Radio Frequency Identification (RFID), mobile telephone system (Cellular communication), Bluetooth, ZigBee, and global positioning systems (GPS)~\cite{8027264, zamanian2005electromagnetic}. However, any vulnerability in the hardware at the core of a data system might seriously compromise its security. Three general categories can be used to group EM radiation-related vulnerabilities~\cite{hayashi2019introduction, ott2011electromagnetic}:

\begin{enumerate}
    \item Unintentional EM radiation generated during information processing 
    \item Intentional EM interference with electronic devices
    \item Unintentional alteration of the system's hardware causing variations in their EM radiation patterns
\end{enumerate}

The EM radiation signals from these parts have a substantial amount of side-channel information about the data processing and software implementation events. 
The microcontroller (MCU) chip is the most significant hardware component onboard that acts as an EM radiation source on the majority of IoT devices because it stores both the CPU and RAM~\cite{SAYAKKARA201943, 7857107}.

\subsection{EM Noise}

EM noise is a term used to describe unintentional EM radiation that is generated by various electrical and electronic devices. This type of noise can result from a variety of sources including electronic components, power supplies, cables, and connectors~\cite{poulin2002interference, Longo2015}. The operation of other electronic devices in the same environment can be hampered by the emission of EM noise, which is often encountered in electrical device-busy environments. The Industrial Internet of Things (IIoT) has emerged as a critical competency of a country and is a significant area of attention due to its potential to improve efficiency in various industrial processes. However, the EM noise generated in industrial environments is diverse and different from other locations like office spaces~\cite{7433518} and this diversity can result in increased levels of EM noise. The presence of EM noise can negatively affect the functionality of electronic components in the environment, due to the electrical and magnetic fields of these components interfering with each other. This noise is a result of the large amounts of data generated by various sensors used to track processes and anticipate errors. This data flow can result in EM interference (EMI), which can negatively affect the functionality of electronic components in the environment. 

The performance and security of electrical systems can be impacted by two types of EMI: unintentional and intentional. Unintentional EMI refers to the emissions from electrical equipment as a by-product of normal operation, while intentional EMI refers to deliberate emissions with the intention of disrupting equipment. Both types of EMI are critical areas of concern and require ongoing research efforts to minimize emissions and develop mitigation techniques. The performance and security of electrical systems are affected by two types of EMI: unintentional and intentional~\cite{lapinsky2006electromagnetic, Sayakkara2018}. Intentional EMI is defined as deliberate emissions with the intent to interfere with equipment, whereas unintentional EMI refers to unintentional emissions from electrical equipment. Through EM compatibility (EMC) design strategies, jamming countermeasures~\cite{song2014modeling}, and secure communication protocols~\cite{radasky2004introduction}, research in both areas is focused on reducing emissions, raising susceptibility, and developing mitigation strategies. Radio frequency (RF) radiation is a significant contributor to unintentional EMI~\cite{rahimpour2021deep}. RF radiation is produced by digital devices that use clocks, oscillators, or other high-frequency pulses, and these devices are known to generate unintended emissions. The leaked state information from these emissions can be used to determine the internal status of the emission-producing device, potentially compromising confidential and sensitive information, including cryptographic key material. The security measures of vulnerable electronic devices are significantly affected by this realization, since the \emph{leaked} state information is frequently essential for determining precisely what operations the device is performing and what data it is processing. Such information may comprise very confidential and sensitive information, including cryptographic key material. 

However, attackers are constantly using intentional EM interference due to EM waves that can disrupt or even harm digital devices and are produced by attackers at strengths above those covered by EMC. In addition, any source of extra electrical load requires energy and produces EM noise, both of which are detectable from the outside. Based on the well-known side-channel attacks approach, the side-channel monitoring device system is being developed with an emphasis on power supply tracking and information extraction for spotting unusual activity on Internet-connected devices~\cite{SAYAKKARA201943}.

\citet{VANECK1985269} developed the first eavesdropping method that took advantage of unintended emissions. EM emissions from Cathode Ray Tube (CRT) video interfaces can be remotely collected and reassembled by an eavesdropper at a significant distance. The attack is conducted by passively seeing a real-time duplicate of what is being displayed on a remote target computer using low-cost commercially available receiver technology. Van Eck's attack has now been expanded to demonstrate that it is still applicable to flat panel technology today~\cite{10.1007/11423409_7}. The following section describes how EM signals are produced by various electronic device components, what data they might include, and what techniques and instruments could be employed to acquire these signals.

\subsubsection{Electronic Device that emits EM radiation}

Electronic devices emit EM energy, which is widely recognized to cause interference, i.e., EMI, with surrounding devices. Government agencies control such unintentional EM radiation, and consumer electronics must pass certification tests on complying with the Federal Communications Commission (FCC)~\cite{5898407}. Due to the integrated circuit operations, e.g., clock distribution and transistor switches, currents flowing across the device generate EM fields that are produced by dynamic interaction and transmit as time-varying EM waves via radiation and conduction. It is well known that since their inner operations, electronic devices can emit EM radiation on undesired frequencies, as explained by Maxwell's equations~\cite{10.2307/108892}.

When an electronic device, e.g., a communication device, a personal computer, or random access memory (RAM), operates in a coordinated, sequential manner in accordance with clock signals, it frequently creates inadvertent radiated EM fields that interfere with surrounding devices. In contrast, emissions in the area around such equipment frequently reduce its performance~\cite{Lavaud2021}. Due to a target device's dependence on different hardware features, its EM emission frequencies are unpredictable. Consequently, it is challenging to create a general-purpose device with the capability to capture EM emissions from a variety of systems and retrieve side-channel information. The following subsections discuss a number of tools that can be utilized to detect EM emissions from electronic devices.

\subsubsection{Software-defined Radio}

A radio communication system classified as a software-defined radio (SDR) system processes various EM signals on the software domain, e.g., modulation, demodulation, and decoding~\cite{5404400}. These radio devices are much more adaptable and versatile than typical radio communication systems. Research and development related to mobile communications make up the majority of SDR use cases. An analogue front end and a digital back end constitute the majority of SDR systems~\cite{7086416}. An analogue front-end controls a radio communication system's send-and-receive operations. The highest bandwidth SDR platforms typically go as high as the DC-18 GHz range and are built to operate over a high sampling rate as well. The Realtek RTL2832U controller and tuner integrated circuit is a common (and affordable) SDR receiver for Digital Video Broadcast (DVB-T) transmissions. They were originally designed to receive video, but they have since been converted to receive radio signals and are now referred to as RTL-SDR devices.

One example SDR is the HackRF One. This device was created to facilitate the testing and development of contemporary and next-generation radio technologies. It can transmit or receive radio signals between 1 MHz and 6 GHz~\cite{GreatScottGadgets}. The device can sample signals at a rate up to a maximum of 20 MHz. HackRF One is categorized as a mid-range system due to its capabilities. For educational purposes and convenient research projects, it serves as transceivers with a full-featured system and suitable sample rates and spectral bands, which comprise the frequencies of FM radios, GSM, and WiFi systems~\cite{9084568}. The HackRF One is capable of acting as a USB peripheral to be controlled by software running on a PC, or it can be configured for autonomous operation. In the former case, HackRF One serves as the computer's sound card. It converts between analog EM waves and digital samples, making it possible to incorporate extensive communication networks. It is intended to evaluate, improve, improvise, and change modern radio frequency systems.

\subsubsection{Oscilloscope}

An oscilloscope is a type of electronic test tool that, according to its definition, enables the analysis and observation of continuously differing signal voltages as a two-dimensional graph of one or more electrical potential variations using the Y-axis and depicted as a function of time on the X-axis. When an electrical signal changes over time, the oscilloscope is applied to monitor it. As a result, voltage and time describe a structure that is continually graphed against a calibrated scale~\cite{kularatna2003digital}. An oscilloscope is a piece of electronic test equipment that makes it possible to observe waveforms, which makes it much more efficient to notice any issues that might be present in an electronic circuit. Simulating an electrical signal of any complexity is possible with a simple modification of values for amplitude, frequency, phase, and other variables.

\subsection{Side-Channel Attacks}

\begin{figure*}[htp]
    \centering
    \includegraphics[width=0.8\textwidth]{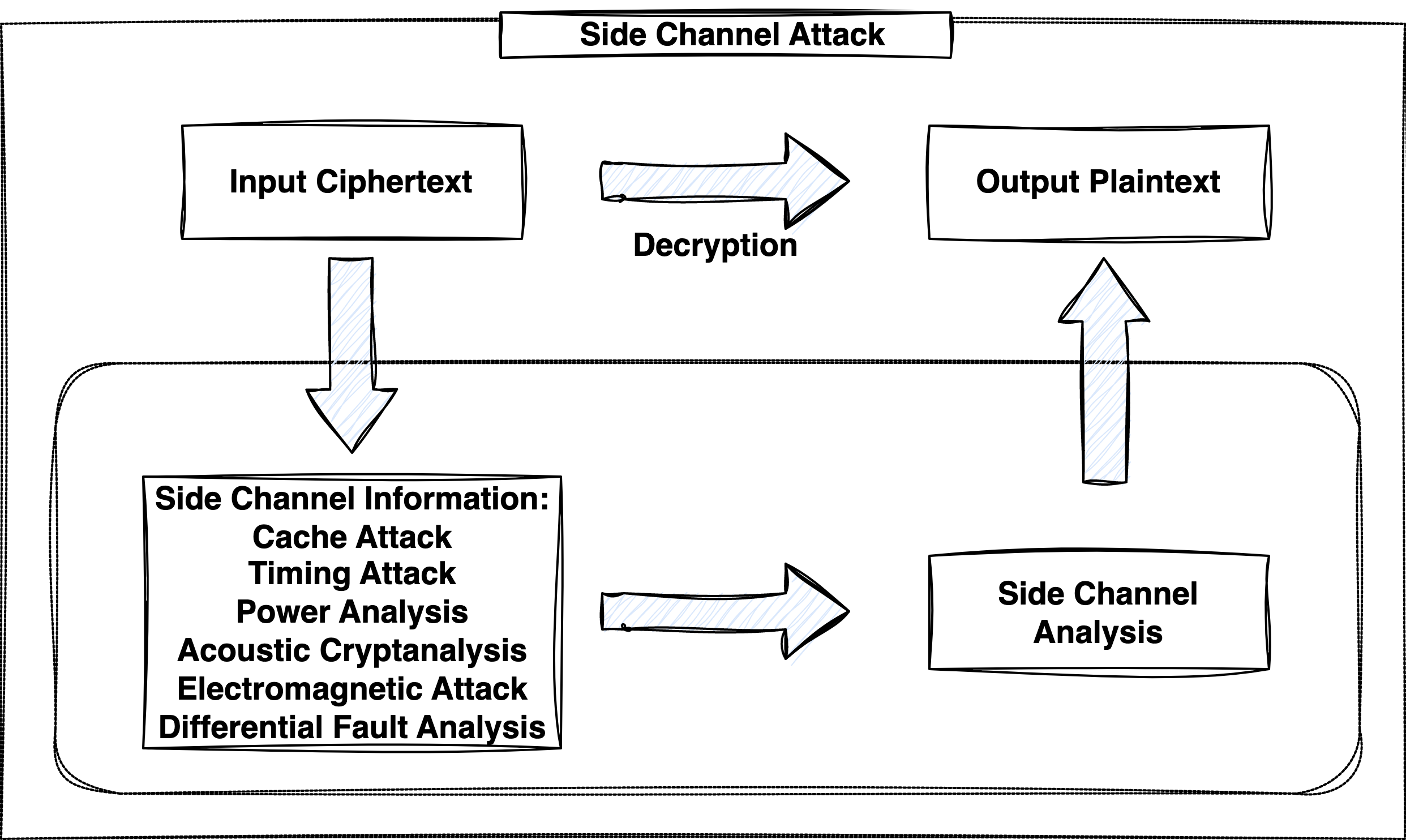}
    \caption{Side Channel Attack (adapted from~\cite{VanathiSidechannelIaas})}
    \label{fig:sidechannelattacks}
\end{figure*}

In 1996, \citet{kocher1996timing} developed various types of side-channel attacks. These attacks are essentially a form of physical attack that hackers use to access cryptographic devices. Embedded electronics are vulnerable to complex hardware-based security threats. Without using the system's ordinary interface, side-channel attacks gather data on a system's operational activity. SCA are a subset of the more general class of implementation attacks which exploit vulnerabilities in a device’s physical implementation rather than attacking, for instance, the mathematical strength of a cryptographic algorithm. To put it differently, by examining physical characteristics of electronic devices, SCA is a technique for acquiring information from them. In order to discover internal computations, side channel attack uses external representations including processing time, power consumption, and EM emission~\cite{4798277}. Figure~\ref{fig:sidechannelattacks} depicts how a side-channel attack could be used to discover the specifics of a typical application workflow and exploit side information to obtain the output (e.g., the plaintext from a ciphertext).

Side-channel attacks (also called implementation attacks or sidebar attacks), which were traditionally considered challenging to execute, have become increasingly widespread~\cite{Won2021, Ueno2022, lisovets2021let}. It is now possible to collect incredibly precise data about a system during operation due to the ever-increasing sensitivity of measuring devices. SCA are often passive, which allows the attacker to use them without drawing attention to themselves or physically harming the system of interest. This sets them apart from many other implementation attacks~\cite{anderson2020security}. This is in contradiction to many other implementation attacks that are either active, which involve the threat of disclosing the attacker's presence or intent, or invasive, which run the risk of damaging a circuit or activating its tamper-resistant mechanisms. Increased processing capacity and machine learning also allow attackers to understand the raw data better. Since they are more familiar with the systems being attacked, attackers are adequately equipped to exploit subtle system changes. SCA vectors are becoming more widely spread. Among the frequent attacks are:\\

\textbf{Timing attack:} Depending on the inputs, the key values, and the mathematical operation itself, distinct mathematical calculations in cryptographic systems may calculate in a range of times~\cite{batina2019csi}. These temporal differences are the target of timing attacks, which aim to extract information.\\

\textbf{Electromagnetic attack:} Measures EM radiation released by a device and analyses the signals it sends out. These attacks, which are often known by the name Van Eck phreaking, are a more focused iteration with the aim of obtaining encryption keys. Attacks involving electrical radiation are frequently non-intrusive and passive, which allows them to be carried out while the target device is still operating normally and without performing actual physical harm~\cite{SAYAKKARA201943}.\\


\textbf{Power analysis attack:} A power analysis attack is a method of compromising security in cryptographic systems by exploiting variations in the electrical power consumption of a device during cryptographic processes~\cite{10.1145/2046707.2046722}. Simple power analysis (SPA) and Differential power analysis (DPA) are two variants of this attack category that rely on observing and analyzing changes in a device's power consumption~\cite{10.1007/3-540-44499-8_6}. SPA involves directly evaluating the power consumption data collected during a cryptographic process, typically through visual inspection of graphs showing the current consumed by the device over time. In contrast, DPA uses statistical techniques to derive encryption keys from crypto-algorithms by analyzing changes in power consumption. While the specific techniques used in SPA and DPA may differ, both attacks involve exploiting the information revealed by a device's power consumption during cryptographic processes.

\subsection{EM Side-Channel Attacks (EM-SCA)}

\begin{figure*}[htp]
    \centering
    \includegraphics[width=0.8\textwidth]{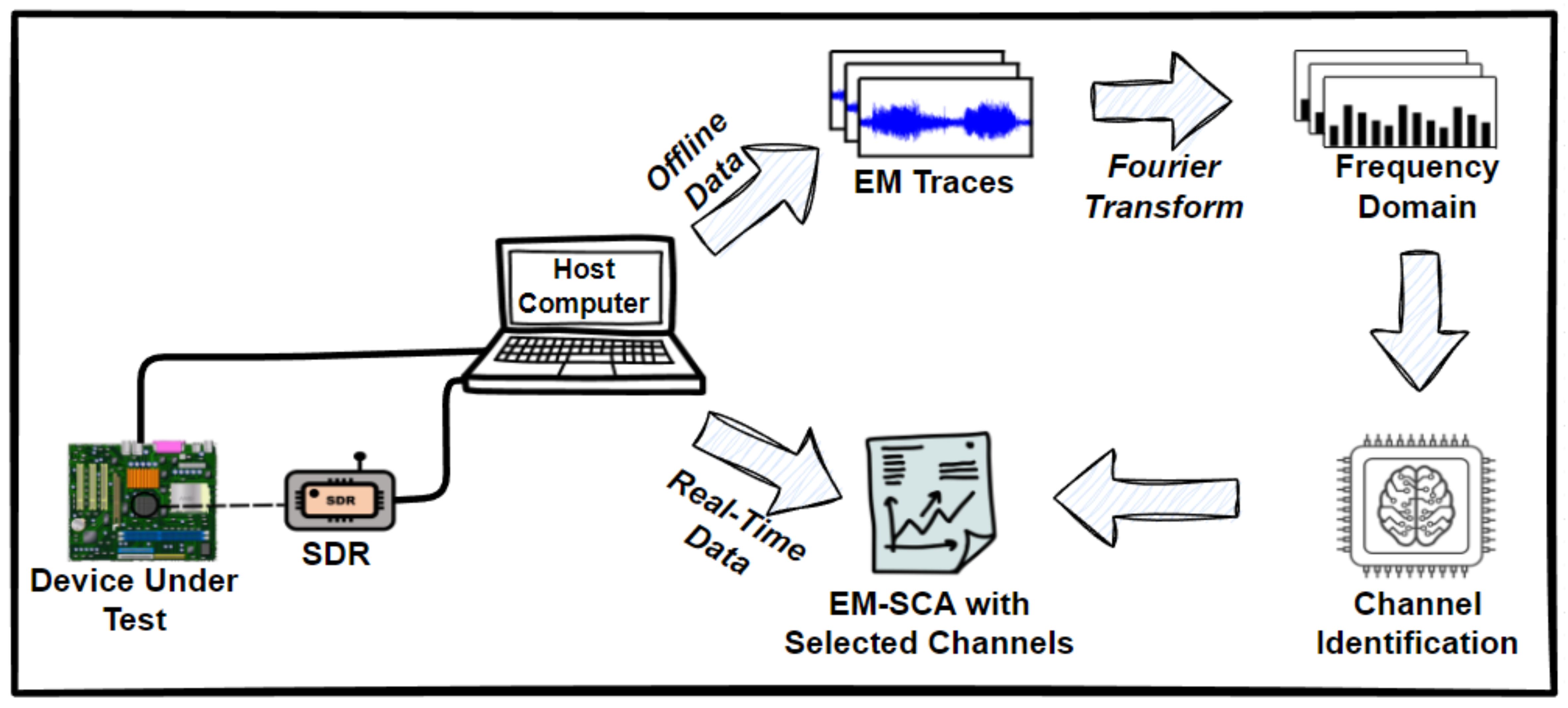}
    \caption{The Workflow to Generate EM traces, Identify Channels, and Subsequently Perform EM-SCA with Pertinent Channels (adapted from~\cite{Sayakkara2020})}
    \label{fig:emsidechannel}
\end{figure*}

Electric currents in conductors that change over time emit EM waves into the surrounding space. Electronic device unintentionally produces EM emissions during their internal systems since they are made up of electrical circuits. EM-SCA makes use of the EM radiation that electronic devices unavoidably leak throughout various computations~\cite{Aghaie2020, Japa2021}. Even cryptographic algorithms like RSA and AES, which are thought to be theoretically impenetrable, can be compromised by analyzing the patterns of EM radiation they cause~\cite{Sayakkara2018a}. Chip cards can be used for EM-SCA, which makes use of antennas that are positioned extremely near to the electrical component. As opposed to power analysis, which requires a resistor in the power supply channel, this method makes use of the EM radiation that every embedded device emits. It just required a simple improvised loop antenna to demonstrate that a superior attack could be carried out with some analytical fundamentals as needed for power. A single EM sensor might record a variety of EM signals, since the EM emission may vary depending on the physical parameters of the active circuits. These signals may each be isolated and examined separately. Regardless of the embedded system, the chip often includes a CPU, clock unit, memory, Flash, and other components.

EM side-channel attacks can give the attacker a smooth entry point without the target device executing any specialized software or allowing access to its internal hardware. An attacker has to be capable of accurately collecting EM radiation in order to utilize them as a side-channel source of information. Oscilloscopes and spectrum analyzers are the main equipment used by experts in radio frequency (RF) engineering and related disciplines to monitor EM emissions from electronic devices for tasks like EMC testing~\cite{Sayakkara2018a}. However, due to their cheaper cost and simplicity of usage with programmable software components, software-defined radios (SDR) are becoming the preferred equipment among EM side-channel security researchers~\cite{SDR123300/2394827779}. It is feasible to visually differentiate individual CPU activities when a target device's obtained EM signal or EM trace which it is shown as a waveform or as a spectrogram. Figure~\ref{fig:emsidechannel} illustrates the workflow to generate EM traces, identify channels, and finally perform EM-SCA with selected channels.

\subsection{Alternative Approaches to Breaking Encryption}

While EM-SCA is one viable approach to attacking encryption on a range of devices, a number of alternative approaches exist. While not the focus of this paper, this section provides a brief overview of these alternatives to appreciate the context for the systematic literature review that follows.

\subsubsection{RowHammer}
Row Hammer attacks allow a software-based attacker to execute bit flips in DRAM places that are not directly accessible. As they undermine the assurances of process isolation offered by existing operating systems, these attacks pose significant vulnerabilities~\cite{7927156}. Kim et al.~\cite{10.1145/2678373.2665726} claim that they establish the RowHammer issue in DRAM, which is a remarkable and likely the initial example of how a circuit-level failure mechanism can lead to a widespread and applicable system security issue. 

The phenomena known as RowHammer occurs when a row in a contemporary DRAM chip is repeatedly accessed, causing bit flips at reliably predictable bit locations in physically nearby rows~\cite{8708249, 9407238}. The attacks are effective because capacitors that are close together release electrical charges that contain the bits more rapidly. These bit flips were once merely an uncommon crashing phenomenon that was only known to be brought on by cosmic rays. But Rowhammer may have negative implications on the security of the devices that employ the vulnerable chips when it is surgically induced~\cite{9138944}. It is brought on by a hardware flaw known as DRAM disturbance defects, which is a symptom of scalable memory technology's circuit-level cell-to-cell interference.

\subsubsection{Cold Boot Attacks/Liquid Nitrogen Memory Analysis}

Dynamic random-access memory (DRAMs) are assumed to lose their data as soon as the system shuts down, however, studies have shown that they are capable of preserving data for several seconds after power loss, with only a small portion of data being discarded~\cite{gupta1108442/1500422}. There is evidence that such data preservation in DRAMs poses a security risk~\cite{10.1145/1506409.1506429}, since systems that rely on data encryption and passwords frequently store sensitive data in DRAM with the expectation that a restart or removal of the DRAM will delete the data. However, in 2008, \citet{10.1145/1506409.1506429} demonstrated that by moving memory modules from a secured system onto an attacker's machines, disc cryptographic keys could be obtained from DDR and DDR2 DRAMs. They used commonly available compressed air spray cans to cool the DRAMs before moving them to another machine, since charge degradation in capacitors slows down dramatically at lower temperatures. The term ``cold boot attack'' refers to this method. A cold boot attack takes place when an attacker gains access to the computer and uses a cold reboot technique to exploit the running operating system in order to acquire the user's specific sensitive information. The cold reboot method is a technique used to restart the computer from an entirely ``cold'', or powered-down, status. The focus of a cold boot attack is typically on extracting information from DRAM and static random-access memory (SRAM), which are accessible for a short period of time after a power outage in the system.

\subsubsection{Acoustic Attacks}

An acoustic attack is a kind of side-channel attack that takes advantage of the sounds produced by computers or other electronic devices. Acoustic attacks are concentrated on computer keyboards~\cite{Halevi2015, 10.1145/2660267.2660296} or other devices, such as the keypad of an ATM (Automated Teller Machine), where the user needs to enter a private password or PIN (Personal Identification Number)~\cite{1301311}. In~\cite{10.1145/3211960.3211973}, the attack used an optical microphone that monitored a computer cover's vibration during typing. The classification technique used in this attack was focused on the contrast of unique acoustic ``fingerprints'' in the temporal domain. This method makes use of the fact that the space bar's acoustic fingerprint is simple to recognize, making it simple to discriminate between typed text and correlate the distinctive acoustic traces to the individual letters.

Additionally, \citet{Shahrad2018} have shown that attackers can physically damage hard drives and crash PCs simply by emitting sounds through a speaker. Magnetic hard disc drives, which are frequently used in desktop and laptop computers, have platters and read/write heads that can vibrate when audible sound waves are involved. Too intense Vibrations can harm software and hardware, corrupting file systems and prompting restarts. The researchers even demonstrated that an attacker could use a device's constructed speakers or nearby speakers to generate such continuous vibrations~\cite{Shahrad2018}. Furthermore, other approaches to compromise systems have been discovered by \citet{savage2015visualizing}, which may apply to the Enterprise Internet of Things (E-IoT). For example, \citet{savage2015visualizing} demonstrated that an attacker can employ passive sound recovery to eavesdrop on talks using recorded video, i.e., from a CCTV system or intercom system. Additional research by \citet{davis2014visual} showed that in certain circumstances, i.e., those involving visible glass or water, an attacker might leverage vibrations on object surfaces for eavesdropping~\cite{davis2014visual}.

\subsubsection{Password Attacks}

Implementing a strong password strategy is one of the most effective ways to prevent attackers from getting unauthorized access to a system or user account~\cite{KANTA2020301075}. However, because it is frequently easy to launch attacks against and effectively breach, it is also one of the security measures that attackers target the most when trying to gain unauthorized access to a system~\cite{Kanta2021, 9138870}. Once an attacker has gained access to the system, they can either acquire or damage sensitive or valuable data, alter any data or system configurations, or simply disable the system by destroying or encrypting data with ransomware. Additionally, after they have gained access to a system, attackers can employ methods to increase their credentials so that they have access to more valuable data. They can also use the system they have already compromised to access other networks' systems and seize more precious information. In order for everyone to comprehend and be cautious of unauthorized access or password attacks, a variety of password attacks are outlined below. Many of these approaches can now be enhanced further through leveraging large language models for candidate key/password generation~\cite{SCANLON2023301609}.

\textbf{Brute Force Attacks:} A brute force attack is a form of trial-and-error attack approach whereby attackers attempt all or most of the password or passphrase possibilities with the intention of successfully breaking user accounts. Brute force attacks can be carried out in a wide range of ways, starting from techniques that randomly or sequentially evaluate every password combination to more efficient methods that employ the most commonly used or previously obtained passwords. The brute force technique is commonly used to break encrypted passwords, where the credentials are stored as encrypted text~\cite{CHO201558}.

\textbf{Dictionary Attack:} A dictionary attack is a form of brute force attack in which a malicious attacker attempts to log in to one or more accounts employing a dictionary attack list of frequently used passwords used by individuals and corporations alike. These dictionary lists often stem themselves from data leaks, and as a result, are based on real-world examples from users. Dictionary attacks are brutally efficient due to common poor security user behaviors such as repeating passwords, variations thereof or utilizing common ones. Attackers also might examine individual users' online presence (from websites to social media accounts) to ascertain their interests, identify prominent terms or phrases, and add them to dictionary attack lists in more focused dictionary attacks~\cite{KANTA2020301075}.

\textbf{Keyloggers:} Keyloggers, also known as keystroke loggers, are pieces of software that record each keystroke a user makes and transmit the data back to the attacker. The attacker has to use a phishing attack, a Trojan horse, a drive-by download, or another malware in order to exploit the victim's device with malicious code~\cite{6481194}. The malware exploits the system without being spotted when the user clicks on the link or attachment. Keyloggers record everything the user inputs, including usernames, passwords, PINs, and credentials. Credit card information, probably personal information used to answer security questions, might all be sent to the attacker.\\

\subsection{Encryption Algorithms}


\begin{figure*}[htp]
    \centering
    \includegraphics[width=0.8\textwidth]{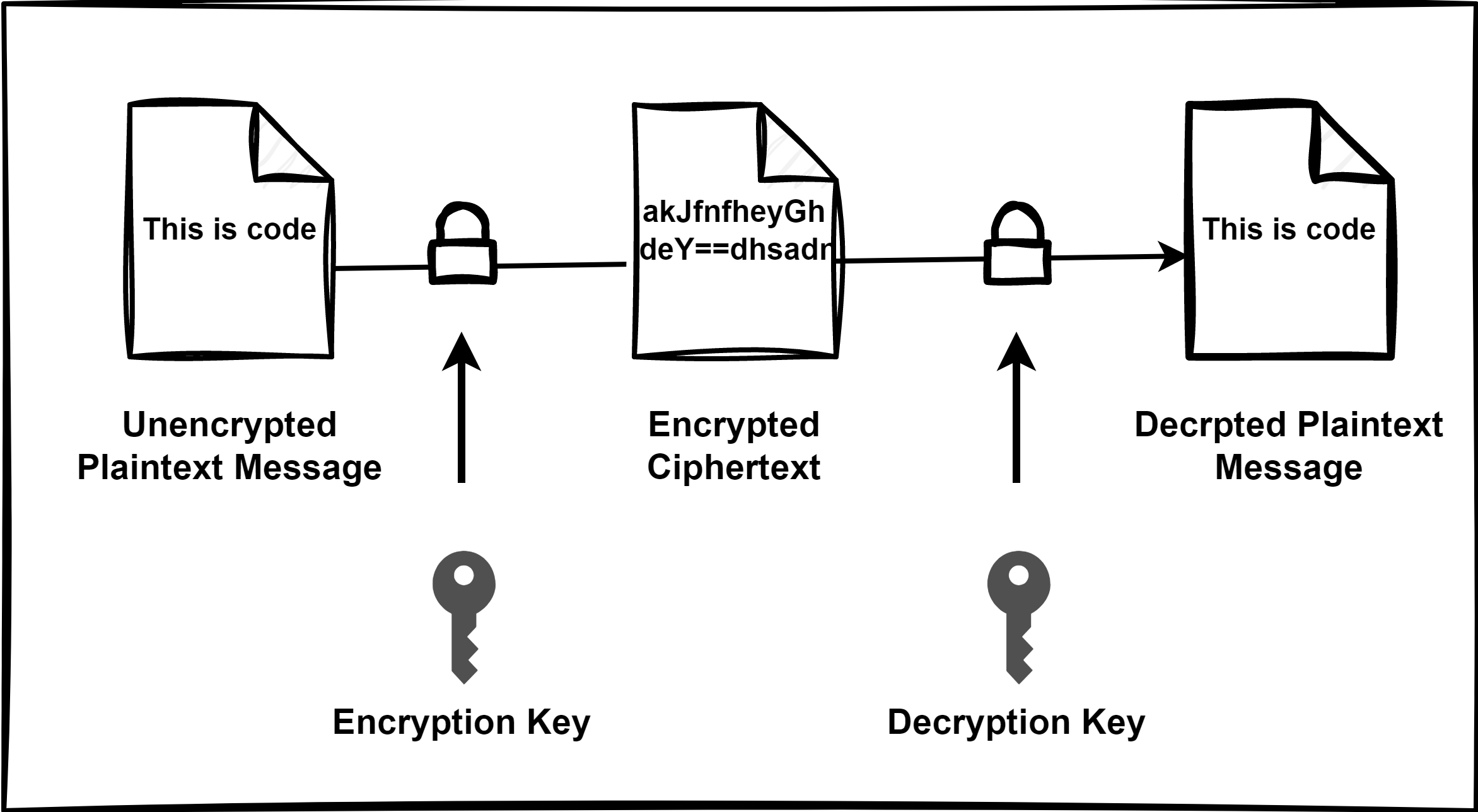}
    \caption{General Principles of How Encryption Operates}
    \label{fig:emscaonencryption}
\end{figure*}

Cryptographic algorithms are mathematical functions used for converting plaintext into ciphertext and vice versa in order to secure electronic data from unauthorized access and tampering. The process involves the use of key management, where secret keys are shared between communicating parties. There are two types of keys used in cryptography: shared keys and public and private key pairs~\cite{7163029}. Shared keys, used in symmetric encryption, are known keys shared by both parties. Public-key encryption employs private and public key pairs, with the private key used to decrypt messages and produce digital signatures and the public key used to encrypt messages and validate digital signatures. Figure~\ref{fig:emscaonencryption} illustrates how the information is transformed by the application of encryption methods and decryption keys from plaintext to ciphertext and back to plaintext. The keys may be either symmetrically identical or unique (asymmetric). In this section, five widely used encryption algorithms will be discussed: Advanced Encryption Standard (AES), Data Encryption Standard (DES), Elliptic Curve Cryptography (ECC), ElGamal Encryption Algorithm (EEA), and Rivest-Shamir-Adleman (RSA).

\subsubsection{AES}

The cryptographic algorithms can be symmetric, in which case the sender and receiver exchange the secret key, or asymmetric, employing a public/private key pair. However, due to performance requirements, data encryption typically depends on symmetric algorithms, even when asymmetric techniques are employed. The Advanced Encryption Standard (AES), which was established by NIST in 2001~\cite{pub2001announcing}, is a common cryptographic algorithm that can be used to encrypt sensitive data in order to protect it from the adversaries~\cite{Yu2018a,lerman2015machine}.

\subsubsection{DES}

The Data Encryption Standard (DES) is a symmetric encryption algorithm that was developed by IBM in the 1970s. It operates on 64-bit data blocks and uses a 56-bit user key for encryption and decryption. According to \citet{yang2004scan}, the DES encryption process is carried out in three phases. The first phase involves bit permutation of the 64-bit plaintext block and storing it in two 32-bit registers (L and R). The second phase involves the application of a round operation, composed of function f and exclusive-or, 16 times on the 32-bit R and the 48-bit round key. The function f performs various operations, including expansion, exclusive-or, substitution, and permutation, to generate a 32-bit output. The final phase involves concatenating and permuting the two 32-bit outputs of round 16 to form the encrypted output. The sixteen round keys required for the encryption process are generated from the 56-bit user key through a simple bit-permutation and shift operation~\cite{standaert2010introduction}.

\subsubsection{ECC}

Elliptic Curve Cryptography (ECC) is a public-key encryption algorithm that uses a set of mathematical operations over a finite field to produce a public and private key pair. The ECC approach is advised for authentication, cryptographic signature, certificates, and other applications in resource-constrained situations due to their effective processing and short key size~\cite{ azarderakhsh2014efficient, mahmood2018elliptic, Fournaris2017}. Although these methods are mathematically and theoretically solid, poor implementation can expose users' data to side-channel attacks.

\subsubsection{EEA}

For use in the European Payments System (EPS), the symmetric encryption algorithm EPS Encryption Algorithm (EEA) was developed. The EEA is a 128-bit block-based substitution-permutation network (SPN) encryption technique. It is made to be quick and secure and uses a 128-bit key for encryption and decryption. \citet{6336699} provides a comprehensive security analysis of the EEA, which highlights the strength of the EEA against various known attacks and shows that the algorithm is secure. Also, \citet{sulaiman2014comparative} provides an overview of the design and security of the EEA. This paper provides valuable insights into the design of the algorithm and its suitability for use in secure payment systems.

\subsubsection{RSA}

The Rivest-Shamir-Adleman (RSA) algorithm is a widely used public-key cryptography algorithm that uses two large prime numbers to produce a public and private key pair. The public key is used for encrypting messages, while the private key is used for decrypting them. RSA is widely used for digital signatures and secure online transactions. RSA is considered to be secure, but it is also relatively slow compared to other encryption algorithms~\cite{barrett2000implementing}. \citet{genkin2014rsa} showed that it is possible to distinguish between CPU operations by listening to acoustic emanations resulting in an attack on the cryptographic keys of the RSA algorithm.

\section{Comparative study of EM-SCA attacks on common encryption algorithms}

\begin{figure*}[htp]
    \centering
    \includegraphics[width=1\textwidth]{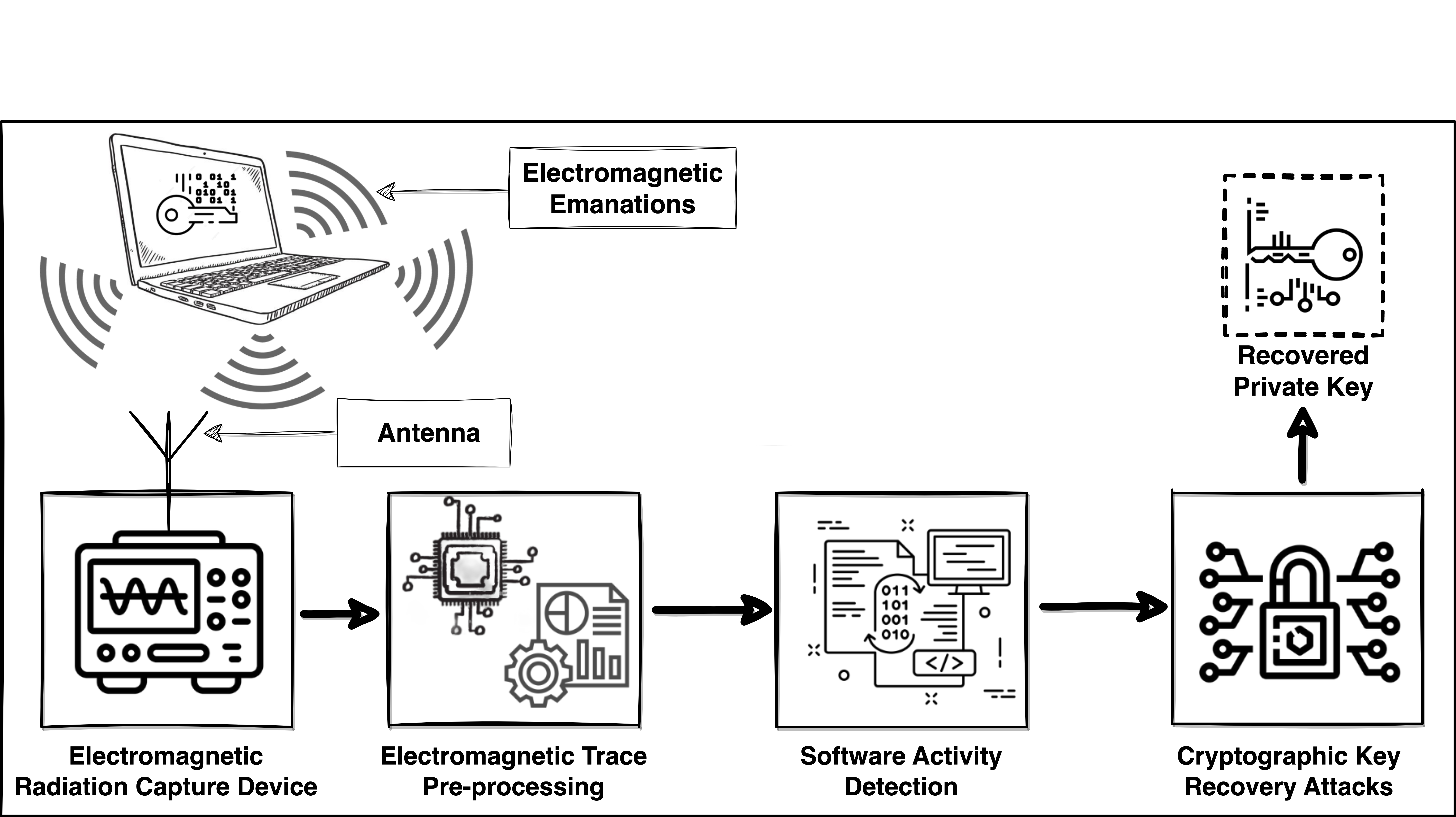}
    \caption{Illustration of EM-SCA Approach for Encryption Recovery (adapted from~\cite{Sayakkara2019a})}
    \label{fig:EM-SCAonEncryption}
\end{figure*}

This section examines how the literature has used EM-SCA to attack encryption in recent years, as well as which encryption algorithms and encrypted devices are most vulnerable to and resistant against EM-SCA. This analysis incorporates sample EM-SCA implementations for several encryption algorithms, e.g., AES, DES, ECC, EEA, and RSA, that must be considered when using EM-SCA on various devices and encryption algorithms.  Figure~\ref{fig:EM-SCAonEncryption} illustrates an example of an EM-SCA approach to breaking an encrypted device, which is a common technique used in the literature to analyze the vulnerabilities of encryption algorithms and devices to EM-SCA. 

\subsection{EM-SCA on AES Implementations}

The Sub-byte (S-box) layer's modification in the final round is the primary target of side-channel attacks on the Advanced Encryption Standard (AES). Such attacks generally target only one key byte at a time. In each key value, the attacker generates a hypothesis of the observable signal behavior, of which there are 256 possible cases~\cite{li2019securing}. Furthermore, In 2018, \citet{Yu2018a} has shown that the AES cryptographic circuit's secret key can be retrieved via a deep learning-based side-channel attack. The researchers were able to model the correlation between EM and power noise using a deep neural network (DNN), based on an analysis of the collected EM emissions and power degradation. This allowed them to effectively discover the secret key by examining approximately 32,500 plaintexts. It is significant to highlight that the security of the encryption technique depends on the modification made to the S-box layer during the most recent round of AES. As shown by the DNN-based attack approach suggested by \citet{Yu2018a}, any vulnerabilities in this layer can be used by attackers to reveal the secret key.

To quickly and accurately anticipate EM-SCA resilience during the design phase, \citet{kumar2017efficient} provide an effective simulation setup for DEMA of cryptographic modules. Without losing predictive value, the simulation cost is decreased. The suggested solution makes use of industry-standard computer-aided design (CAD) tools to do highly efficient transistor-level simulations only during crucial time periods when information leakage occurs. The EM radiation is restricted to the currents dispersed on the top metallization layer power/ground (P/G) interconnects in order to shorten the wall-clock time further, and parallel traces are formed for various encryption operations. \citet{kumar2017efficient} simulate various differential EM attacks on the AES block encryption using this approach. Their study indicates that a uniformly spaced P/G network is a bad design choice and more vulnerable to AES execution than a non-uniformly spaced P/G network. For instance, the number of encryption operations needed to crack the key is cut in half for a probe placed 100 micrometers from the chip surface. In order to improve the EM SCA robustness of cryptographic modules and discover trade-offs during the physical implementation of ciphers, developers might benefit from their technique.

Furthermore, \citet{Vasselle2019} explains how a secret AES key was acquired from a smartphone's hardware crypto processor. Several parts of a smartphone's entire surface can include sensitive information about the computation: Power Management Integrated Circuit (PMIC), external decoupling capacitors, sense resistors, and system-on-chip die. A physical side-channel attack on an ARMv8 Cryptographic Extension (ARM CE)-based AES implementation is presented in a publication by \citet{Haas2022}. The usual source of cryptographic capability for iPhones, Apple's CoreCrypto library, which uses the ARM CE code, was the target of the attack. However, employing iPhones as a device for hardware security research is not simple. Hence, they bootstrap the investigation using the iTimed toolkit~\cite{ haas2021itimed}. The authors highlight that their attack can be associated with this approach for all iPhone models up to the iPhone X due to the flexibility of the iTimed toolkit. The test infrastructure has been constructed to enable transmitting inputs and immediately seeing outputs on the main processor after reverse-engineering the iPhone's printed circuit board (PCB). The secret keys can be successfully extracted via an EM-based side-channel attack. This approach must operate on between 5 and 30 million traces or around 80 and 480 MB of data to be successful.

\citet{Iyer2021c} propose a Fischer statistic (F-statistic) analysis of variance (ANOVA) indicator to discover optimal probe configurations for EM-SCA attacks efficiently. The potential of the presented indication to accelerate EM-SCA attacks is evaluated by retrieving the secret key from an FPGA implementation of AES-128 utilizing several probes with different directions, heights, and sizes. The authors demonstrate that AES implementation experiments revealed a substantial reduction in the search area for noisy measurement configurations, aiding developers in quickly identifying, validating, and fixing security vulnerabilities caused by unintended EM emissions. However, \citet{Won2021} employed a high-sensitivity low noise EM sensor with a broad enough bandwidth (100kHz - 2.5GHz) to collect the target activity in order to measure the EM emission. The points specific to AES-128 operation alone are ascertained by executing correlation with side-channel traces and information sharing like plaintext and ciphertext. Therefore, Key retrieval and architecture estimation have both been accomplished using correlation power analysis(CPA). Both the device intrinsic key successfully obtained from PUF with AES-256 and the user-defined key with AES-128 are used to demonstrate the key recovery.

\subsection{EM-SCA on DES Implementations}

In 2008, \citet{Burnside2008} demonstrated the use of EM-SCA to retrieve the secret key of cryptographic operations, such as Triple-DES (3DES) and AES, from contactless smartcards. They propose a novel method for isolating and amplifying the SCA leakage of cryptographic RFIDs through analog demodulation circuitry. The authors demonstrate the efficiency of the setup by analyzing the Mifare DESFire MF3ICD40 smartcard and other real-world devices. Additionally, they evaluate the effect of the proposed signal processing and re-emissions techniques in terms of correlation power analysis (CPA) results, which demonstrate the susceptibility of the DESFire MF3ICD40 to SCA. Advancing the argument, it is clear that EM-SCA poses a severe threat to the security of RFID systems that implement DES cryptographic operation. The work of \citet{Burnside2008} significantly contributes to the field, demonstrating the potential for EM-SCA to compromise DES encryption systems.

Advancing the argument, it is clear that EM-SCA poses a severe threat to the security of RFID systems that implement DES cryptographic operations. This threat is further supported by \citet{Sauvage2009}, which emphasizes the advantages of employing EM Analysis (EMA) over traditional DPA. The research indicates that EM imaging has the advantage of exposing active regions, which can be helpful in pinpointing a specific processor, apparent when active but invisible when inactive. The paper notably illustrates the first images of deep-submicron FPGAs. The research emphasizes that the attacker can be directed to the leakiest places by using the coarse resolution that can be acquired via EM imaging, rather than an accurate localization of the cryptographic target. In less than 6,300 measurements, the authors claim to have successfully attacked a DES module, which is the best-breaking performance against this encryption technique used in FPGAs at the time. The authors used the HZ-15 Probe Set from Rohde \& Schwarz, which is typically used to investigate EM emissions in the near-field to locate ``EMC non-compliant'' components of a PCB. The study describes the precise preprocessing of the traces that must be carried out in order to resynchronize the traces, since a realistic attack lacks a reliable synchronization signal. The research concludes by presenting the most local measurements that very closely match the energy usage of the element that leaks the most data.

Furthermore, \citet{Lellis2017} demonstrates how DPA and DEMA attacks effectively reveal sensitive data stored and processed by cryptosystems. The authors suggest a novel attack flow that can realign traces thrown off by countermeasures like random frequency and temporal shift. Signature extraction, subsampling, and energy computation are the three processes in the proposed flow. The DEMA attack is employed against a hardware implementation of the DES algorithm prototyped in FPGA, and the authors analyze the appropriate segment length to increase the attack's effectiveness. The findings show that when segments are sized to roughly half the clock cycle of the cryptosystem operating frequency, the amount of traces necessary for a successful attack can be reduced by up to 93.69\%. The findings of two case studies the authors conducted to further support their suggested attack flow show a notable decrease in the number of traces needed for a successful DEMA attack. Both studies \citet{Sauvage2009} \& \citet{Lellis2017} highlight the importance of precise preprocessing of traces and the potential vulnerabilities of cryptographic systems to DPA and DEMA attacks.

\subsection{EM-SCA on ECC Implementations}

The subject of ECC side-channel attacks has drawn a significant amount of attention. They benefit from the information that unintentionally leaked from a device that was designed to prevent tampering. \citet{mukhtar2021edge} used a side-channel EM leakage trace collected from an FPGA to conduct the Elliptic curve scalar multiplication (ECSM). The EM probes are employed to collect the leakages, which are then recorded properly to create a leakage dataset for further side-channel investigation. The authors suggest a sophisticated evaluation technique based on deep learning, i.e.,  a multi-layer perceptron (MLP) neural network that can effectively find side-channel leakages. In order to find leaks in the deployed system, the idea is to choose an appropriate attack model, train it via the cloud, and then load the learned model onto the edge device. The study furthermore assesses the viability of machine-learning-based attacks in a real-world setting on an embedded edge chip, where there are few available traces and significant generated noise from adjacent components, i.e., IoT devices. However, their findings demonstrate that for an unprotected version of the ECC algorithm, most of the trained models perform poorly for all key classes except for one, in which the secret key can be obtained with an accuracy of 70.6\% using a Support Vector Machine (SVM). The results of the study demonstrate that EM side-channel attacks can still be effective against ECC algorithms. The authors suggest that to improve the effectiveness of these attacks; more data can be collected from the device or generated using techniques e.g., Generative Adversarial Networks (GANs).

Additionally, the SEMA method has revealed that two additional ECC-based algorithms, Elliptic Curve-based Diffie Hellman (ECDH) and Elliptic Curve-based Digital Signature Algorithm (ECDSA), are susceptible to EM side-channel attacks~\cite{Genkin2016}. Many IoT and mobile devices frequently use ECC algorithms to protect the data, since of their minimal computing overhead~\cite{SAYAKKARA201943, mukhtar2021edge}. This reveals that such devices can be analyzed through EM side channels to obtain access to data that is encrypted. In 2016, \citet{Genkin2016a} used low-cost tools and extra signal processing techniques to eliminate EM emissions from a mobile device (iPhone 3GS) at very low frequencies. The attack, directed at the OpenSSL wNAF implementation, successfully recovered a small number of ECDSA nonces. The authors then successfully established a lattice attack for full key recovery with approximately 100 signatures.

\subsection{EM-SCA on EEA Implementations}


The effectiveness of this method against well-known encryption algorithms like AES, DES, ECC, and RSA has been the subject of numerous studies. However, the comprehensive analysis revealed that there had not been much work done on EM-SCA for EEA encryption. Despite intensive searches for pertinent literature on the subject, no studies that specifically examine EM-SCAs on EEA encryption were identified. The fact that EEA encryption is widely utilized in numerous applications, including mobile phones, smart cards, and sensors, among others, highlights this gap in the study. It is essential to emphasize that this knowledge gap offers a chance for future studies to examine EM-effectiveness SCAs on EEA encryption, which may offer insightful information about the security of this encryption algorithm. This review thus emphasizes the need for additional study in this field to improve the comprehension of the limitations and restrictions of EEA encryption in the face of EM-SCA attacks.

\subsection{EM-SCA on RSA Implementations}

Numerous research has shown how EM-SCA has the potential to break encryption. \citet{goller2015side} demonstrated attacks at clock-rate frequencies on public key cryptography on Android smartphones using a naive square-and-occasionally-multiply RSA algorithm, with the phone's shielding plate frequently removed. Their results demonstrate that it is possible to recover the secret key of a Square-and-Multiply algorithm using a far-field antenna by averaging several traces. In addition, \citet{nakano2014pre} exhibited significant attacks on smartphones using naive implementations of square-and-sometimes-multiply RSA and double-and-sometimes-add ECC with the battery cover open, the battery removed, and the probe placed directly over the vulnerable part. \citet{kenworthy2012mobile} illustrates how to attack naive square-and-occasionally-multiply RSA without using any force. These investigations show how susceptible encryption algorithms—especially RSA—are to EM-SCA attacks. Some attacks included physically entering the device, but others were non-invasive, making them challenging to identify. These investigations have revealed how susceptible encryption algorithms, especially RSA, are to EM-SCA attacks, and some of these attacks are non-invasive, making them difficult to detect.

Furthermore, significant attention has been paid to the issue of RSA side-channel attacks, such as the study by \citet{10.1007/978-3-662-44709-3_14}. The authors demonstrated physical side-channel attacks on a well-known software implementation of RSA and ElGamal running on laptop computers. Based on the discovery that the electric potential on the chassis of laptop computers changes in a computation-dependent manner, the study illustrates the application of novel side channels. The authors exploit a variety of side channels, i.e., the chassis potential, EM emanations, and power analyses, to demonstrate how to extract 4096-bit RSA and 3072-bit ElGamal keys from laptops of various models. Additionally, the authors examine two methods for obtaining the prime modulus used in the CRT-based RSA modular exponentiation algorithm. The adaptive chosen-ciphertext side-channel attack is the first method, and acoustic emanations are the second method. The authors additionally examine the mathematical library code in GnuPG and demonstrate why the particular ciphertext used in this implementation results in exploitable, key-dependent leakage. According to the authors, measuring the whole attacks takes a few seconds with medium-frequency transmissions and an hour with low-frequency signals. To achieve an acceptable signal-to-noise ratio in the presence of significant noise, delicate grounding, and impedance-matching issues, the results necessitate careful selection and tuning of the signal capture equipment. The study comes to the conclusion that key-extraction attacks against GnuPG's implementation of RSA and ElGamal can leverage any of these channels. These results highlight the need for greater comprehension of the constraints and weaknesses of encryption algorithms, particularly RSA, in the context of EM-SCA attacks. This information can greatly improve law enforcement's ability to look into and prosecute crimes involving encryption, as well as digital forensic investigations.

\subsection{Recovery Techniques}

Cryptographic keys are among the various data types that can be recovered using EM-SCA techniques and are of major forensic interest. Numerous methods, such as simple EM analysis (SEMA), differential EM analysis (DEMA), and CEMA, have been developed for this purpose.

A fully automated EM-SCA attack framework, SCNIFFER, was developed by \citet{Danial2020a}. It accelerates the process through employing a gradient-search technique to find high-leakage areas on the target device. SCNIFFER is a system that incorporates an EM leakage scanning platform and correlation EM analysis, and can conduct all phases of an attack automatically. The authors formulate hypotheses about secret values, predict the EM leakage of an intermediate variable depending on the key, and then use CEMA to retrieve the secret key. While the hardware performs encryption, measurements (or traces) are taken. After that, the measurements are compared to the anticipated leakage for each hypothesis. For a range of microcontroller architectures and cryptographic methods, the attack only utilizes a small number of traces. The hypothesis with the highest correlation determines the estimate for the secret value.  The authors stated that the minimum traces to disclosure (MTD) is the number of traces required to recover the key in this approach successfully. SCNIFFER maintains effectiveness even as chip size grows or as measures that lower signal-to-noise ratios(SNR), like masking, are implemented~\cite{Danial2020a}.

In 2016, \citet{Pammu2016} an interceptive SCA based on CEMA is presented to attack the 16-byte secret key of the AES-128 engine for IoT applications by constructing a communication link using two Arduino boards with ATmega processors. The key is successfully discovered with the support of HDM and 20,000 EM traces. The implementation under consideration accesses data from a static random access memory, an 8-bit data bus, and FLASH memory, which results in the highest levels of EM radiation.

In the broader literature, \citet{Ding2009} retrieved AES keys on a Stellaris LM4F232H5QD MCU with a 32-bit ARM Cortex-M4F by using a prototype 128-bit AES-ECB software implementation (50MHz). The authors used the Hamming weight attack model to focus on intermediate processes during the initial AES round. A digital television receiver that cost 20 (USD) at the time was modified to gather EM waveforms using a near-field probe and software-defined radio. The AES key bits were then extracted using 100,000 traces, and the DEMA attack was launched without integrating any triggers.

In 2016, \citet{Genkin2016a} discovered how to obtain encryption keys from PCs using their unintended radio emissions. They utilize non-intrusive wireless approaches to quickly and conveniently recover encryption keys from laptops running GNU Privacy Guard (GnuPG). Software called GnuPG makes it possible to authenticate and encrypt data. According to them, the unintentional frequency characteristic that is emitted changes each time the CPU on a target PC executes a new activity. They are able to decipher the activities that GnuPG is carrying out when decrypting data from these emissions by using the distinct RF signature to discover what activities the CPU is carrying out. They indicate that a low-cost RTL-SDR with an upconverter could also perform in their studies, where they used a Funcube Dongle Pro+ to analyze the unintended RF emissions emanating from a laptop computer at about 1.6–1.75 MHz. 

In 2018, \citet{Frieslaar2018} discovered that the libcrypto++ library's implementation of AES-128 cryptography on a Raspberry Pi is susceptible to SCA attacks. The data for this procedure was recovered using digital filtering techniques, and the cryptographic process can be seen clearly in the EM spectrum. The information obtained was used in the DEMA attack to identify the 16 sub-keys needed to decrypt the AES-128 secret key. Meanwhile, in 2021, \citet{Alam2021} showed a notable EM-SCA on different ECDSA (Elliptic Curve Digital Signature Algorithm) implementations, which were used in fully functional smartphones and IoT devices. Based on differences in operand values during the conditional swap operation, the attack targeted the secret scalar used in signing operations in implementations of Libgcrypt, OpenSSL, HACL*, and curve25519-donna. These differences in the EM signal could be used to recover each bit of the secret. When tested on two Android-based mobile phones and an IoT device with affordable equipment, the attack was effective in obtaining the entire secret key in a matter of seconds for all implementations. The findings showed that the signal snippets corresponding to the two swap circumstances could be correctly recovered, along with the swap condition and key. The results also revealed signal quality variances among the tested devices, with the two Android-based devices showing more signal quality fluctuations than the Linux-based device. These results show that ECDSA implementations are susceptible to EM-SCA.

\citet{Lellis2017} demonstrated an energy-based attack flow against the \citet{soares2011robust} architecture that was proposed. This attack used time alignment and subsampling based on current traces and successfully located sub-keys of two-stage pipelined Globally asynchronous locally synchronous (GALS) systems. The authors evaluated the DEMA attack against an embedded system of the DES algorithm prototyped in FPGA and determined the optimum segment length to enhance the attack's effectiveness. Their findings demonstrate that scaling segments to nearly half the operation frequency of the cryptosystem reduces the number of traces required for a successful attack by up to 93.69\%. The authors highlight that the fundamental metric used to determine an attack's effectiveness is the minimum number of traces required to reveal the cryptographic secret.

\subsection{Artificial Intelligence}

Recent developments in the field of artificial intelligence (AI), which incorporates both machine learning (ML) and deep learning (DL), have shown promising performance across numerous computer science fields. ML/DL-powered systems are now replacing several tasks that formerly required human intuition to make decisions. Methods for EM side-channel analysis that formerly required human involvement can be automated due to the advancement of AI systems. AI techniques can aid in two possible ways for EM-SCA: gaining valuable insights into what a running device is doing without the need to directly access the encrypted data, and conducting successful cryptographic key retrieval attacks~\cite{Du2020}.

In recent years, researchers have explored the use of deep learning techniques for side-channel attacks on encryption systems. A method for retrieving the DNN model via an EM side channel has recently been presented by \citet{MITTAL2021102163}. They consider a Binarized Neural Network (BNN) the victim network and attempt to obtain information about its architecture and characteristics. They use near-field probes to enhance the EM signals and reduce noise. They also extract the signal from the noise using frequency-domain filtering. According to the author, a layer's transition and computation rates are correlated. Furthermore, a layer's temporal behavior is related to its parameter count because a layer's parameter count determines its calculation latency. They point out that while recovering layering borders,  and layer depth are simple using the EM side channel, recovering other variables like layer dimension, ifmap/ofmap width, and filter kernel sizes take considerable time.

Furthermore, another two notable examples of deep learning approach for a side-channel attack on encryption systems approach has been demonstrated by~\cite{Xiao2020} and~\cite{Wang2020}. \citet{Xiao2020} introduces a novel approach to side-channel attacks using a convolutional neural network (CNN) to target the AES cipher on the HI Silicon Kirin 620 SoC, a popular mobile device platform. The proposed method acquires side-channel traces through a small capacitor connected to the core power supply line, which captures electromagnetic emissions from the chip surface. The proposed CNN-based side-channel attack method trains the CNN on 100,000 traces and utilizes a label set to the target operation to predict the output of the S-Box during the first round, leading to the recovery of real key bytes in under 1,000 traces despite slight misalignment. The first round's keys can be recovered using a novel approach that builds a differential equation from the chosen plaintext and intermediate value, which increases the likelihood that the key is correct. The paper concludes that CNN-based side-channel attacks pose a threat to embedded devices with gigahertz clock frequencies and demonstrate the effectiveness of CNNs in recovering real key bytes even with slight misalignment, using a relatively small number of acquired traces. In a separate study, \citet{Wang2020} used deep learning-based SCA on AES-128 and leveraging far-field EM emissions as a side channel. Their tests demonstrate that deep learning can retrieve the AES-128 key from a Bluetooth device that implements it, even if only one measurement is made for each encryption. The authors also display the outcomes for recorded traces with 100 and 1K repetition. For the instance of 1K repetitions, one of their models can typically recover the key from traces taken in a workplace at a 15 m distance from the target with fewer than 400 traces. Compared to the template attack (TA) described in~\cite{camurati2020understanding}, which calls for 5K traces, the author states that their method has been improved.

In addition, another deep learning for EM analysis has been conducted by \citet{benadjila2020deep}. The authors thoroughly evaluated the implementation of CNNs to accomplish AES key recovery using the Hamming weight model. They evaluated masked and unmasked reference software AES implementations on an 8-bit AVR microcontroller, and found that the VGG-16 image recognition network design performed well for EM-based side-channel analysis. Similarly, \citet{Yu2021} used deep learning which is a DNN model to obtain the AES key from just a few hundred EM traces with poor Signal-to-Noise Ratio (SNR), using Keysight N2894A and Langer LF-3 probes to measure the power consumption and EM emissions from the microprocessors. They showed that side-channel attacks can be performed with fewer than 230 EM traces, making cross-device and cross-domain Meta-Transfer Learning based Side Channel Attack (MTL-SCA) implementable. In contrast, \citet{Jap2020} used EM side-channel information for parameter recovery in decision tree-based algorithms for the Industrial Internet of Things (IIoT). The authors used a divide-and-conquer strategy to recover the secret parameters by focusing on each function or component of the building block. In real-world trials, the attack on a 32-bit ARM microcontroller demonstrated the retrieval of Bonsai's secret parameters, i.e., node predictors, branching function, and sparse projection parameters.

While the focus of~\cite{Xiao2020,Wang2020,benadjila2020deep,Yu2021,Jap2020} is on hardware devices, the study conducted by \citet{Pan2021} highlights the potential privacy risks associated with the EM side-channel leakage of information about mobile app usage behavior and in-app services. This suggests that EM-SCA may be relevant in not only hardware devices, but also software applications, which could have significant implications for the security of personal information. To demonstrate EM-SCA on the software application, \citet{Pan2021} proposes a novel method called EM-SCA to extract the comprehensive information about app usage by analyzing EM signals that mobile devices produce when carrying out app-related actions, e.g., in WeChat and Taobao. The proposed method, MAGTHIEF, uses multi-label classification based on deep learning to identify apps and in-app services based on magnetometer readings, enabling the continuous collection of fine-grained mobile app usage data without the requirement for user authorization. The authors trained two classification models for each of the two most popular mobile operating systems (Android and iOS) and obtained high average macro F1 scores of 0.87 and 0.95 when classifying multiple apps and in-app services, respectively, and up to 89.5\% time duration accuracy when recognizing app trajectory in the real-world scene. Interestingly, the study discovered that while devices with the same operating system produce generally comparable EM patterns, devices with different operating systems produce varied EM patterns when executing the same app activity.

\subsection{EM fault injection}

Besides EM-SCA attacks, a system's behavior under extraordinary stress is the goal of the fault injection technique~\cite{moradi2018model}. The fault injection attack for hardware implementations is a subtype of active attack where an attacker can actively change the chip's behavior using specialized experimental setups. Based on differential fault analysis, safe errors, fault sensibility analysis, collisions, and round alterations, the errors caused by the aforementioned fault injection may be used to recover the secret information. Additionally effective at revealing cryptographic devices and processor control flow is the invasive fault injection attack, which involves inserting faults into a cryptographic device with the assistance of a laser or focused ion beam (FIB) and then observing the outputs~\cite{BHUNIA2019245}. It has been discovered that utilizing an inexpensive EM probe allows for simple data eavesdropping from a distance~\cite{Sayakkara2019a}. Moreover, through the use of glitch injection, methods like differential fault attacks (DFAs)~\cite{Lim2020} have been able to determine private information. Due to their ease of use and low cost, fault injection attacks (FIAs) and EM-SCA have both gained enormous popularity~\cite{Gunathilake2021, Danial2020a, Hell2020}.

According to \citet{Chusseau2014}, adding ferrite material increases the EM flux's intensity, while several turns provide stronger coupling between the injection probe and the target at low frequencies. They make highlight how EM fault injection can be generated to crack secured ICs' cryptographic algorithms. Such an objective inevitably calls for multidisciplinary research on the physics of EM near-field probes, signal management in timing, and localization to cause the fault in the most vulnerable area of the IC while evading defenses, and finally, a precise logical comprehension of fault reproduction to ultimately ensure future ICs.

On the other hand, \citet{Nakamura2015} reveals a technique for calculating the timing of fault occurrences brought on by the aforementioned intentional EM interference (IEMI)-based method that can be utilized to determine whether a faulty output that has been received is suitable for fault analysis. The purpose of this technique is to monitor side-channel data, such as EM leakage, and calculate the time of fault injection by identifying a distinctive shift in the observed waveform. They ran an IEMI fault injection experiment on real cryptographic hardware (a side-channel attack standard evaluation board) to evaluate this approach. The authors noted that it is possible to perform DFA by employing this estimating method; however, they get an inaccurate output from fault injection. Furthermore, \citet{dehbaoui2013electromagnetic} launched an EM fault injection-based key recovery attack on a 32-bit ARM Cortex-M3 MCU running a reference software AES implementation (24MHz). The effort focused on the AES round counter at rounds nine and ten to successfully (up to 100\%) generate an extra round of execution. This enabled workable cryptanalysis that used two pairings of positive and negative ciphertexts to retrieve the encryption key.

In 2019, \citet{10.1145/3319535.3354201} propose VoltJockey, a novel software-controlled hardware fault-based attack, to TrustZone, the safe environment for ARM-based processors. In order to bypass TrustZone security measures, the exploit modifies the voltages of multicore processors that support DVFS.  By altering the victim core's frequency and selecting a low voltage for the core, the attacker can cause hardware errors in the victim procedure while posing no risk to the attacker core. The VoltJockey attack's approach, difficulties, and technical specifics are presented by the authors, who also show how it may be used to extract AES keys and change RSA output in TrustZone. A Google Nexus 6 smartphone with a Qualcomm APQ8084AB SoC was the target of proof-of-concept attacks by the authors, which had reported rates of 4.6\% (RSA verification) and 2.2\%. (AES key recovery). 

In 2020, \citet{Ravi2020} have discovered a vulnerability that allows information about individual bits of the decryption output to leak through the message decoding function in a number of second-round lattice-based public key encryption and key encapsulation mechanisms (PKE/KEMs). The shared secret of five lattice-based PKE/KEMs, including NewHope, Kyber, Saber, Round5, and LAC, has been recovered by performing message recovery combining practical EM emanation-based side-channel attacks and EM-based fault injection attacks. The authors of this paper introduced new clustering-based attack methods, found a hidden vulnerability in the NewHope KEM implementation, and demonstrated the real-world application of EM-based combined SCA and FIA over lattice-based PKE/KEMs. The side-channel attacks covered in this research are carried out in two stages, the pre-processing and attack phases, which entail creating profiles or templates for various message values, querying the device with attack ciphertexts modified from the target ciphertext, and observing the corresponding side-channel traces. According to experimental validation of the attacks on the NewHope512 implementation in the pqm4 library, the entire message can be retrieved with a 100\% success rate in as little as 256 traces.

Similarly, \citet{Jap2020} explored the vulnerability of decision tree-based machine learning algorithms to EM side-channel attacks, using the publicly available Bonsai library as an example. The authors demonstrated that an attacker could rebuild the complete architecture by using a divide-and-conquer strategy to extract hidden algorithmic parameters such as sparse projection, branching functions, and node predictors. In particular, the author used a technique called TA to recover the secret index parameter (Zidx) used in the multiplication operation of the Bonsai decision tree algorithm. The author points out that invasions of privacy could result from the potential of secret parameter recovery and that security issues surrounding machine learning algorithms are becoming more and more significant, particularly as they are employed for industrial applications. The authors demonstrate that even security systems designed to withstand faults can be vulnerable to SCA. Specifically, TA have proven to be successful in cracking security systems

\subsection{Fine-grained EM-SCA attacks}
The SCNIFFER framework has been used to show a low-cost, end-to-end automated EM-SCA attack~\cite{Danial2020a}. Nevertheless, since EM fields degrade quickly at a distance, fine-grained EM-SCA attacks lose even more effective than their coarse-grained equivalents when probes are placed distant from the sources of interest. Thus, the ``acquisition cost'' of fine-grained EM-SCA attacks is a term used to describe the measurement costs associated with attempts to localize vulnerabilities~\cite{Wang2021}. Since the number of nodes for effective configurations comprises a probe's transverse position, height, and direction, the acquisition cost of fine-grained EM-SCA attacks can be significant~\cite{Sayakkara2021, Chatterjee2019, Yoshida2019}. In general, the number of measurements taken for each probe configuration must be kept to a minimum for attacks utilizing extensive search techniques~\cite{Sosa2021}, which scan the whole chip/board at a very high resolution with various probe orientations. Several methods have been put forth in the past few years to speed up the discovery and lower acquisition costs~\cite{Belaid2020, Vosoughi2020}. \citet{Dey2022} used repeated measurements and averaged the collected signals to increase the signal-to-noise ratio. Nonetheless, these discovery procedures for fine-grained EM-SCA attacks should be carefully evaluated. This assumes that attackers have some degree of authority over the inputs to the cryptosystem. Additionally, the repetition of the measurements can be utilized as a criterion to eliminate out configurations~\cite{camurati2020understanding} and drastically lower the acquisition cost if the threat model allows attackers to repeat particular inputs.


\subsection{Data processing}

In the field of side-channel analysis and cryptography, preprocessing is a vital stage in increasing attack effectiveness since it is frequently utilized to boost attack success~\cite{Abdellatif2019, Wang2020, 9400816}. The sample size of the EM-SCA can be continuously expanded to increase the success rate, but this may result in a lengthy cracking time, limiting the viability of the EM-SCA. To illustrate, the 128-bit AES algorithm, which must be processed 16 times for each byte's sub-keys, is one example of an algorithm that must be processed data once for each byte's sub-keys in order to be cracked~\cite{Iyer2021, Vasselle2019, Kumar2020}. EM traces may not correctly encompass the cryptographic operation within its perimeter and have varying lengths for various reasons. \citet{Sayakkara2019a} identified two reasons why labelled EM trace data is unsuitable for direct use in machine learning-based classification: the intrinsic variation in the amount of time needed to complete each cryptographic computation and the delays in the HackRF data collecting software to initiate and terminate EM sampling. However, by converting EM traces into the frequency domain using Fast Fourier Transformation (FFT)~\cite{smith1997scientist, Frieslaar2018}, the discrepancies in lengths can be minimized. 

Meanwhile, \citet{Abdellatif2019} propose a pre-processing step using wavelet analysis to improve the success rate of side-channel attacks by increasing the SNR of the EM traces. Discrete Wavelet Transforms (DWT) is used to decompose the signal into a set of terms of a basis set function, which is compared to the commonly used Fourier transform in terms of the basic functions they use. The use of wavelet analysis for side-channel attacks is not new and has been previously proposed in the literature, as mentioned in the paper~\cite{souissi2011novel}. However, the focus of~\citet{Abdellatif2019} is addressing the challenge of low SNR, which is a common problem in side-channel attacks. By combining the two methods, a more effective approach for analyzing EM trace data can be developed. This demonstrates that the use of pre-processing steps can be beneficial in addressing challenges associated with EM trace data and improving the effectiveness of side-channel attacks.

Furthermore, EM-SCA is a crucial technique for analyzing the unintended EM emissions from electronic devices i.e., IoT devices. The process of data gathering and processing is challenging due to the large file sizes of EM trace data and the need for real-time analysis. SDR devices are commonly used for capturing EM data and provide the ability to capture a large bandwidth surrounding the target frequency, allowing for signal differentiation in the frequency domain. However, the high sampling rates used by SDRs result in extremely large file sizes for EM trace data. To address this issue, \citet{Sayakkara2019a} suggest down-sampling the data while maintaining the maximum possible bandwidth, which does not negatively impact the performance of classification algorithms. The authors highlight the need for real-time analysis in live forensic analysis, where data preprocessing and classification must be performed within a tight time frame to keep up with the real-time I-Q data stream. The processing delay remains well below the TCP retransmission timeout, even at the highest SDR device sampling rate.

The conventional method of performing a side-channel attack requires a significant amount of data gathering and processing. To address this issue, \citet{Zhou2019} propose a fast EM side channel attack approach that drastically cuts the time needed for an EM bypass attack. Their investigation reveals that various frequency ranges significantly affect the side channel attack experimental results. Consequently, the revised method's first step is to use FFT to remove noise from the original acquired data. The revised approach allows for a maximum sample size of 256. For each of the 1000 key cracks, the experiment applied the improved approach with different sample sizes. The sample size is reduced by filtering, and the number of data processing operations is decreased by adopting plaintext for encryption. The results showed that the improved approach is 50 times faster than conventional methods, as it cuts down the number of data processing operations.

\citet{Bu2018} proposed a new method of preprocessing energy traces collected from an AES circuit to enhance the efficiency of the CEMA attack. The authors established a hardware platform to collect EM traces and applied two preprocessing techniques, Haar wavelet reconstruction and low-pass filtering, to the collected data. The Haar wavelet reconstruction is a time-frequency analysis technique that can decompose signals into high and low-frequency coefficients. The high-frequency coefficients were reconstructed using a soft threshold, resulting in the removal of unnecessary glitch signals and the preservation of the original signal's trend and characteristics. Low-pass filtering, on the other hand, is a common preprocessing technique used to reduce high-frequency noise in signals~\cite{10.1007/978-3-642-10838-9_13, longo2015soc, 9417659}. In this study, the authors designed and implemented a low-pass filter on MATLAB, which was applied to the EM traces to reduce noise while preserving the original signal's trend and characteristics. The preprocessing of EM traces was deemed necessary due to the weak EM leakage that can result in a significant amount of noise in the collected data~\cite{10.1007/978-3-540-85053-3_26, 10.1145/3195970.3196042, Haas2022}. The implementation of this preprocessing technique improved the efficiency of the CEMA attack.

\subsection{Other Encryption Algorithms}

In this systematic literature review, a total of 114 research papers were collected and tagged based on their focus on cryptographic algorithms, i.e., AES, DES, ECC, EEA, and RSA, as illustrated in Table~\ref{tab:legend}. However, 13 papers did not fall into any of these categories, as they focused on other encryption algorithms, e.g., Lightweight Encryption Algorithm cypher (LEA), PRESENT (lightweight block cyphers), Verify Pin algorithms and Pin encryption algorithm. Therefore, while these 13 papers do not fit into the main categories of cryptographic algorithms that are the focus of this review, they offer valuable insights into the vulnerabilities of less commonly used encryption techniques relevant to specific applications and devices.

The development of lightweight block cyphers for low-power microcontrollers and RFIDs has recently received great attention~\cite{4397176}. LEA (Lightweight Encryption Algorithm) is one such encryption that has been proposed and can be examined using various techniques, including DFA. \citet{Lim2020} proposes a novel DFA method on the ARX-based lightweight block cypher LEA. The proposed method reduces the number of required fault-injected ciphertexts by approximately 70.97\% compared to previously proposed attacks in~\cite{park2014differential, 10.1007/978-3-319-24315-3_27} and uses a relaxed fault model, namely Random Word Error. Compared to earlier proposed attacks, the proposed method requires a substantially less number of fault-injected ciphertexts and the fewest key candidates to complete the exhaustive search. Additionally, the proposed method is experimentally demonstrated to be applicable to real IoT devices by constructing an electromagnetic fault injection environment on an actual microcontroller and revealing the secret key through the attack. The author claims that other cryptographic algorithms that use modular addition operations can also employ the proposed DFA approach. The paper discusses the experimental environment and results in detail, showing the number of candidates for the last round key (i.e., the key used to encrypt the message in the final round of the cryptographic system) and the number of required fault-injected ciphertexts for the proposed method. The paper compares the results obtained from the proposed method to those obtained from previously proposed attacks to evaluate the effectiveness of the proposed method in breaking the cryptographic system. 

On the other hand, the vulnerability of another lightweight block cipher, PRESENT, which is the lightweight block cipher algorithm that is frequently used in an embedded device, was analyzed using EM-SCA by \citet{nozaki2017differential}. The proposed approach classified electromagnetic waves using selection functions, calculated intermediate values, calculated differential values for key analysis, and derived Hamming distance using DEMA. The findings demonstrated that by eliminating the second Hamming distance value calculated between two rounds of the Present cipher from their analysis, they have increased the accuracy and success rate of their EM-SCA. 

Furthermore, \citet{Bouder2016} focused on the vulnerability of PIN codes to template attacks, which can accurately predict the PIN code entered into a device using only a small number of EM traces. The VERIFY PIN algorithms, which are meant to withstand fault injection attacks, are the target of what is described in the study as the first EM-SCA. The experimental equipment used to carry out the attack is described in full in the study and includes an ARM-based STM32-F100RB microcontroller operating at 24MHz and a Picoscope to monitor EM leakage. In the attack's profiling phase, EM traces for every possible PIN value are physically modeled, and in the attack phase, EM traces from the targeted device are obtained and compared to the templates produced in the profiling phase. The study outlines a six-step attack strategy that tests three distinct PIN codes while using two identical devices—one for profiling and the other for the attack. The paper concludes that template attacks are still effective for cracking PIN numbers even when there are minimal EM traces accessible. Moreover, the authors also point out that template attacks can be carried out using affordable and portable experimental setups, and some countermeasures against FIA can develop new side-channel attack vulnerabilities.

Moreover, \citet{Simon2015} have developed a statistical analysis method for assessing EM information leakage from Subscriber Identity Module (SIM) cards. The suggested method employed statistical and differential SCA to provide a spatial leakage map for the SIM card's surface as well as a time window of peak activity. The authors employed an EM probe to gather near-field EM radiation from the SIM card by using an oscilloscope set to 25 mega-samples/second over a 50ms frame. The secret key and PIN encryption algorithm for SIM cards were extracted using a combination of spatial and temporal analytics. Without prior knowledge of operational frequencies, the proposed approach offers a better way to locate operationally important spots on the SIM card.

\subsection{Instrumented/Uninstrumented}

Depending on how closely a system or process is monitored or measured, it is said to be \textit{instrumented} or \textit{uninstrumented}. Systems that have several sensors and monitoring tools to gather information about their performance and behavior are referred to as instrumented systems. The examination of system behavior, performance improvement, and debugging are only a few uses for this data. On the other hand, uninstrumented systems lack any monitoring or measurement components. This means that it is impossible to assess or examine the system's behavior or performance directly. Comparing instrumented versus uninstrumented systems in a research setting can be useful for understanding how monitoring and measurement affect system performance or assessing the efficacy of various monitoring or measurement strategies.
The analysis of the papers considered as part of this systematic review revealed that only one study is uninstrumented: \citet{Yoshida2019}. The vast majority of papers utilize instrumented approaches to gather, measure, and analyze data relevant to their research interests.

\section{Encrypted Devices Targeted}

The EM field that is emitted by electronic devices has the potential to compromise internally managed sensitive data. Lightweight electronic devices have recently become prevalent in people's day-to-day lives. Many examples of RFID tags and wireless sensor nodes manage systems for electronic payments, product identification, IoT, alarm information, habitat monitoring, structural monitoring, and emergency medical response. These lightweight services are widely used, which leads to privacy and security concerns. Lightweight cryptographic methods have been used in several application-specific integrated circuits (ASICs) to date. ASICs result in high prices linked with high non-recurring engineering (NRE) expenses and lengthy manufacturing delays, which may be the fatal factor for low-volume goods despite being highly desirable for mass production and low power consumption. For this market segment, field programmable gate arrays (FPGAs) offer appealing alternatives owing to their inexpensive or zero NRE costs and quick time-to-market characteristics. EM-SCA attacks are increasingly focusing on IoT devices and components, including cryptographic hardware on System-on-a-Chips, FPGA, and ASIC, as well as on microprocessors.

According to a study by \citet{li2015ultra}, KLEIN is side-channel resilient to first-order attacks, although it might still be susceptible to higher-order attacks. Lightweight cryptography recently gained a re-keying option that improves guarding against side-channel attacks~\cite{Patranabis2019, Vosoughi2020, Naija2017}. Other important features, including EM emission, have yet to be completely observed. \citet{nozaki2017differential} proposed a DEMA approach for the PRESENT lightweight block cipher. For the experiments, a shielded loop antenna, an oscilloscope, and a Spartan-6 XC6SLX150 FPGA implemented on a SASEBO-W evaluation board were employed. The outcomes demonstrated that the proposed analysis approach worked well in discovering the correct key. The study of EM-SCA for a lightweight block cipher, i.e., TWINE and PRINCE, are discussed by \citet{yoshikawa2016twine} and \citet{yoshikawa2016prince}, respectively. This section aims to assess the potential of EM-SCA in various practical areas of information acquisition from an electronic device, such as FPGA, IoT, Integrated circuits (IC), microcontrollers, mobile phones, CPUs, and Raspberry Pi.

\subsection{Access Card/RFID}

Numerous large-scale, security-related applications, such as public transportation, wireless payment, access control, and digital identity, are now based on contactless smartcards that use RFID technology~\cite{kasper2011side, Plos2012, Naija2018a, kasper2011side}. In essence, passive RFID systems are made up of an antenna and a tiny microchip. This antenna employs an RFID reader's EM field to collect energy. The energy generated by this field powers all aspects of passive RFID technology. The majority of RFID devices today include cryptographic mechanisms, for example, to perform a mutual authentication or to encrypt the data sent over the air interface, due to the sensitivity of the stored and transmitted data (such as personal information, cash balance, etc.) and the fact that accessing the wireless interface is practically impossible to monitor.

\citet{carluccio2005electromagnetic} exposed a contactless smart card's weakness in terms of security because they showed how to attack the smart card using an EM side-channel. Standardized cryptographic algorithms are frequently used in contactless smart cards, i.e., AES~\cite{Naija2018}, ECC~\cite{de2005electromagnetic}, DES~\cite{kasper2009side}, and RSA~\cite{xu2018side}. In 2007, \citet{oren2007remote} demonstrated that remote side-channel attacks are still effective even when the attacker is several meters from the victim card. Smart cards are prevalent in smart devices, which are crucial components of the IoT and sensor networks. The security concerns in these areas have been examined in several representative papers, involving key management~\cite{du2007effective,du2009transactions}, access control~\cite{hei2013pipac}, defend from phishing attack~\cite{wu2015effective}, and hardware security~\cite{cheng2017lightweight}. At the same time, side-channel attacks against smart cards have drawn the attention of an increasing number of academics.

The susceptibility of smart cards to SCA was further demonstrated by \citet{KIM20122899} in a study of commercial smart cards used in a real-world electronic payment system's SEED block cipher to statistical side-channel analysis attacks. SEED, developed by the Korea Internet \& Security Agency (KISA), is a well-known block cipher used for encrypting sensitive information~\cite{Lu2010}. The findings demonstrate that unprotected SEED implementations enable retrieval of the secret key using low power or EM traces, whereas concealing countermeasures are ineffective. The demodulation technique of indirect EM leakage data at the carrier frequency CEMA can address these attacks. Both protected and unprotected SEED implementations were susceptible to AM demodulation attacks, leading to successful attacks on combination smart cards with certifications.

Using a specially designed RFID tag, \citet{hutter2007power} provided proof of concept for the ability to carry out EM side-channel analysis on RFID-based systems. The authors successfully recovered the AES key that was utilized in the challenge-response protocol between the RFID tag and the reader performing AES encryption operations for various plaintexts and measuring side-channel information leakage. The power harvesting antenna is kept inside the reader's RF field, but the RFID circuitry was put outside of it to prevent interference from the reader's RF field. A wire of suitable length was used to link the two parts. Due to this, it was possible to monitor the EM emissions from the RFID circuitry without any interruption. The antenna and RFID circuitry are inseparable by any logical means, hence it is not viable to use a comparable strategy in a standard RFID tag.

In addition, the security of the data sent between the server/reader and tags in RFID systems is supported by mutual access control based on symmetric challenge-response procedures. For instance, Mifare DESFire and Mifare DESFire MF3ICD40 integrate 3DES block cipher~\cite{kasper2010cloning, kasper2011side}, whereas Mifare DESFire EV1 incorporates AES block cipher. These are examples of high-frequency commercial tags that employ symmetric cryptographic functions in mutual authentication protocols. \citet{oren2007remote} make use of wireless power delivery to RFID devices to detect energy usage patterns and deduce sensitive data. Radio receivers can reradiate the local oscillator under certain circumstances~\cite{behzad2007wireless}, which can be exploited to discover radio receivers~\cite{chaman2018ghostbuster,camurati2018screaming}.
 
\subsection{FPGA, Chip, IC, Microcontroller and Microprocessor}

Without modifying a PCB or IC, EM radiation can be easily detected and even observed from a distance. The fall in SNR brought on by core voltage reductions, on-chip decoupling capacitors, system-on-chip implementations, and other factors makes monitoring a device's power usage harder. Consequently, EM analysis now poses a bigger risk to cryptographic devices than power analysis~\cite{Duan2015, hori2012electromagnetic, Das2020a}. In 2001, when \citet{Gandolfi2001} employed tiny EM probes to retrieve key material from three different types of microcontrollers completely, the effectiveness of EM side-channel attacks against secure systems was proven. More recently, FPGAs~\cite{Nakamura2015, Chusseau2014, Bu2018, Sauvage2009} and PCs~\cite{Yilmaz2020a, Genkin2016} have been the targets of EM side-channel attacks that recovered cryptographic keys. The operator capability and information flow management are closely related to the dynamic power consumption of an FPGA implementation. While the temporary current flow generates the EM radiations, side-channel information about the algorithm's running and the data structure can be found in its temporal and magnitude variations. As a result, the attacker can use this EM leakage to infer the FPGA device's execution details by locating the leak point and collecting data with an oscilloscope and an EM probe~\cite{Sauvage2009}.

\citet{hori2012electromagnetic} created the Sasebo-GIII board (fitted with Xilinx 28-nm Kintex-7 FPGA) to carry out CEMA using the Hamming-distance model. The EM radiation emissions from the AES circuit in the Kintex-7 FPGA were evaluated and correlated to those from the 65nm Virtex-5 FPGA (Sasebo-GII). With a lower core voltage and hence less side-channel information, the Kintex-7 FPGA was constructed using improved fabrication techniques. The waveforms of the produced EM radiation were retrieved using a Langer LF-B 3 EM probe, an Agilent DSO6104A oscilloscope, a fifth-order Bessel low-pass filter, a Miteq AU-3A-0150 amplifier (50 dB, 0.3-600 MHz). Contrary to predictions and despite a 5x lower observed voltage on the more recent platform, only 7k traces were necessary to recover a key, as opposed to 19k traces on the more previous platform. The authors speculate that the physical locations of the subkey bytes within the AES structure could be one of the causes of this. On the other hand, \citet{batina2019csi} illustrates a technique for attacking an FPGA-based neural network implementation that employs CEMA. Particularly, the adversary uses the hamming weight model to retrieve the weight values of the neural network. In another study, \citet{Yu2020} utilizes EM side-channel analysis to attack binary neural networks (BNNs). In this research, the adversary can sample EM leaks to extract the architecture of the BNN and then use the adversarial active training technique to create a model with comparable inference accuracy. Although this attacking approach may reconstitute the BNN model with greater inference accuracy, it still necessitates training and only applies to BNN.

Furthermore,\citet{Iyer2019} provides an adaptable acquisition protocol for EM-SCA on cryptographic devices. In accordance with the protocol, a multi-step acquisition entails a greedy search in a 4-D configuration space made up of the probe coordinates, orientation, and quantity of signals received. The ideal attack configuration for an 18mm by 18mm FPGA implementation of AES is selected to test the protocol. Compared to the exhaustive acquisition strategy, the protocol can minimize the search time for the ideal configurations for obtaining the secret key from the AES-128 realization by up to 35. Their discovery offers an effective method for determining the optimum near-field measurement setups for EM-SCA on physically implemented cryptography modules.

Moreover, \citet{Montminy2013} presented a method to undertake differential attacks on cryptographic systems by gathering side-channel data using an SDR. The author explains how to evaluate a 32-bit microprocessor using a correlation-based frequency-dependent leakage mapping technique, and discovers that different key bytes leak at various frequencies. The target cryptographic device is an LM4F232H5QD evaluation kit running AES-128 in Electronic Code Book (ECB) mode on a Stellaris ARM Cortex-M4F-based microcontroller.  The author emphasizes two important aspects of obtaining key bytes at various center frequencies. Firstly, the author acknowledges that not every key byte can be extracted at every center frequency. Second, the author states that pooled standard error provides the foundation for calculating confidence. The author further stresses that when the quantity of traces grows, even slight alterations in the correlation coefficients can increase confidence levels. This suggests the technique efficiently determines the frequencies to focus on with SDRs. Combining these ideas, the author highlights the need for a comprehensive review of the available data and statistical analysis to extract key bytes at particular center frequencies. The author explicitly states that the confidence metric has limitations when only a few samples are obtained for each encryption operation, leading to failure to identify the proper key byte despite high confidence.

\subsection{Computer}

Electric currents in conductors that change over time emit EM waves into the surrounding space. Computer systems unavoidably produce EM emissions during internal operations since they are made up of electronic circuits~\cite{guri2016usbee}. The consequent EM emission may unintentionally hold information about the operations connected with that component, depending on the specific component on the device that contributes. As an illustration, computer displays are known to be the root of powerful EM emissions that make image reconstruction simpler~\cite{Sayakkara2018, VANECK1985269, 10.1007/11423409_7}. Similarly, the central processing units (CPUs) of computers are also known to indicate the CPU activities being carried out~\cite{SAYAKKARA201943}. Furthermore, \citet{Sayakkara2018} investigated the challenges faced in EM side-channel attacks on video displays and aimed to improve image reconstruction accuracy in low-cost RF signal acquisition hardware. The authors evaluated different approaches to enhance the quality of reconstructed images, including noise reduction of RF signals and blending multiple reconstructed images. The experiment utilized HackRF SDR hardware with a sample rate up to 20MHz, connected to the target computer with a small antenna. A Butterworth band-pass filter was used to extract the relevant EM signal, and the impact of using narrowband signals and blending multiple images was evaluated. The authors concluded by discussing possible future improvements in EM side-channel attacks on computer monitors.

The initial Van Eck phreaking attacks on computer Cathode Ray Tube (CRT) monitors exemplify a non-intrusive EM side-channel attack. These are simple to eavesdrop on since they generate a lot of radiation when in use. A CRT monitor's image can be replicated from behind a wall and in another room, making this side-channel attack a useful tool for eavesdropping on the target. The attack is still possible, as shown by \citet{10.1007/11423409_7}, although modern LCD computer monitors switch much less current, limiting the quantity of radiation emitted.

One can visually differentiate individual CPU activities when an obtained EM signal from a target device, or EM trace, is shown as a waveform or as a spectrogram. Simple EM analysis (SEMA) is the simplest use of these examples to listen in on CPU activity. It has been commonly used to describe how computer systems work~\cite{kocher1999differential}. An attacker can acquire numerous skills by keeping an eye on the instructions being executed on the CPU, including the ability to reverse-engineer unknown software and to keep an eye on the control flow of known software.

The effectiveness of EM analysis as a viable option to the power analysis attack on computer CPUs was practically proved by~\cite{quisquater2001electromagnetic}. The authors successfully created a precise 3-dimensional EM signature of the chip executing an idle loop by placing the EM probe across a microcontroller. The radiation spectrum of each processor was demonstrated to be sufficiently distinctive to be used as a recognizable feature for processor identification. In order to reduce the impacts of outside noise, these tests were conducted in a Faraday cage. A small magnetic loop antenna was used to collect the EM emissions (diameter 3 mm). In addition, according to \citet{SAYAKKARA201943}, combining numerous side-channel attacks targeted at a single computer system can be more effective than utilizing a single side-channel attack alone. Power and EM-SCA can be integrated to provide superior outcomes, as has been demonstrated~\cite{agrawal2003multi}. Some CPU actions may have a more pronounced impact on the device's power consumption than on its EM emission, and conversely. By purposefully manipulating data into the EM emission of the CPU or the monitor, malware operating on a victim machine might help an EM side-channel attacker retrieve more data over the EM side-channel alone~\cite{cheddad2010digital,yang2017comm, Sayakkara2018a}.

\subsection{Mobile Phones}

Modern digital devices, i.e., mobile devices, are developed and distributed with built-in security due to the rising concerns about security and privacy across communities. Popular smartphones with operating systems like iOS and Android encrypt their internal storage to shield user data from outsiders~\cite{ahmad2013comparison}. \citet{SAYAKKARA201943} claims that various mobile device platforms frequently use ECC algorithms to protect the data due to their low processing expense. This demonstrates that such devices can be examined through EM side channels to gain access to data that is encrypted.

In 2011, the first EM-based SCA on a mobile device was published by \citet{aboulkassimi2011electromagnetic}. It was designed to target benchmark AES software implementations on the Java Platform, Micro Edition (Java ME). A commercially available EM probe and oscilloscope were used to obtain the measured data, which were prompted using the device's microSD card interface. The authors provide strategies for overcoming the Java Virtual Machine's (JVM) just-in-time (JIT) compilation and garbage collection-induced temporal distortions in EM traces. Two approaches, a spectral density-based approach (SDA) and a template-based resynchronization approach (TRA), were developed to navigate these problems statistically. With the second method, a 32-bit RISC-based device might recover one AES key byte in an hour with 250 traces (370MHz).

EM-based DPA attacks were shown by \citet{balasch2015dpa} against a Texas Instruments AM3358 Sitara SoC on a Beaglebone Black single-board computer (SBC) with an ARM Cortex-A8 (1GHz) processor and a Linux operating system.  The authors specifically target a slightly adapted version of \citet{konighofer2008fast} intended to prevent side-channel analysis as well as an unencrypted 128-bit AES software implementation. After evaluation, a key recovery using the bit-sliced implementation and a first-order DPA required 1.2 million traces, but a second-order DPA was feasible with 400,000 traces. In contrast, the unprotected algorithm just needed 10,000 traces. On five unidentified mobile phones, \citet{goller2015side} used the square-and-multiply approach to attack a reference software RSA implementation.  A software-defined radio, a probe, and a high-gain amplifier at a capacitor close to the main CPU were utilized for the measurement collection after the insulating layer of the device was removed to minimize EM attenuation. The waveforms had a strong correlation with specific bits of the secret key using simple power analysis. Full key recovery with high confidence (0.999 correlation) required 276 traces, whereas without the shielding plate, 170 traces were needed.

In 2016, \citet{belgarric2016side} performed EM analysis to pinpoint ECC addition and multiplication processes in the Bouncy Castle ECDSA implementation for Android. Full key recovery was accomplished through a lattice attack on an unidentified smartphone with a Qualcomm MSM7225 SoC. An EM probe was positioned on the SoC after the device's external case was opened, and traces were activated through the USB port. Full key recovery required 39 ECDSA signature traces and took 102 seconds; as an illustrative use case, the authors were also able to retrieve the key for an Android Bitcoin wallet. \citet{Genkin2016a} published simultaneously with their work in which they successfully full key retrieval against OpenSSL's ECDSA implementation for iOS and Android. Compared to \citet{belgarric2016side}, the approach was less intrusive and simply needed a probe to be put close to the target device; there were no hardware or software triggers. 5000 signature traces from the Sony-Ericsson Xperia X10 and the iPhone 3GS were used in the exploit, of which two (0.04

In 2017, the study on the EM characteristics of program execution in the ARM TrustZone was done by \citet{Bukasa2017}. On a Raspberry Pi 2 with a Broadcom BCM2836 SoC (quad-core ARM Cortex-A7 at 900MHz), they examined an unencrypted reference AES software implementation and a dummy PIN verification method. The impact of execution in a secure world against a non-secure world and multicore versus single core was investigated. Targeting the first-round S-box of AES and gathering 150,000 EM traces might be performed to recover the key through template analysis. Depending on the setup of the system, a success rate of 17.81\%–38.30\% might be achieved in recovering the key. The success rate for multicore execution in the secure world was the lowest (17.81\%), whereas the success rate for single-core execution in the non-secure world with the MMU turned off was the highest (38.30\%).

The first physical SCA on an Apple iPhone 4 was recently demonstrated by~\cite{lisovets2021let} for retrieving the 256-bit hardware-based user identifier (UID) key. The UID is combined with the user's passcode to produce a passcode key from a password-based KDF that unwraps keys from the system key bag, which holds per-file data encryption keys for FBE utilizing the SoC's AES encryption engine. An attack in two parts was carried out: the AES algorithm was attacked using EM correlation power analysis following — 1) utilizing a known bootloader vulnerability to add an unauthorized component and 2) using EM correlation power analysis. A trigger from a GPIO pin that the unauthorized bootloader had modified was used to activate the measurement of EM traces from a probe on the Apple A4 SoC. A LeCroy WaveRunner 8254M oscilloscope (2.5GHz bandwidth, 40GS/s sampling rate) and a Langer EMV-Technik RF-B 0,3-3 EM probe, as well as a Langer EMV-Technik PA 303 SMA amplifier, were employed for the specific equipment. After 300 million traces were acquired over two weeks, it took up to three hours of analysis utilizing two Nvidia RTX 2080 TI GPUs before the key could be recovered.

In another study, \citet{Hu2018} aimed to develop a new signal acquisition system for detecting EM side-channel leakage signals from mobile devices that use the AES encryption algorithm. They developed a technique for processing the raw side-channel leakage signal and used time-frequency analysis and filter technology to recover the encryption characteristics of AES. After carrying out a short-time Fourier Transform (STFT), the authors observed that the frequency range of the AES encryption traces is 50 MHz to 80 MHz and the whole ten-round encryption features are clearly visible after the second filter. This study significantly contributes to the field as it demonstrates a method for locating mobile devices' side-channel EM leakage signals.

\subsection{Internet of Things}

Since the invention of computers, Moore's law has allowed engineers to advance their capabilities continuously. The development of smaller, more energy-efficient computing equipment, i.e., mobile devices, which have less processing power but can operate for longer on battery power, has been made possible by this trend in addition to making computers more prevalent and resource-rich. The end outcome of these developments is the rise of the IoT, which consists of small computational devices with sensors and network connectivity that have already begun to permeate several aspects of daily life, such as smart home systems, sports, and healthcare~\cite{lin2017survey, Boozer2021, Balogh2021, rizvi2022network}. IoT device emissions can originate from a variety of components, with the processor playing the largest role. The software activity running on the device has a direct impact on changes in the processor's EM emissions~\cite{Tirumaladass2020}. 

Recently, EM-SCA was put forth as a technique for gathering information from IoT devices that is relevant for forensic purposes~\cite{Sayakkara2021, Sayakkara2020a, Sayakkara2018}. Law enforcement may be capable of performing EM-SCA procedures on an IoT device as soon as it is seized if it is powered on. It has been claimed that various information, such as modified firmware, software behavior, and cryptographic techniques, can be identified with EM-SCA-based methodologies~\cite{Sayakkara2019a}. There is no need to physically interact with the equipment under investigation because collecting and analyzing EM emission signals is non-invasive.

According to \citet{camurati2018screaming} and \citet{Sayakkara2019a}, the use of EM-SCA to detect cryptographic activities in IoT devices has gained significant attention in the research community. \citet{camurati2018screaming} discovered that mixed-signal computers, like system-on-chips (SoCs), which combine a CPU and a radio transceiver on a single silicon die, can result in long-distance EM leakages known as "screaming channels". This has greatly expanded the possible attack surface as SoC adoption on IoT devices grows in popularity. On the other hand, \citet{Sayakkara2019a} explored the use of EM-SCA to detect cryptographic activity in IoT devices by focusing on the EM emissions produced by a Raspberry Pi. The study aimed to automatically detect the data encryption operations performed by the device using AES-128, AES-256, and 3DES cryptographic algorithms. The results showed that a neural network classifier could discriminate between these encryption techniques with 80\% accuracy, demonstrating the potential of EM-SCA as a tool for detecting cryptographic operations and suggesting its application to detect encryption algorithms on less capable hardware devices. These findings contribute to the growing body of literature on the use of EM-SCA in the context of IoT security.

Additionally, several case studies have investigated the vulnerability of IoT devices to SCA, particularly through the use of EM analysis. \citet{Levina2021} presents a case study that investigates how side-channel attacks can target pet wearables. This study's authors concentrated on a Jagger \& Lewis dog activity tracker~\cite{jaggerlewis} that was no longer produced. The authors conducted an EM attack on the device to demonstrate the SCA. They were able to retrieve the original JSON text by analyzing traces taken at the time of Base64 encoding. In order to analyze the form visually and estimate its amplitude and temporal properties, they recorded data from the targeted device. According to the results of the case study, the authors make the assumption that the device employs Base64, the most straightforward data encoding scheme, to ensure that the fewest operations are required to account for the tracker's low power.

Meanwhile, \citet{Dinu2018} conducted a case study to assess the vulnerability of Thread, a networking stack developed for secure communication between IoT devices, to side-channel attacks. They identified several attack vectors to bypass the security mechanisms of the networking stack, focusing on manipulating security material such as cryptographic keys. Their fully implemented attack employed network-level mechanisms and electromagnetic side-channel analysis techniques to gain unauthorized access to an existing Thread network. Although Thread has complex security mechanisms, the study found that such attacks are still feasible and pose a relatively higher threat to the commercial setting. The full attack was unsuccessful due to a side effect of a feature not related to security, namely packet fragmentation.

Furthermore, \citet{Durvaux2020} explores how SCA against IoT devices can be carried out using relatively inexpensive components. The author demonstrates how to construct a full side-channel test bench using an EM probe and a Red Pitaya STEMlab platform for sampling to measure power leakages. The authors implement the AES-256 algorithm in C programming language using a simplistic 8-bit architecture, which is carried out using the Arduino Uno board's integrated 8-bit Atmel ATmega328P microcontroller in one use case, and the 32-bit microcontroller Atmel ATSAM3X8E in the Arduino Due board in another. The authors demonstrate that pre-processing techniques can increase the success rate of these attacks. They also outline how to carry out a comprehensive key recovery attack, focusing on specific key bytes from each round. They address the use of key enumeration techniques to make up for the lack of information that can be used in the attack, yet they point out that this strategy is more difficult for AES-256 due to the dependence between targeted rounds. The results of their study demonstrate that even with relatively inexpensive components, SCA can still be carried out effectively on IoT devices.

\section{Countermeasures}



\begin{figure*}[htp]
    \centering
    \includegraphics[width=\textwidth]{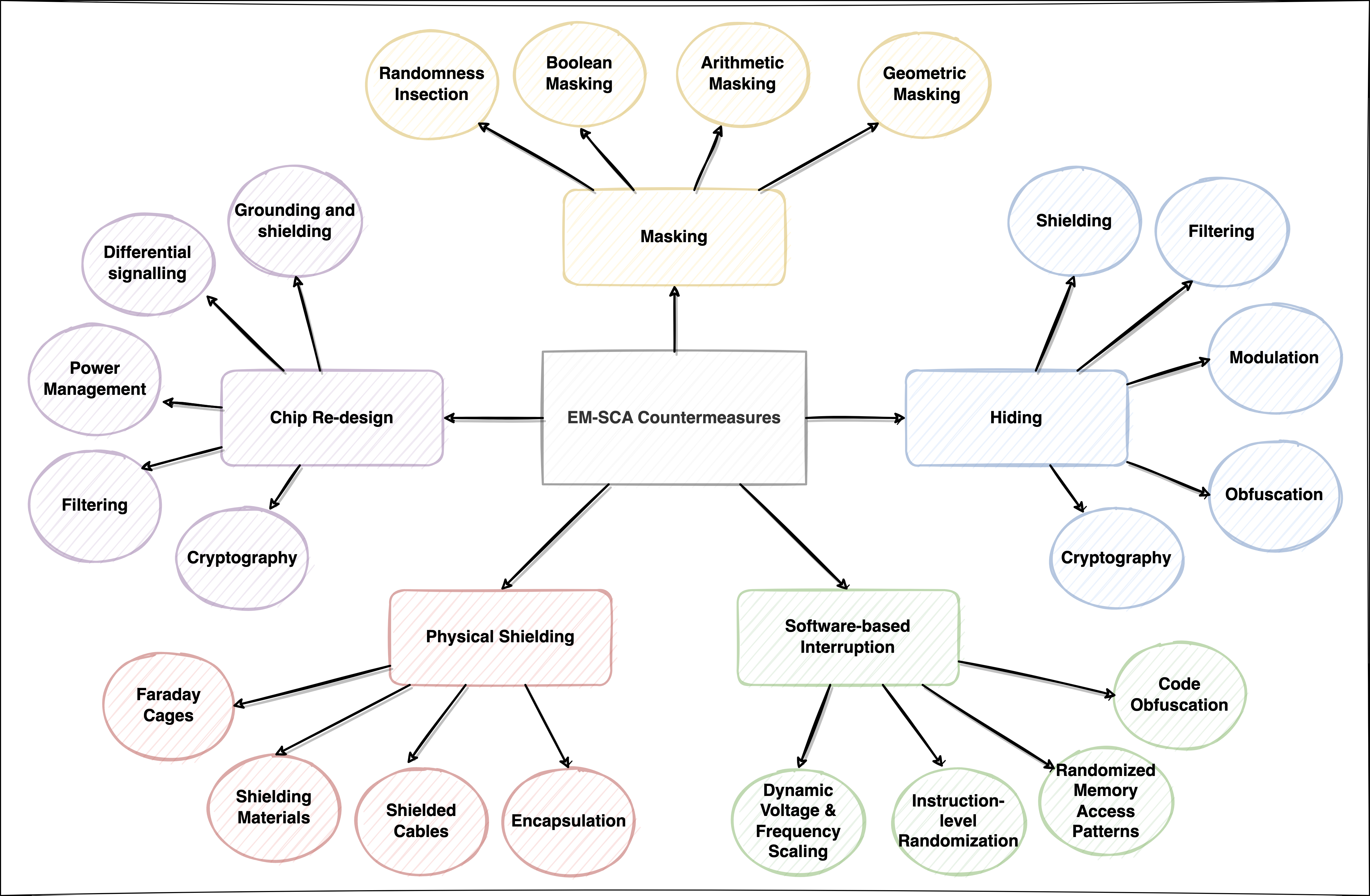}
    \caption{Variety of EM-SCA countermeasures identified as part of the literature review}
    \label{fig:CountermeasureEM-SCA}
\end{figure*}

While the primary focus of this literature review is not on countermeasures to EM-SCA, it would be remiss to omit a brief discussion on the topic. As a result, this section provides a brief overview of EM-SCA countermeasures in the literature. Due to the serious implications of EM-SCA in retrieving data from computing systems, numerous software and hardware countermeasures have been developed~\cite{Naija2017, Agosta2018, Jevtic2021, Singh2019, Das2021a, Japa2021}. \citet{nassar2012rsm} proposed a masking scheme for AES called Rotating S-boxes Masking (RSM). The RSM is used to protect the AES design against SCAs (DPA, CPA). Generally, the masking techniques used to protect block cyphers against SCA are based on the internal modification of the block cypher architectures~\cite{nassar2012rsm}. 


Figure~\ref{fig:CountermeasureEM-SCA} illustrates the various countermeasures identified in the literature review, and their relation to the different categories of EM-SCA attacks. While these countermeasures can help to mitigate EM-SCA, it is important to note that they may not be insurmountable, and determined attackers with sufficient resources and expertise may still be able to circumvent them. 

In order to resist primary leaks and make SCA more challenging, one strategy is hiding and masking~\cite{camurati2018screaming, Chong2021, Shan2014}. \citet{camurati2018screaming} demonstrated that the physical separation of analogue and digital components or the use of a separate processor for cryptographic operations are two alternative methods for preventing information from leaking into radio signals. The paper also emphasizes the significance of countermeasure incorporation into chip design, such as the adoption of digital radio technologies that are less vulnerable to side-channel attacks and/or the integration of several dies inside a System in Package (SiP) technology. The research notes that these countermeasures might have a negative effect on performance and might be challenging to put into practice without jeopardizing other functional requirements, cost, or chip size. 

In a separate study, \citet{khan2017implications} aimed to investigate the effectiveness of shielding techniques in reducing EM emanations from local power grids. He focused on the use of metal-insulator-metal (MIM) capacitors and the effect of upper metal layers on EM emanations from lower metal layers. Sheet metal is commonly used for EM shielding due to its ability to absorb radio and magnetic waves, however, the authors pointed out that gaps in the shield and electrical resistivity of the conductor can reduce its shielding capability. The MIM capacitor, which is composed of two metal plates and a dielectric layer, was tested as a shielding solution. The results showed that while the MIM shielding reduced EM emission by almost 3 dB, it may also increase emanations through inductive coupling from the local power grid. 

In 2017, \citet{Naija2017} suggested a parallel design for the PRESENT block cypher as a single method of concealing defenses against EM-SCA.
While 10,000 EM traces are utilized to target a present-day serial architecture, 200,000 EM traces are used to attack the suggested design. The authors then suggested that the mutual authentication protocol implement a countermeasure by gradually restricting the quantity of EM traces. This restriction makes it impossible for the attacker to use EMA. When compared to current countermeasures at primitive block cyphers (2,471 gate equivalents (GEs)), the suggested countermeasure is based on a time delay function and only needs 960 GEs. 

Furthermore, \citet{Patranabis2019} addressed the issue of design-for-security approaches to create compact block cyphers with combined defenses against side-channel and fault attacks for IoT applications. They chose FPGAs as their target platforms because of the variety of programmable features that make them ideal for IoT. Three key design ideas were brought forward, the first dealing with side-channel protection and the other with fault attack defense. Recursive design strategies for MDS linear layers with lightweight roots are suggested with regard to fault attacks. Such linear layers serve the dual functions of establishing Fault Space Transformation (FST) for defending against fault attacks with characteristics similar to DFA and differential fault intensity attacks (DFIA) and providing diffusion for resistance against classical cryptanalysis. Combining two lightweight strategies—masking with periodic refresh and shuffling over several rounds—was assessed in the context of side-channel protection.

In 2022, \citet{judy2022electromagnetic} look at the most recent advancements in SCA and their defenses. The prominent encryption system AES is implemented in hardware, and the research primarily focuses on the direct interpretation of the EM output. The authors examined two strategies for preventing side-channel attacks—physical shielding and software-based interruption—using an ATMEGA8 AES implementation in their studies. The findings demonstrated that physical shielding provided some signal attenuation, but that the attack might still succeed with higher trace counts. On the other hand, the software-based interruption strategy improved the mean time of failure by more successfully interfering with the synchronization of the side-channel attack and obscuring its characterization. 

In 2023, \citet{gao2023eo} presented a new defensive method for guarding against both intrusive and non-intrusive EM-SCA. An active shield and a module to obfuscate information leaking make up the suggested framework. The top metal layer of the chip is used to implement a random Hamiltonian topology method known as the Artificial Fish-Swarm Random Hamiltonian Algorithm (AFSRHA), which is used to construct the active shield. The information leakage obfuscation module keeps track of intrusive attacks and sends out signals to the wire mesh to increase EM emission and strengthen the wire mesh's defense against SCA. Through simulated studies, the suggested EO-shield system's security was proven, and it was demonstrated to be capable of deterring intrusive attacks and lowering the correlation between EM emanations and data processing. The authors argue that in order to provide enough protection, the active shield does not necessarily need to comprise the entire top metal layer. The EO-shield scheme presents a promising solution for protecting against EM-SCA and provides a multidimensional approach by combining an active shield with an information leakage obfuscation module.

\section{Discussion}

Having conducted a thorough review of the extant literature pertaining to EM-SCA on encryption, it is imperative to determine the various conditions under which EM-SCA can be employed in different scenarios. This section delves into an in-depth examination of the three research questions that have been identified under different themes.

\subsection{RQ1: What is the state of the art of EM-SCA with respect to attacking encryption?}

The state of the art of EM-SCA with respect to attacking encryption is constantly evolving. Researchers continue to develop new methods and techniques for performing EM side-channel attacks, and they are becoming increasingly sophisticated. Research on EM-SCA on encryption algorithms began in the 1990s. However, the first published research paper on the topic is ``Power Analysis of Key Schedules'' by Paul Kocher, Joshua Jaffe, and Benjamin in 1999~\cite{kocher1999differential}. This paper presented a method for performing a side-channel attack on the key schedule of the DES algorithm using EM emissions. EM-SCA exploit the EM emissions that are generated by a device as it performs cryptographic operations. These emissions can include EM radiation, electric currents, and magnetic fields. By analyzing these emissions, an attacker can infer information about the cryptographic operations being performed, such as the secret key being used. 

Power analysis can measure the electrical power consumption of a device as it encrypts data~\cite{Gandolfi2001, Won2021, Yu2021}. Information about the secret key being used can be deduced by analyzing the power consumption patterns. Similarly, the EM radiation that a device emits as it encrypts data has been measured via EM-SCA analysis, which has been used extensively in publications. The analysis of EM emissions can reveal information about the secret key.

One of the most advanced EM-SCA techniques mention previously is called ``template attacks'', where an attacker uses a set of known plaintext-ciphertext pairs to create a \textit{template} that can be used to infer the secret key used for encryption~\cite{Bukasa2017, camurati2020understanding}. These template attacks are often quite successful against hardware-implemented advanced encryption algorithms, as shown by~\citet{Bukasa2017}. Machine learning-based attacks have been applied as a cutting-edge method that involves utilizing machine learning algorithms to examine the EM emissions produced by a device as it executes cryptographic operations~\cite{Yu2018a, mukhtar2021edge, Xiao2020, Yu2021}. These machine learning-based attacks can be particularly successful since they automatically recognize EM emission patterns that signify the use of the encryption key. Additionally, ``power analysis attacks'', whereby the attacker monitors the electrical power usage of a device while it encrypts data and utilizes this data to deduce information about the secret key, remains a well-liked technique for conducting EM-SCA.

Finally, it should be noted that the state of the art for EM-SCA on encryption is constantly changing and that new approaches and strategies are continually being developed. Even the most cutting-edge encryption algorithms and implementations can be subject to EM-SCA, thus it's crucial to bear this in mind while designing and putting into practice cryptographic systems.

\subsection{RQ2: Which encryption algorithms and encrypted devices are most susceptible to, and resilient against, EM-SCA attacks?}

Since EM-SCA attacks take advantage of the EM emissions produced by a device as it executes cryptographic operations, all encryption algorithms and devices are theoretically vulnerable to them. Nevertheless, some encryption methods and devices are more prone to them than others.

Symmetric encryption algorithms, i.e., AES and DES, are particularly vulnerable to EM side-channel attacks since they require a secret key to both encrypt and decrypt data~\cite{kumar2017efficient, Haas2022, kasper2009side}. EM-SCA can be used by attackers to extract information about the secret key, and examples of devices that implement DES and AES, and thus are vulnerable, e.g., smartphones~\cite{Haas2022}, laptops/computers~\cite{Sayakkara2018}, and other portable devices that use encryption to protect sensitive data. The AES algorithm is more effective compared to DES and can process larger block sizes of plaintext with low power dissipation and fast processing time. The AES has a higher potential for implementation in IoT systems~\cite{He2011a, Frieslaar2017a, Tirumaladass2020}, especially in sensor-based applications. However, the AES implementation is still vulnerable to SCA, particularly the CEMA attack~\cite{Pammu2016, Pammu2016a, Kong2019}. The AES-128 algorithm requires 10 rounds of iterations to encrypt the plaintext, making it highly effective compared to other encryption algorithms such as DES and triple-DES. Despite its effectiveness, the AES implementation may still be vulnerable to side-channel attacks due to the leakage of physical parameters that correlate with intermediate data, as demonstrated in the literature.

On the other hand, EM-SCA can also be carried out against public-key encryption algorithms i.e., RSA and ECC, albeit these attacks are typically more challenging to carry out than attacks against symmetric encryption algorithms. However, \citet{goller2015side} demonstrate the square-and-multiply approach to attack a reference software RSA implementation and \citet{belgarric2016side} performed EM analysis to pinpoint ECC addition and multiplication processes in the Bouncy Castle ECDSA implementation for Android. Examples of these types of devices include smartcards~\cite{Gandolfi2001, carluccio2005electromagnetic}, hardware security modules~\cite{carlier2005generalizing, Nakamura2015}, and other devices that use public-key encryption to protect sensitive data. In addition, there is a gap in the literature on the use of EM-SCA for EEA encryption, providing an avenue for future research to evaluate the effectiveness of EM-SCA attacks. However, no studies specifically assess the efficacy of EM-SCA attacks on EEA encryption. It is crucial to emphasize the necessity of further research to comprehend better the limitations and vulnerabilities of EEA encryption against EM-SCA attacks. The lack of information hinders law enforcement and digital forensic investigations, which struggle to access EEA-encrypted devices due to the limited literature on EM-SCA. More investigation is necessary to improve the understanding of EM-SCA attacks on EEA encryption and develop better countermeasures.

In addition, when compared to hardware-based systems, software-based devices are typically more vulnerable to EM side-channel attacks~\cite{goller2015side, Bukasa2017}. This is thus because software implementations frequently produce higher EM emissions than hardware does. Due to their frequent insufficient physical security features and restricted computing capacity, smartcards, RFID tags, and other small, portable devices are also especially susceptible to EM side-channel attacks.

Furthermore, since EM-SCA attacks take advantage of EM emissions produced by a device while it executes cryptographic operations, no encryption algorithm can be claimed to be totally resistant to them. Every encryption process produces some form of EM emissions. To be more resistant to EM-SCA than others, several encryption algorithms and devices have been developed. The \textit{masking} technique, which is utilized to hide the delicate cryptographic processes in the circuit, is one case of an encryption algorithm that has been developed to be more resistant to EM-SCA~\cite{camurati2018screaming, singh2018improved, 9424617}. By reducing the quantity of EM emissions produced by the cryptographic processes, this strategy can make it more challenging for an attacker to deduce information regarding the secret key. In terms of devices, hardware implementations are typically thought to be more resistant to EM-SCA than software implementations, since they typically produce fewer EM emissions. Devices that have been built with physical security features, including tamper-proof enclosures, can also be more resistant to EM-SCA.

The vulnerability of encryption algorithms to EM-SCA can fluctuate over time, since the state-of-the-art in these attacks is always advancing. Furthermore, a cryptographic system's precise implementation characteristics may have an impact on how vulnerable a certain encryption technique is. Similarly, the resilience of a device can also depend on the precise implementation details of the cryptographic system. It is necessary to note that an encryption algorithm's or a device's resilience to EM-SCA can vary over time.

\subsection{RQ3: Which approaches to EM-SCA on encryption prove most fruitful, or demonstrates the most promise for the future?}

EM-SCA exploit information leaked through EM emanations from a device, i.e., as power consumption or EM radiation, to extract secret information, i.e., as encryption keys. Although there are several EM-SCA techniques that have been demonstrated in the literature, most of them appear to be employable under ideal circumstances, for example, using Faraday cages to suppress noise or selecting the most effective victim devices for an attack. It is important to investigate which EM-SCA approaches are more realistic and practically usable under real-world conditions. One such method utilizes live data forensic techniques to look into encrypted and powered-on devices~\cite{Sayakkara2018a}. A strategy like this may compel the victim device to carry out cryptographic operations, which could then be seen using EM emissions to extract the key. However, this method may not be practical for many types of personal devices, such as those that employ SSL-based online traffic and infrequently execute cryptographic operations. According to the literature, using several side-channel attacks to target a single computer system can be more effective than utilizing one side-channel attack alone. Another practical approach for EM-SCA attacks involves the use of fault injection techniques to introduce faults into the cryptographic device's operation, which can be monitored using EM emissions. This approach has been demonstrated in several studies and has been shown to be successful in recovering encryption keys, e.g., FIA and EM-SCA can be combined to get superior results, as has been demonstrated~\cite{Ravi2020}.

Modern SoC processors used in smartphones include many cores that can execute several threads at once~\cite{SAYAKKARA201943}. As a result, the device can run many exhaustively significant software operations concurrently, such as encrypting data stored in non-volatile storage and carrying out a wireless network connection. The number of concurrent activities grows further as the number of cores increases. EM radiation can be seen, and EM trace data can be gathered as long as the CPU cores operate at the same clock frequency. Similarly, the usage of DVFS methods in contemporary CPUs~\cite{chawla2020securing} is another potential issue that contributes to this predicament. It enables a SoC processor to change its clock rates dynamically in response to workload. These methods require for the simultaneous observation of EM radiation at several frequencies, as well as the ability to spontaneously and dynamically modify the frequency of the signal being observed. One prominent example of the impact of DVFS on side-channel attacks is the VoltJockey exploit, which was proposed by~\cite{10.1145/3319535.3354201}. It represents a cutting-edge, fault-based attack that targets the multicore CPUs' DVFS capabilities. According to these findings, the use of DVFS in modern CPUs poses a serious risk to the security of mobile devices, especially when combined with side-channel attacks such as EM radiation analysis. Thus, the discovery of possible side-channel attack vulnerabilities in contemporary smartphones has significant implications for law enforcement and digital investigators. However, due to the complexity of these attacks, it is essential to have the knowledge and tools required to carry them out successfully. With appropriate training and tools, these agencies can access encrypted data while maintaining the security and privacy of individuals and organizations. It is crucial to stress that these attacks must only be used in appropriate situations and compliance with relevant laws and regulations.

In addition, recent developments in AI and DL have shown significant uses for implementing EM-SCA. To be more precise, the analysis of EM-SCA using deep learning techniques i.e., CNN, DNN, MTL and MLP, has produced encouraging results that have been published. These methods can be trained to detect patterns in EM signals that signify specific encryption keys or algorithms, making them effective in exploiting side-channel leakages. Besides, applying CNN and DNN algorithms that can adjust to the changing EM environment and learn to extract sensitive information from it has previously demonstrated promising results. On top of that, the literature suggests that GAN can be employed to generate more data, thereby improving the results of EM-SCA. Another promising approach involves using transfer learning techniques to pre-train machine learning models on a large dataset of EM signals, and then fine-tuning the model for specific encryption algorithms. To summarize, utilizing deep learning techniques i.e., CNN, DNN, and MTL, in EM-SCA on encryption can be considered the most promising approach for the future. These advancements provide a significant opportunity to enhance the effectiveness of EM-SCA in attacking encryption, making it an effective mechanism for digital forensics and law enforcement.

\section{Conclusion and Future Directions}


The state of the art of EM-SCA with respect to attacking encryption is ever-evolving, and new methods are being constantly developed. EM-SCAs are a major concern for the security of cryptographic systems. These attacks take advantage of the EM emissions generated by a device as it performs cryptographic operations to infer information about the secret key used for encryption. The evolution of EM-SCA techniques has led to a wide range of successful attacks, from power analysis attacks to machine learning-based attacks. Symmetric encryption algorithms like AES and DES are particularly vulnerable to EM-SCA, while public-key encryption algorithms like RSA are also subject to these attacks, albeit to a lesser degree. Power analysis, which counts the amount of electricity used by a device to encrypt data, is one of the earliest techniques for carrying out EM-SCA attacks. The secret key's information can then be determined using this data on power use. Other EM-SCA techniques include template attacks, which use well-known plaintext-ciphertext pairs to create a template from which the secret key can be deduced, and machine learning-based attacks, which use machine learning algorithms to look at the EM emissions generated by a device as it executes cryptographic operations. 


Although many EM-SCA techniques in the literature require ideal conditions, i.e., using Faraday cages or selecting specific victim devices, practical approaches using live data forensic technique can be utilized to investigate encrypted and powered-on devices. This can compel the device to perform cryptographic operations that can be analyzed using EM emissions. Another practical approach is using fault injection techniques to introduce fault into the cryptographic device's operation, which can be monitored using EM emissions and has successfully recovered encryption keys. Additionally, using multiple side-channel attacks to target a single computer or digital device system can be more effective than using a single side-channel attack. Furthermore, machine learning algorithms show promise in the future of EM-SCA research. Future research can focus on developing effective strategies for ransomware disaster recovery and data recover, e.g., EM-SCA can be used to extract the key, eliminating the need to pay the ransom.

Advancing this argument and conducting further research is necessary to improve the understanding of the limitations and vulnerabilities of encryption in the face of EM-SCA attacks. Several open questions require further exploration, i.e., determining the most practical and realistic EM-SCA approaches under real-world conditions, including law enforcement and digital forensic needs. In addition, EEA encryption has received little attention in the literature, and further investigation is necessary to understand its vulnerabilities against EM-SCA attacks better. Moreover, Deep Learning-based attacks, e.g., CNN, DNN, MTL, and MLP, have the ability to identify patterns in EM signals that indicate specific encryption keys or algorithms, showing potential for use in digital forensic and law enforcement applications. To further advance this research area, new techniques and approaches should be developed to investigate the use of EM-SCA for ransomware disaster recovery and data recovery. This approach can be particularly useful in cases where traditional data recovery methods are not feasible, e.g., when the encrypted data has been overwritten or deleted. However, more research is required to investigate this approach's effectiveness and develop techniques for practical implementation in real-world scenarios. Additionally, exploring machine learning-based attacks and utilizing EM-SCA in law enforcement and digital forensics are critical areas of research. Therefore, further research is needed to address unanswered questions and challenges in these areas to enhance the effectiveness of EM-SCA in combating encryption and supporting law enforcement and digital forensics efforts.

As EM-SCA techniques continue to advance, it is crucial to consider the security of cryptographic systems during design and implementation. To this end, numerous software and hardware countermeasures have been proposed and developed to protect against these attacks, including masking techniques, cryptographic countermeasures, and the physical separation of analogue and digital components. However, these countermeasures may have a negative effect on performance and may be challenging to implement without compromising other requirements like cost and chip size. Consequently, the future direction of EM-SCA research should focus on developing more effective countermeasures that are secure and efficient. This includes the exploration of new hardware-based countermeasures, the improvement of existing software-based countermeasures, and the integration of these countermeasures into cryptographic systems in a way that does not negatively impact performance. Moreover, the development of new EM-SCA attack techniques will likely continue, making it important for the community to stay updated on the latest advancements and continuously improve the security of cryptographic systems.

In addition, quantum computing can potentially transform the computing industry due to its ability to solve issues that conventional computers are currently unable to handle. Quantum computers will be able to break the encryption used to protect sensitive data as they develop in capability, which will jeopardize system security~\cite{buchanan2017will}. This has ramifications for those who depend on encryption to protect their personal information as well as for sectors including finance, healthcare, and national security. The influence of quantum computing on public-key cryptography, the foundation of contemporary internet security, is one of its most important consequences. The foundation of public-key cryptography is the inefficiency of finding discrete logarithms or large-number factors—both of which can be solved exponentially more quickly on a quantum computer. This indicates that a sufficiently powerful quantum computer could decrypt data using the popular AES, DES, and ECC encryption techniques. An extensive body of literature has been dedicated to the pursuit of developing quantum-resistant encryption techniques. This is due to the advent of quantum computing and its potential to disrupt traditional encryption methods. Several quantum-resistant encryption algorithms are currently under investigation, including Lattice-based cryptography~\cite{mavroeidis2018impact}, Multivariate cryptography~\cite{fernandez2020towards}, and code-based cryptography~\cite{sendrier2017code}. These algorithmic developments are aimed at providing a secure encryption mechanism in the post-quantum era, specifically designed to withstand attacks from quantum computers. As a result, the field of cryptography and system security will be significantly impacted by the development of quantum computing. As the advancements in quantum computing pose a threat to the current encryption algorithms, it is imperative for the future development of new quantum-resistant encryption methods and quantum-powered encryption. Nonetheless, it is important to note that even quantum encryption algorithms may still be vulnerable to various EM-SCA approaches~\cite{10.1145/3338467.3358948}. This underscores the need for individuals and organizations to be cognizant of the potential impacts of quantum computing on their systems and data. 

As encryption becomes more prevalent in daily lives, digital forensic investigators face increasing difficulties accessing and recovering digital evidence from encrypted devices. Strong encryption cannot be bypassed without a key or passphrase, leaving investigators with the decision to perform a live forensic acquisition. Therefore, it is essential for researchers and practitioners in the field of cryptography to ensure that future quantum-facilitated cryptographic systems they develop and implement are resistant to EM-SCA. Furthermore, it is imperative to develop new techniques and approaches to break encryption, investigate how EM-SCA can be used for disaster recovery and data recovery, and explore machine learning-based attacks. In this regard, addressing the vulnerability of quantum encryption algorithms to EM-SCA approaches can equip digital forensic investigators with the necessary tools to overcome the challenges posed by strong encryption, enabling them to continue playing a critical role in law enforcement and the justice system. Thus, it is vital to continue research in this area to ensure that digital forensic investigators can fulfill their role in the justice system while maintaining the privacy and security of electronic devices and data.

\bibliographystyle{IEEEtranN}
\bibliography{bibfile}

\begin{thebibliography}{273}
\providecommand{\natexlab}[1]{#1}
\providecommand{\url}[1]{#1}
\csname url@samestyle\endcsname
\providecommand{\newblock}{\relax}
\providecommand{\bibinfo}[2]{#2}
\providecommand{\BIBentrySTDinterwordspacing}{\spaceskip=0pt\relax}
\providecommand{\BIBentryALTinterwordstretchfactor}{4}
\providecommand{\BIBentryALTinterwordspacing}{\spaceskip=\fontdimen2\font plus
\BIBentryALTinterwordstretchfactor\fontdimen3\font minus \fontdimen4\font\relax}
\providecommand{\BIBforeignlanguage}[2]{{%
\expandafter\ifx\csname l@#1\endcsname\relax
\typeout{** WARNING: IEEEtranN.bst: No hyphenation pattern has been}%
\typeout{** loaded for the language `#1'. Using the pattern for}%
\typeout{** the default language instead.}%
\else
\language=\csname l@#1\endcsname
\fi
#2}}
\providecommand{\BIBdecl}{\relax}
\BIBdecl

\bibitem[Zhou and Feng(2005)]{cryptoeprint:2005/388}
\BIBentryALTinterwordspacing
Y.~Zhou and D.~Feng, ``Side-channel attacks: Ten years after its publication and the impacts on cryptographic module security testing,'' Cryptology ePrint Archive, Paper 2005/388, 2005, \url{https://eprint.iacr.org/2005/388}. [Online]. Available: \url{https://eprint.iacr.org/2005/388}
\BIBentrySTDinterwordspacing

\bibitem[Sayakkara et~al.(2019{\natexlab{a}})Sayakkara, Le-Khac, and Scanlon]{SAYAKKARA201943}
\BIBentryALTinterwordspacing
A.~Sayakkara, N.-A. Le-Khac, and M.~Scanlon, ``A survey of electromagnetic side-channel attacks and discussion on their case-progressing potential for digital forensics,'' \emph{Digital Investigation}, vol.~29, pp. 43--54, 2019. [Online]. Available: \url{https://www.sciencedirect.com/science/article/pii/S1742287618303840}
\BIBentrySTDinterwordspacing

\bibitem[Das et~al.(2020)Das, Nath, Ghosh, and Sen]{Das2020c}
\BIBentryALTinterwordspacing
D.~Das, M.~Nath, S.~Ghosh, and S.~Sen, ``{Killing EM side-channel leakage at its source},'' \emph{2020 IEEE 63rd International}, 2020. [Online]. Available: \url{https://ieeexplore.ieee.org/abstract/document/9184657/}
\BIBentrySTDinterwordspacing

\bibitem[Sayakkara et~al.(2019{\natexlab{b}})Sayakkara, Le-Khac, and Scanlon]{Sayakkara2019a}
\BIBentryALTinterwordspacing
A.~Sayakkara, N.-A. Le-Khac, and M.~Scanlon, ``{Leveraging Electromagnetic Side-Channel Analysis for the Investigation of IoT Devices},'' \emph{Digital Investigation}, vol.~29, 2019. [Online]. Available: \url{https://www.sciencedirect.com/science/article/pii/S1742287619301616}
\BIBentrySTDinterwordspacing

\bibitem[Getz and Moeckel(1996)]{GetzRMoeckel}
R.~Getz and B.~Moeckel, ``Understanding and eliminating emi in microcontroller applications,'' \emph{National Semiconductor}, 1996.

\bibitem[Quisquater and Samyde(2001)]{quisquater2001electromagnetic}
J.-J. Quisquater and D.~Samyde, ``Electromagnetic analysis (ema): Measures and counter-measures for smart cards,'' in \emph{Proceedings of the International Conference on Research in Smart Cards: Smart Card Programming and Security}, ser. E-SMART '01.\hskip 1em plus 0.5em minus 0.4em\relax Berlin, Heidelberg: Springer-Verlag, 2001, p. 200–210.

\bibitem[Du et~al.(2020)Du, Hargreaves, Sheppard, Anda, Sayakkara, Le-Khac, and Scanlon]{Du2020}
\BIBentryALTinterwordspacing
X.~Du, C.~Hargreaves, J.~Sheppard, F.~Anda, A.~Sayakkara, N.-A. Le-Khac, and M.~Scanlon, ``Sok: Exploring the state of the art and the future potential of artificial intelligence in digital forensic investigation,'' in \emph{Proceedings of the 15th International Conference on Availability, Reliability and Security}, ser. ARES '20.\hskip 1em plus 0.5em minus 0.4em\relax New York, NY, USA: Association for Computing Machinery, 2020. [Online]. Available: \url{https://doi.org/10.1145/3407023.3407068}
\BIBentrySTDinterwordspacing

\bibitem[De~Mulder et~al.(2005)De~Mulder, Buysschaert, Ors, Delmotte, Preneel, Vandenbosch, and Verbauwhede]{de2005electromagnetic}
E.~De~Mulder, P.~Buysschaert, S.~Ors, P.~Delmotte, B.~Preneel, G.~Vandenbosch, and I.~Verbauwhede, ``Electromagnetic analysis attack on an fpga implementation of an elliptic curve cryptosystem,'' in \emph{EUROCON 2005 - The International Conference on "Computer as a Tool"}, vol.~2, 2005, pp. 1879--1882.

\bibitem[Gandolfi et~al.(2001)Gandolfi, Mourtel, and Olivier]{Gandolfi2001}
K.~Gandolfi, C.~Mourtel, and F.~Olivier, ``Electromagnetic analysis: Concrete results,'' in \emph{Cryptographic Hardware and Embedded Systems --- CHES 2001}.\hskip 1em plus 0.5em minus 0.4em\relax Springer Berlin Heidelberg, 2001, pp. 251--261.

\bibitem[Danial et~al.(2020)Danial, Das, Ghosh, Raychowdhury, and Sen]{Danial2020a}
J.~Danial, D.~Das, S.~Ghosh, A.~Raychowdhury, and S.~Sen, ``Scniffer: Low-cost, automated, efficient electromagnetic side-channel sniffing,'' \emph{IEEE Access}, vol.~8, pp. 173\,414--173\,427, 2020.

\bibitem[Sayakkara et~al.(2018{\natexlab{a}})Sayakkara, Le-Khac, and Scanlon]{Sayakkara2018a}
\BIBentryALTinterwordspacing
A.~Sayakkara, N.-A. Le-Khac, and M.~Scanlon, ``{Electromagnetic Side-channel Attacks: Potential for Progressing Hindered Digital Forensic Analysis},'' in \emph{Companion Proceedings for the ISSTA/ECOOP 2018 Workshops}.\hskip 1em plus 0.5em minus 0.4em\relax Association for Computing Machinery, 2018, p. 138. [Online]. Available: \url{https://doi.org/10.1145/3236454.3236512}
\BIBentrySTDinterwordspacing

\bibitem[Casey et~al.(2011)Casey, Fellows, Geiger, and Stellatos]{CASEY2011129}
\BIBentryALTinterwordspacing
E.~Casey, G.~Fellows, M.~Geiger, and G.~Stellatos, ``The growing impact of full disk encryption on digital forensics,'' \emph{Digital Investigation}, vol.~8, no.~2, pp. 129--134, 2011, standards, professionalization and quality in digital forensics. [Online]. Available: \url{https://www.sciencedirect.com/science/article/pii/S1742287611000727}
\BIBentrySTDinterwordspacing

\bibitem[Lillis et~al.(2016)Lillis, Becker, O'Sullivan, and Scanlon]{David2016}
\BIBentryALTinterwordspacing
D.~Lillis, B.~Becker, T.~O'Sullivan, and M.~Scanlon, ``Current challenges and future research areas for digital forensic investigation,'' 2016. [Online]. Available: \url{https://arxiv.org/abs/1604.03850}
\BIBentrySTDinterwordspacing

\bibitem[Lou et~al.(2021)Lou, Zhang, Jiang, and Zhang]{10.1145/3456629}
\BIBentryALTinterwordspacing
X.~Lou, T.~Zhang, J.~Jiang, and Y.~Zhang, ``A survey of microarchitectural side-channel vulnerabilities, attacks, and defenses in cryptography,'' \emph{ACM Comput. Surv.}, vol.~54, no.~6, jul 2021. [Online]. Available: \url{https://doi.org/10.1145/3456629}
\BIBentrySTDinterwordspacing

\bibitem[Kitchenham(2004)]{kitchenham2004procedures}
B.~Kitchenham, ``Procedures for performing systematic reviews,'' \emph{Keele University Technical Report TR/SE-0401}, 2004.

\bibitem[Harzing(2023)]{HarzingPOP}
A.-W. Harzing, ``Publish or perish,'' \url{https://harzing.com/resources/publish-or-perish}, 2023, accessed on Jun 27, 2022.

\bibitem[Dechammakl(2018)]{EMDiagram}
Dechammakl, ``Electromagnetic waves,'' \url{https://commons.wikimedia.org/wiki/File:Electromagnetic_waves.png}, 2018, accessed: 2022-09-07.

\bibitem[Naren et~al.(2020)Naren, Elhence, Chamola, and Guizani]{9016183}
Naren, A.~Elhence, V.~Chamola, and M.~Guizani, ``Notice of retraction: Electromagnetic radiation due to cellular, wi-fi and bluetooth technologies: How safe are we?'' \emph{IEEE Access}, vol.~8, pp. 42\,980--43\,000, 2020.

\bibitem[Ishimaru(2017)]{ishimaru1144/3294442}
A.~Ishimaru, \emph{Electromagnetic wave propagation, radiation, and scattering: from fundamentals to applications}.\hskip 1em plus 0.5em minus 0.4em\relax John Wiley and Sons, 2017.

\bibitem[Rizvi et~al.(2022)Rizvi, Scanlon, McGibney, and Sheppard]{rizvi2022network}
S.~Rizvi, M.~Scanlon, J.~McGibney, and J.~Sheppard, ``Application of artificial intelligence to network forensics: Survey, challenges and future directions,'' \emph{IEEE Access}, vol.~10, pp. 110\,362--110\,384, 2022.

\bibitem[Asthana et~al.(2017)Asthana, Megahed, and Strong]{8027264}
S.~Asthana, A.~Megahed, and R.~Strong, ``A recommendation system for proactive health monitoring using iot and wearable technologies,'' in \emph{2017 IEEE International Conference on AI and Mobile Services (AIMS)}, 2017, pp. 14--21.

\bibitem[Zamanian and Hardiman(2005)]{zamanian2005electromagnetic}
A.~Zamanian and C.~Hardiman, ``Electromagnetic radiation and human health: A review of sources and effects,'' \emph{High Frequency Electronics}, vol.~4, no.~3, pp. 16--26, 2005.

\bibitem[Hayashi and Homma(2019)]{hayashi2019introduction}
Y.-i. Hayashi and N.~Homma, ``Introduction to electromagnetic information security,'' \emph{IEICE Transactions on Communications}, vol. 102, no.~1, pp. 40--50, 2019.

\bibitem[Ott(2011)]{ott2011electromagnetic}
\BIBentryALTinterwordspacing
H.~W. Ott, \emph{Electromagnetic Compatibility Engineering}.\hskip 1em plus 0.5em minus 0.4em\relax Wiley, 2011. [Online]. Available: \url{https://books.google.ie/books?id=2-4WJKxzzigC}
\BIBentrySTDinterwordspacing

\bibitem[Bito et~al.(2017)Bito, Bahr, Hester, Nauroze, Georgiadis, and Tentzeris]{7857107}
J.~Bito, R.~Bahr, J.~G. Hester, S.~A. Nauroze, A.~Georgiadis, and M.~M. Tentzeris, ``A novel solar and electromagnetic energy harvesting system with a 3-d printed package for energy efficient internet-of-things wireless sensors,'' \emph{IEEE Transactions on Microwave Theory and Techniques}, vol.~65, no.~5, pp. 1831--1842, 2017.

\bibitem[Poulin and Amiot(2002)]{poulin2002interference}
F.~Poulin and L.-P. Amiot, ``Interference during the use of an electromagnetic tracking system under or conditions,'' \emph{Journal of Biomechanics}, vol.~35, no.~6, pp. 733--737, 2002.

\bibitem[Longo et~al.(2015{\natexlab{a}})Longo, Mulder, Page, and Tunstall]{Longo2015}
\BIBentryALTinterwordspacing
J.~Longo, E.~D. Mulder, D.~Page, and M.~Tunstall, ``{SoC it to EM: electromagnetic side-channel attacks on a complex system-on-chip},'' 2015. [Online]. Available: \url{https://link.springer.com/chapter/10.1007/978-3-662-48324-4_31}
\BIBentrySTDinterwordspacing

\bibitem[Cheffena(2016)]{7433518}
M.~Cheffena, ``Propagation channel characteristics of industrial wireless sensor networks [wireless corner],'' \emph{IEEE Antennas and Propagation Magazine}, vol.~58, no.~1, pp. 66--73, 2016.

\bibitem[Lapinsky and Easty(2006)]{lapinsky2006electromagnetic}
\BIBentryALTinterwordspacing
S.~E. Lapinsky and A.~C. Easty, ``Electromagnetic interference in critical care,'' \emph{Journal of Critical Care}, vol.~21, no.~3, pp. 267--270, 2006. [Online]. Available: \url{https://www.sciencedirect.com/science/article/pii/S0883944106000499}
\BIBentrySTDinterwordspacing

\bibitem[Sayakkara et~al.(2018{\natexlab{b}})Sayakkara, Le-Khac, and Scanlon]{Sayakkara2018}
\BIBentryALTinterwordspacing
A.~Sayakkara, N.-A. Le-Khac, and M.~Scanlon, ``{Accuracy Enhancement of Electromagnetic Side-Channel Attacks on Computer Monitors},'' in \emph{Proceedings of the 13th International Conference on Availability, Reliability and Security}.\hskip 1em plus 0.5em minus 0.4em\relax Association for Computing Machinery, 2018. [Online]. Available: \url{https://doi.org/10.1145/3230833.3234690}
\BIBentrySTDinterwordspacing

\bibitem[Song et~al.(2015)Song, Jeong, and Yook]{song2014modeling}
T.-L. Song, Y.-R. Jeong, and J.-G. Yook, ``Modeling of leaked digital video signal and information recovery rate as a function of snr,'' \emph{IEEE Transactions on Electromagnetic Compatibility}, vol.~57, no.~2, pp. 164--172, 2015.

\bibitem[Radasky et~al.(2004)Radasky, Baum, and Wik]{radasky2004introduction}
W.~A. Radasky, C.~E. Baum, and M.~W. Wik, ``Introduction to the special issue on high-power electromagnetics (hpem) and intentional electromagnetic interference (iemi),'' \emph{IEEE Transactions on Electromagnetic Compatibility}, vol.~46, no.~3, pp. 314--321, 2004.

\bibitem[Rahimpour et~al.(2021)Rahimpour, Kiyani, Hodges, and Turner]{rahimpour2021deep}
\BIBentryALTinterwordspacing
S.~Rahimpour, M.~Kiyani, S.~E. Hodges, and D.~A. Turner, ``Deep brain stimulation and electromagnetic interference,'' \emph{Clinical Neurology and Neurosurgery}, vol. 203, p. 106577, 2021. [Online]. Available: \url{https://www.sciencedirect.com/science/article/pii/S0303846721001049}
\BIBentrySTDinterwordspacing

\bibitem[{van Eck}(1985)]{VANECK1985269}
\BIBentryALTinterwordspacing
W.~{van Eck}, ``Electromagnetic radiation from video display units: An eavesdropping risk?'' \emph{Computers and Security}, vol.~4, no.~4, pp. 269--286, 1985. [Online]. Available: \url{https://www.sciencedirect.com/science/article/pii/016740488590046X}
\BIBentrySTDinterwordspacing

\bibitem[Kuhn(2005)]{10.1007/11423409_7}
M.~G. Kuhn, ``Electromagnetic eavesdropping risks of flat-panel displays,'' in \emph{Privacy Enhancing Technologies}, D.~Martin and A.~Serjantov, Eds.\hskip 1em plus 0.5em minus 0.4em\relax Springer Berlin Heidelberg, 2005, pp. 88--107.

\bibitem[Cobb et~al.(2012)Cobb, Laspe, Baldwin, Temple, and Kim]{5898407}
W.~E. Cobb, E.~D. Laspe, R.~O. Baldwin, M.~A. Temple, and Y.~C. Kim, ``Intrinsic physical-layer authentication of integrated circuits,'' \emph{IEEE Transactions on Information Forensics and Security}, vol.~7, no.~1, pp. 14--24, 2012.

\bibitem[Maxwell(1865)]{10.2307/108892}
\BIBentryALTinterwordspacing
J.~C. Maxwell, ``A dynamical theory of the electromagnetic field,'' \emph{Philosophical Transactions of the Royal Society of London}, vol. 155, pp. 459--512, 1865. [Online]. Available: \url{http://www.jstor.org/stable/108892}
\BIBentrySTDinterwordspacing

\bibitem[Lavaud et~al.(2021)Lavaud, Gerzaguet, Gautier, Berder, and ...]{Lavaud2021}
\BIBentryALTinterwordspacing
C.~Lavaud, R.~Gerzaguet, M.~Gautier, O.~Berder, and ..., ``{Whispering devices: A survey on how side-channels lead to compromised information},'' \emph{Journal of Hardware and}, 2021. [Online]. Available: \url{https://link.springer.com/article/10.1007/s41635-021-00112-6}
\BIBentrySTDinterwordspacing

\bibitem[Xu et~al.(2010)Xu, Su, and Zhou]{5404400}
J.~L. Xu, W.~Su, and M.~Zhou, ``Software-defined radio equipped with rapid modulation recognition,'' \emph{IEEE Transactions on Vehicular Technology}, vol.~59, no.~4, pp. 1659--1667, 2010.

\bibitem[Machado and Wyglinski(2015)]{7086416}
R.~G. Machado and A.~M. Wyglinski, ``Software-defined radio: Bridging the analog–digital divide,'' \emph{Proceedings of the IEEE}, vol. 103, no.~3, pp. 409--423, 2015.

\bibitem[Osmann(2020)]{GreatScottGadgets}
\BIBentryALTinterwordspacing
M.~Osmann, ``Software defined radio with hackrf,'' 2020, accessed: 2022-07-21. [Online]. Available: \url{https://greatscottgadgets.com/sdr}
\BIBentrySTDinterwordspacing

\bibitem[Martoyo et~al.(2018)Martoyo, Setiasabda, Kanalebe, Uranus, and Pardede]{9084568}
I.~Martoyo, P.~Setiasabda, H.~Y. Kanalebe, H.~P. Uranus, and M.~Pardede, ``Software defined radio for education: Spectrum analyzer, fm receiver/transmitter and gsm sniffer with hackrf one,'' in \emph{2018 2nd Borneo International Conference on Applied Mathematics and Engineering (BICAME)}, 2018, pp. 188--192.

\bibitem[Kularatna(2003)]{kularatna2003digital}
N.~Kularatna, \emph{Digital and Analogue Instrumentation: Testing and Measurement}.\hskip 1em plus 0.5em minus 0.4em\relax Institution of Engineering and Technology, 2003.

\bibitem[R.~Vanathi(2019)]{VanathiSidechannelIaas}
\BIBentryALTinterwordspacing
S.~C. R.~Vanathi, ``Side channel attacks in iaas and its defense mechanisms,'' \emph{International Journal of Engineering and Advanced Technology (IJEAT)}, vol.~8, no.~3S, 2019. [Online]. Available: \url{https://www.amrita.edu/publication/side-channel-attacks-in-iaas-and-its-defense-mechanisms/}
\BIBentrySTDinterwordspacing

\bibitem[Kocher(1996)]{kocher1996timing}
P.~C. Kocher, ``Timing attacks on implementations of diffie-hellman, rsa, dss, and other systems,'' in \emph{Advances in Cryptology --- CRYPTO '96}, N.~Koblitz, Ed.\hskip 1em plus 0.5em minus 0.4em\relax Springer Berlin Heidelberg, 1996, pp. 104--113.

\bibitem[Kong et~al.(2009)Kong, Aciicmez, Seifert, and Zhou]{4798277}
J.~Kong, O.~Aciicmez, J.-P. Seifert, and H.~Zhou, ``Hardware-software integrated approaches to defend against software cache-based side channel attacks,'' in \emph{2009 IEEE 15th International Symposium on High Performance Computer Architecture}, 2009, pp. 393--404.

\bibitem[Won and Bhasin(2021)]{Won2021}
Y.-S. Won and S.~Bhasin, ``A systematic side-channel evaluation of black box aes in secure mcu: Architecture recovery and retrieval of puf based secret key,'' in \emph{2021 IEEE International Symposium on Circuits and Systems (ISCAS)}, 2021, pp. 1--5.

\bibitem[Ueno et~al.(2022)Ueno, Xagawa, Tanaka, Ito, Takahashi, and Homma]{Ueno2022}
\BIBentryALTinterwordspacing
R.~Ueno, K.~Xagawa, Y.~Tanaka, A.~Ito, J.~Takahashi, and N.~Homma, ``{Curse of Re-encryption: A Generic Power/Em Analysis on Post-quantum Kems},'' \emph{IACR Transactions on Cryptographic Hardware and Embedded Systems}, no.~1, Nov. 2022. [Online]. Available: \url{https://tches.iacr.org/index.php/TCHES/article/view/9298}
\BIBentrySTDinterwordspacing

\bibitem[Lisovets et~al.(2021)Lisovets, Knichel, Moos, and Moradi]{lisovets2021let}
O.~Lisovets, D.~Knichel, T.~Moos, and A.~Moradi, ``Let’s take it offline: Boosting brute-force attacks on iphone’s user authentication through sca,'' \emph{IACR Transactions on Cryptographic Hardware and Embedded Systems}, vol.~3, pp. 496--519, 2021.

\bibitem[Anderson(2020)]{anderson2020security}
R.~Anderson, \emph{Security Engineering: A Guide to Building Dependable Distributed Systems}.\hskip 1em plus 0.5em minus 0.4em\relax Wiley, 2020.

\bibitem[Batina et~al.(2019)Batina, Bhasin, Jap, and Picek]{batina2019csi}
L.~Batina, S.~Bhasin, D.~Jap, and S.~Picek, ``{CSI NN: Reverse Engineering of Neural Network Architectures through Electromagnetic Side Channel},'' in \emph{Proceedings of the 28th USENIX Conference on Security Symposium}, ser. SEC'19.\hskip 1em plus 0.5em minus 0.4em\relax USA: USENIX Association, 2019, p. 515–532.

\bibitem[Moradi et~al.(2011)Moradi, Barenghi, Kasper, and Paar]{10.1145/2046707.2046722}
\BIBentryALTinterwordspacing
A.~Moradi, A.~Barenghi, T.~Kasper, and C.~Paar, ``On the vulnerability of fpga bitstream encryption against power analysis attacks: Extracting keys from xilinx virtex-ii fpgas,'' in \emph{Proceedings of the 18th ACM Conference on Computer and Communications Security}, ser. CCS '11.\hskip 1em plus 0.5em minus 0.4em\relax New York, NY, USA: Association for Computing Machinery, 2011, p. 111–124. [Online]. Available: \url{https://doi.org/10.1145/2046707.2046722}
\BIBentrySTDinterwordspacing

\bibitem[Mayer-Sommer(2000)]{10.1007/3-540-44499-8_6}
R.~Mayer-Sommer, ``Smartly analyzing the simplicity and the power of simple power analysis on smartcards,'' in \emph{Cryptographic Hardware and Embedded Systems --- CHES 2000}, {\c{C}}.~K. Ko{\c{c}} and C.~Paar, Eds.\hskip 1em plus 0.5em minus 0.4em\relax Springer Berlin Heidelberg, 2000, pp. 78--92.

\bibitem[Sayakkara et~al.(2020{\natexlab{a}})Sayakkara, Miralles-Pechu{\'{a}}n, Le-Khac, and Scanlon]{Sayakkara2020}
\BIBentryALTinterwordspacing
A.~Sayakkara, L.~Miralles-Pechu{\'{a}}n, N.-A. Le-Khac, and M.~Scanlon, ``{Cutting Through the Emissions: Feature Selection from Electromagnetic Side-Channel Data for Activity Detection},'' \emph{Forensic Science International: Digital Investigation}, vol.~32, p. 300927, 2020. [Online]. Available: \url{https://www.sciencedirect.com/science/article/pii/S2666281720300226}
\BIBentrySTDinterwordspacing

\bibitem[Aghaie and Moradi(2020)]{Aghaie2020}
A.~Aghaie and A.~Moradi, ``Ti-puf: Toward side-channel resistant physical unclonable functions,'' \emph{IEEE Transactions on Information Forensics and Security}, vol.~15, pp. 3470--3481, 2020.

\bibitem[Japa et~al.(2021)Japa, Majumder, Sahoo, Vaddi, and Kaushik]{Japa2021}
A.~Japa, M.~K. Majumder, S.~K. Sahoo, R.~Vaddi, and B.~K. Kaushik, ``Hardware security exploiting post-cmos devices: Fundamental device characteristics, state-of-the-art countermeasures, challenges and roadmap,'' \emph{IEEE Circuits and Systems Magazine}, vol.~21, no.~3, pp. 4--30, 2021.

\bibitem[Tuttlebee(2002)]{SDR123300/2394827779}
W.~H. Tuttlebee, \emph{Software Defined Radio: Enabling Technologies}.\hskip 1em plus 0.5em minus 0.4em\relax John Wiley and Sons, 2002.

\bibitem[Mutlu(2017)]{7927156}
O.~Mutlu, ``The rowhammer problem and other issues we may face as memory ecomes denser,'' in \emph{Design, Automation \& Test in Europe Conference \& Exhibition (DATE), 2017}, 2017, pp. 1116--1121.

\bibitem[Kim et~al.(2014)Kim, Daly, Kim, Fallin, Lee, Lee, Wilkerson, Lai, and Mutlu]{10.1145/2678373.2665726}
\BIBentryALTinterwordspacing
Y.~Kim, R.~Daly, J.~Kim, C.~Fallin, J.~H. Lee, D.~Lee, C.~Wilkerson, K.~Lai, and O.~Mutlu, ``Flipping bits in memory without accessing them: An experimental study of dram disturbance errors,'' \emph{SIGARCH Comput. Archit. News}, vol.~42, no.~3, p. 361–372, jun 2014. [Online]. Available: \url{https://doi.org/10.1145/2678373.2665726}
\BIBentrySTDinterwordspacing

\bibitem[Mutlu and Kim(2020)]{8708249}
O.~Mutlu and J.~S. Kim, ``Rowhammer: A retrospective,'' \emph{IEEE Transactions on Computer-Aided Design of Integrated Circuits and Systems}, vol.~39, no.~8, pp. 1555--1571, 2020.

\bibitem[Yağlikçi et~al.(2021)Yağlikçi, Patel, Kim, Azizi, Olgun, Orosa, Hassan, Park, Kanellopoulos, Shahroodi, Ghose, and Mutlu]{9407238}
A.~G. Yağlikçi, M.~Patel, J.~S. Kim, R.~Azizi, A.~Olgun, L.~Orosa, H.~Hassan, J.~Park, K.~Kanellopoulos, T.~Shahroodi, S.~Ghose, and O.~Mutlu, ``Blockhammer: Preventing rowhammer at low cost by blacklisting rapidly-accessed dram rows,'' in \emph{2021 IEEE International Symposium on High-Performance Computer Architecture (HPCA)}, 2021, pp. 345--358.

\bibitem[Kim et~al.(2020)Kim, Patel, Yağlıkçı, Hassan, Azizi, Orosa, and Mutlu]{9138944}
J.~S. Kim, M.~Patel, A.~G. Yağlıkçı, H.~Hassan, R.~Azizi, L.~Orosa, and O.~Mutlu, ``Revisiting rowhammer: An experimental analysis of modern dram devices and mitigation techniques,'' in \emph{2020 ACM/IEEE 47th Annual International Symposium on Computer Architecture (ISCA)}, 2020, pp. 638--651.

\bibitem[Gupta and Nisbet(2016)]{gupta1108442/1500422}
K.~Gupta and A.~Nisbet, ``Memory forensic data recovery utilising ram cooling methods,'' \emph{Proceedings of 14th Australian Digital Forensics Conference}, pp. 11--16, 2016.

\bibitem[Halderman et~al.(2009)Halderman, Schoen, Heninger, Clarkson, Paul, Calandrino, Feldman, Appelbaum, and Felten]{10.1145/1506409.1506429}
\BIBentryALTinterwordspacing
J.~A. Halderman, S.~D. Schoen, N.~Heninger, W.~Clarkson, W.~Paul, J.~A. Calandrino, A.~J. Feldman, J.~Appelbaum, and E.~W. Felten, ``Lest we remember: Cold-boot attacks on encryption keys,'' \emph{Commun. ACM}, vol.~52, no.~5, p. 91–98, may 2009. [Online]. Available: \url{https://doi.org/10.1145/1506409.1506429}
\BIBentrySTDinterwordspacing

\bibitem[Halevi and Saxena(2015)]{Halevi2015}
T.~Halevi and N.~Saxena, ``\BIBforeignlanguage{English}{Keyboard acoustic side channel attacks: Exploring realistic and security-sensitive scenarios},'' \emph{\BIBforeignlanguage{English}{International Journal of Information Security}}, vol.~14, no.~5, pp. 443--456, 2015.

\bibitem[Zhu et~al.(2014)Zhu, Ma, Zhang, and Liu]{10.1145/2660267.2660296}
\BIBentryALTinterwordspacing
T.~Zhu, Q.~Ma, S.~Zhang, and Y.~Liu, ``Context-free attacks using keyboard acoustic emanations,'' in \emph{Proceedings of the 2014 ACM SIGSAC Conference on Computer and Communications Security}, ser. CCS '14.\hskip 1em plus 0.5em minus 0.4em\relax New York, NY, USA: Association for Computing Machinery, 2014, p. 453–464. [Online]. Available: \url{https://doi.org/10.1145/2660267.2660296}
\BIBentrySTDinterwordspacing

\bibitem[Asonov and Agrawal(2004)]{1301311}
D.~Asonov and R.~Agrawal, ``Keyboard acoustic emanations,'' in \emph{IEEE Symposium on Security and Privacy, 2004. Proceedings. 2004}, 2004, pp. 3--11.

\bibitem[Sabra et~al.(2018)Sabra, Maiti, and Jadliwala]{10.1145/3211960.3211973}
\BIBentryALTinterwordspacing
M.~Sabra, A.~Maiti, and M.~Jadliwala, ``Keystroke inference using ambient light sensor on wrist-wearables: A feasibility study,'' in \emph{Proceedings of the 4th ACM Workshop on Wearable Systems and Applications}, ser. WearSys '18.\hskip 1em plus 0.5em minus 0.4em\relax New York, NY, USA: Association for Computing Machinery, 2018, p. 21–26. [Online]. Available: \url{https://doi.org/10.1145/3211960.3211973}
\BIBentrySTDinterwordspacing

\bibitem[Shahrad et~al.(2018)Shahrad, Mosenia, Song, Chiang, Wentzlaff, and Mittal]{Shahrad2018}
\BIBentryALTinterwordspacing
M.~Shahrad, A.~Mosenia, L.~Song, M.~Chiang, D.~Wentzlaff, and P.~Mittal, ``{Acoustic Denial of Service Attacks on Hard Disk Drives},'' in \emph{Proceedings of the 2018 Workshop on Attacks and Solutions in Hardware Security}.\hskip 1em plus 0.5em minus 0.4em\relax Association for Computing Machinery, 2018, p.~34. [Online]. Available: \url{https://doi.org/10.1145/3266444.3266448}
\BIBentrySTDinterwordspacing

\bibitem[Savage(2015)]{savage2015visualizing}
\BIBentryALTinterwordspacing
N.~Savage, ``Visualizing sound,'' \emph{Commun. ACM}, vol.~58, no.~2, p. 15–17, jan 2015. [Online]. Available: \url{https://doi.org/10.1145/2693430}
\BIBentrySTDinterwordspacing

\bibitem[Davis et~al.(2014)Davis, Rubinstein, Wadhwa, Mysore, Durand, and Freeman]{davis2014visual}
\BIBentryALTinterwordspacing
A.~Davis, M.~Rubinstein, N.~Wadhwa, G.~J. Mysore, F.~Durand, and W.~T. Freeman, ``The visual microphone: Passive recovery of sound from video,'' \emph{ACM Transactions on Graphics}, vol.~33, no.~4, jul 2014. [Online]. Available: \url{https://doi.org/10.1145/2601097.2601119}
\BIBentrySTDinterwordspacing

\bibitem[Kanta et~al.(2020{\natexlab{a}})Kanta, Coisel, and Scanlon]{KANTA2020301075}
\BIBentryALTinterwordspacing
A.~Kanta, I.~Coisel, and M.~Scanlon, ``A survey exploring open source intelligence for smarter password cracking,'' \emph{Forensic Science International: Digital Investigation}, vol.~35, p. 301075, 2020. [Online]. Available: \url{https://www.sciencedirect.com/science/article/pii/S2666281720303723}
\BIBentrySTDinterwordspacing

\bibitem[Kanta et~al.(2021)Kanta, Coray, Coisel, and Scanlon]{Kanta2021}
\BIBentryALTinterwordspacing
A.~Kanta, S.~Coray, I.~Coisel, and M.~Scanlon, ``How viable is password cracking in digital forensic investigation? analyzing the guessability of over 3.9 billion real-world accounts,'' \emph{Forensic Science International: Digital Investigation}, vol.~37, p. 301186, 2021. [Online]. Available: \url{https://www.sciencedirect.com/science/article/pii/S2666281721000949}
\BIBentrySTDinterwordspacing

\bibitem[Kanta et~al.(2020{\natexlab{b}})Kanta, Coisel, and Scanlon]{9138870}
A.~Kanta, I.~Coisel, and M.~Scanlon, ``Smarter password guessing techniques leveraging contextual information and osint,'' in \emph{2020 International Conference on Cyber Security and Protection of Digital Services (Cyber Security)}, 2020, pp. 1--2.

\bibitem[Scanlon et~al.(2023)Scanlon, Breitinger, Hargreaves, Hilgert, and Sheppard]{SCANLON2023301609}
\BIBentryALTinterwordspacing
M.~Scanlon, F.~Breitinger, C.~Hargreaves, J.-N. Hilgert, and J.~Sheppard, ``Chatgpt for digital forensic investigation: The good, the bad, and the unknown,'' \emph{Forensic Science International: Digital Investigation}, vol.~46, p. 301609, 2023. [Online]. Available: \url{https://www.sciencedirect.com/science/article/pii/S266628172300121X}
\BIBentrySTDinterwordspacing

\bibitem[Cho et~al.(2015)Cho, Jeong, and Park]{CHO201558}
\BIBentryALTinterwordspacing
J.-S. Cho, Y.-S. Jeong, and S.~O. Park, ``Consideration on the brute-force attack cost and retrieval cost: A hash-based radio-frequency identification (rfid) tag mutual authentication protocol,'' \emph{Computers and Mathematics with Applications}, vol.~69, no.~1, pp. 58--65, 2015. [Online]. Available: \url{https://www.sciencedirect.com/science/article/pii/S0898122112001393}
\BIBentrySTDinterwordspacing

\bibitem[Wazid et~al.(2013)Wazid, Katal, Goudar, Singh, Tyagi, Sharma, and Bhakuni]{6481194}
M.~Wazid, A.~Katal, R.~Goudar, D.~P. Singh, A.~Tyagi, R.~Sharma, and P.~Bhakuni, ``A framework for detection and prevention of novel keylogger spyware attacks,'' in \emph{2013 7th International Conference on Intelligent Systems and Control (ISCO)}, 2013, pp. 433--438.

\bibitem[Unger et~al.(2015)Unger, Dechand, Bonneau, Fahl, Perl, Goldberg, and Smith]{7163029}
N.~Unger, S.~Dechand, J.~Bonneau, S.~Fahl, H.~Perl, I.~Goldberg, and M.~Smith, ``Sok: Secure messaging,'' in \emph{2015 IEEE Symposium on Security and Privacy}, 2015, pp. 232--249.

\bibitem[Pub(2001)]{pub2001announcing}
N.~F. Pub, ``Announcing the advanced encryption standard (aes),'' \emph{Federal Information Processing Standards Publication}, vol. 197, pp. 1--51, 2001.

\bibitem[Yu and Chen(2018)]{Yu2018a}
\BIBentryALTinterwordspacing
W.~Yu and J.~Chen, ``Deep learning-assisted and combined attack: A novel side-channel attack,'' \emph{Electronics Letters}, vol.~54, no.~19, pp. 1114--1116, 2018. [Online]. Available: \url{https://ietresearch.onlinelibrary.wiley.com/doi/abs/10.1049/el.2018.5411}
\BIBentrySTDinterwordspacing

\bibitem[Lerman et~al.(2015)Lerman, Bontempi, and Markowitch]{lerman2015machine}
L.~Lerman, G.~Bontempi, and O.~Markowitch, ``A machine learning approach against a masked aes,'' \emph{Journal of Cryptographic Engineering}, vol.~5, no.~2, pp. 123--139, 2015.

\bibitem[Yang et~al.(2004)Yang, Wu, and Karri]{yang2004scan}
B.~Yang, K.~Wu, and R.~Karri, ``Scan based side channel attack on dedicated hardware implementations of data encryption standard,'' in \emph{2004 International Conferce on Test}, 2004, pp. 339--344.

\bibitem[Standaert(2010)]{standaert2010introduction}
\BIBentryALTinterwordspacing
F.-X. Standaert, \emph{Introduction to Side-Channel Attacks}.\hskip 1em plus 0.5em minus 0.4em\relax Boston, MA: Springer US, 2010, pp. 27--42. [Online]. Available: \url{https://doi.org/10.1007/978-0-387-71829-3_2}
\BIBentrySTDinterwordspacing

\bibitem[Azarderakhsh et~al.(2014)Azarderakhsh, J{\"a}rvinen, and Mozaffari-Kermani]{azarderakhsh2014efficient}
R.~Azarderakhsh, K.~U. J{\"a}rvinen, and M.~Mozaffari-Kermani, ``Efficient algorithm and architecture for elliptic curve cryptography for extremely constrained secure applications,'' \emph{IEEE Transactions on Circuits and Systems I: Regular Papers}, vol.~61, no.~4, pp. 1144--1155, 2014.

\bibitem[Mahmood et~al.(2018)Mahmood, Chaudhry, Naqvi, Kumari, Li, and Sangaiah]{mahmood2018elliptic}
\BIBentryALTinterwordspacing
K.~Mahmood, S.~A. Chaudhry, H.~Naqvi, S.~Kumari, X.~Li, and A.~K. Sangaiah, ``An elliptic curve cryptography based lightweight authentication scheme for smart grid communication,'' \emph{Future Generation Computer Systems}, vol.~81, pp. 557--565, 2018. [Online]. Available: \url{https://www.sciencedirect.com/science/article/pii/S0167739X17309263}
\BIBentrySTDinterwordspacing

\bibitem[Fournaris et~al.(2017)Fournaris, Papachristodoulou, and Sklavos]{Fournaris2017}
A.~P. Fournaris, L.~Papachristodoulou, and N.~Sklavos, ``Secure and efficient rns software implementation for elliptic curve cryptography,'' in \emph{2017 IEEE European Symposium on Security and Privacy Workshops (EuroS\&PW)}, 2017, pp. 86--93.

\bibitem[Bikos and Sklavos(2013)]{6336699}
A.~N. Bikos and N.~Sklavos, ``Lte/sae security issues on 4g wireless networks,'' \emph{IEEE Security \& Privacy}, vol.~11, no.~2, pp. 55--62, 2013.

\bibitem[Sulaiman and Al~Shaikhli(2014)]{sulaiman2014comparative}
A.~G. Sulaiman and I.~F. Al~Shaikhli, ``Comparative study on 4g/lte cryptographic algorithms based on different factors,'' \emph{International Journal of Computer Science and Telecommunications}, vol.~5, no.~7, pp. 7--10, 2014.

\bibitem[Barrett(2000)]{barrett2000implementing}
P.~Barrett, ``Implementing the rivest shamir and adleman public key encryption algorithm on a standard digital signal processor,'' in \emph{Advances in Cryptology—CRYPTO’86: Proceedings}.\hskip 1em plus 0.5em minus 0.4em\relax Springer, 2000, pp. 311--323.

\bibitem[Genkin et~al.(2014{\natexlab{a}})Genkin, Shamir, and Tromer]{genkin2014rsa}
D.~Genkin, A.~Shamir, and E.~Tromer, ``Rsa key extraction via low-bandwidth acoustic cryptanalysis,'' in \emph{Advances in Cryptology -- CRYPTO 2014}.\hskip 1em plus 0.5em minus 0.4em\relax Springer Berlin Heidelberg, 2014, pp. 444--461.

\bibitem[Li et~al.(2019)Li, Iyer, and Orshansky]{li2019securing}
G.~Li, V.~Iyer, and M.~Orshansky, ``Securing aes against localized em attacks through spatial randomization of dataflow,'' in \emph{2019 IEEE International Symposium on Hardware Oriented Security and Trust (HOST)}, 2019, pp. 191--197.

\bibitem[Kumar et~al.(2017)Kumar, Scarborough, Yilmaz, and Orshansky]{kumar2017efficient}
A.~Kumar, C.~Scarborough, A.~Yilmaz, and M.~Orshansky, ``Efficient simulation of em side-channel attack resilience,'' in \emph{2017 IEEE/ACM International Conference on Computer-Aided Design (ICCAD)}, 2017, pp. 123--130.

\bibitem[Vasselle et~al.(2019)Vasselle, Maurine, and Cozzi]{Vasselle2019}
\BIBentryALTinterwordspacing
A.~Vasselle, P.~Maurine, and M.~Cozzi, ``Breaking mobile firmware encryption through near-field side-channel analysis,'' in \emph{Proceedings of the 3rd ACM Workshop on Attacks and Solutions in Hardware Security Workshop}, ser. ASHES'19.\hskip 1em plus 0.5em minus 0.4em\relax New York, NY, USA: Association for Computing Machinery, 2019, p. 23–32. [Online]. Available: \url{https://doi.org/10.1145/3338508.3359571}
\BIBentrySTDinterwordspacing

\bibitem[Haas and Aysu(2022)]{Haas2022}
\BIBentryALTinterwordspacing
G.~Haas and A.~Aysu, ``Apple vs. ema: Electromagnetic side channel attacks on apple corecrypto,'' in \emph{Proceedings of the 59th ACM/IEEE Design Automation Conference}, ser. DAC '22.\hskip 1em plus 0.5em minus 0.4em\relax New York, NY, USA: Association for Computing Machinery, 2022, p. 247–252. [Online]. Available: \url{https://doi.org/10.1145/3489517.3530437}
\BIBentrySTDinterwordspacing

\bibitem[Haas et~al.(2021)Haas, Potluri, and Aysu]{haas2021itimed}
G.~Haas, S.~Potluri, and A.~Aysu, ``itimed: Cache attacks on the apple a10 fusion soc,'' in \emph{2021 IEEE International Symposium on Hardware Oriented Security and Trust (HOST)}, 2021, pp. 80--90.

\bibitem[Iyer and Yilmaz(2021)]{Iyer2021c}
V.~V. Iyer and A.~E. Yilmaz, ``Using the anova f-statistic to rapidly identify near-field vulnerabilities of cryptographic modules,'' in \emph{2021 IEEE MTT-S International Microwave Symposium (IMS)}, 2021, pp. 112--115.

\bibitem[Burnside et~al.(2008)Burnside, Erdogan, and Arslan]{Burnside2008}
A.~Burnside, A.~Erdogan, and T.~Arslan, ``{The Re-emission Side Channel},'' in \emph{2008 Bio-inspired, Learning and Intelligent Systems for Security}, 2008, pp. 154--159.

\bibitem[Sauvage et~al.(2009)Sauvage, Guilley, and Mathieu]{Sauvage2009}
\BIBentryALTinterwordspacing
L.~Sauvage, S.~Guilley, and Y.~Mathieu, ``Electromagnetic radiations of fpgas: High spatial resolution cartography and attack on a cryptographic module,'' \emph{ACM Trans. Reconfigurable Technol. Syst.}, vol.~2, no.~1, mar 2009. [Online]. Available: \url{https://doi.org/10.1145/1502781.1502785}
\BIBentrySTDinterwordspacing

\bibitem[Lellis et~al.(2017)Lellis, Soares, and Souza]{Lellis2017}
R.~Lellis, R.~I. Soares, and A.~Souza, ``An energy-based attack flow for temporal misalignment coutermeasures on cryptosystems,'' in \emph{2017 IEEE International Symposium on Circuits and Systems (ISCAS)}, 2017, pp. 1--4.

\bibitem[Mukhtar et~al.(2021)Mukhtar, Mehrabi, Kong, and Anjum]{mukhtar2021edge}
\BIBentryALTinterwordspacing
N.~Mukhtar, A.~Mehrabi, Y.~Kong, and A.~Anjum, ``Edge enhanced deep learning system for iot edge device security analytics,'' \emph{Concurrency and Computation: Practice and Experience}, p. e6764, 2021. [Online]. Available: \url{https://onlinelibrary.wiley.com/doi/abs/10.1002/cpe.6764}
\BIBentrySTDinterwordspacing

\bibitem[Genkin et~al.(2016{\natexlab{a}})Genkin, Pachmanov, Pipman, and Tromer]{Genkin2016}
D.~Genkin, L.~Pachmanov, I.~Pipman, and E.~Tromer, ``Ecdh key-extraction via low-bandwidth electromagnetic attacks on pcs,'' in \emph{Topics in Cryptology - CT-RSA 2016}, K.~Sako, Ed.\hskip 1em plus 0.5em minus 0.4em\relax Cham: Springer International Publishing, 2016, pp. 219--235.

\bibitem[Genkin et~al.(2016{\natexlab{b}})Genkin, Pachmanov, Pipman, Tromer, and Yarom]{Genkin2016a}
\BIBentryALTinterwordspacing
D.~Genkin, L.~Pachmanov, I.~Pipman, E.~Tromer, and Y.~Yarom, ``Ecdsa key extraction from mobile devices via nonintrusive physical side channels,'' in \emph{Proceedings of the 2016 ACM SIGSAC Conference on Computer and Communications Security}, ser. CCS '16.\hskip 1em plus 0.5em minus 0.4em\relax New York, NY, USA: Association for Computing Machinery, 2016, p. 1626–1638. [Online]. Available: \url{https://doi.org/10.1145/2976749.2978353}
\BIBentrySTDinterwordspacing

\bibitem[Goller and Sigl(2015)]{goller2015side}
G.~Goller and G.~Sigl, ``Side channel attacks on smartphones and embedded devices using standard radio equipment,'' in \emph{International Workshop on Constructive Side-Channel Analysis and Secure Design}.\hskip 1em plus 0.5em minus 0.4em\relax Springer, 2015, pp. 255--270.

\bibitem[Nakano et~al.(2014)Nakano, Souissi, Nguyen, Sauvage, Danger, Guilley, Kiyomoto, and Miyake]{nakano2014pre}
Y.~Nakano, Y.~Souissi, R.~Nguyen, L.~Sauvage, J.-L. Danger, S.~Guilley, S.~Kiyomoto, and Y.~Miyake, ``A pre-processing composition for secret key recovery on android smartphone,'' in \emph{Information Security Theory and Practice. Securing the Internet of Things: 8th IFIP WG 11.2 International Workshop, WISTP 2014, Heraklion, Crete, Greece, June 30--July 2, 2014. Proceedings 8}.\hskip 1em plus 0.5em minus 0.4em\relax Springer, 2014, pp. 76--91.

\bibitem[Kenworthy and Rohatgi(2012)]{kenworthy2012mobile}
G.~Kenworthy and P.~Rohatgi, ``Mobile device security: The case for side channel resistance,'' in \emph{Proceedings of the Mobile Security Technologies Conference (MoST)}, 2012.

\bibitem[Genkin et~al.(2014{\natexlab{b}})Genkin, Pipman, and Tromer]{10.1007/978-3-662-44709-3_14}
D.~Genkin, I.~Pipman, and E.~Tromer, ``Get your hands off my laptop: Physical side-channel key-extraction attacks on pcs,'' in \emph{Cryptographic Hardware and Embedded Systems -- CHES 2014}.\hskip 1em plus 0.5em minus 0.4em\relax Berlin, Heidelberg: Springer Berlin Heidelberg, 2014, pp. 242--260.

\bibitem[Pammu et~al.(2016{\natexlab{a}})Pammu, Chong, Ho, and Gwee]{Pammu2016}
A.~A. Pammu, K.-S. Chong, W.-G. Ho, and B.-H. Gwee, ``Interceptive side channel attack on aes-128 wireless communications for iot applications,'' in \emph{2016 IEEE Asia Pacific Conference on Circuits and Systems (APCCAS)}, 2016, pp. 650--653.

\bibitem[Ding et~al.(2009)Ding, Li, Chang, and Zhao]{Ding2009}
G.-l. Ding, Z.-x. Li, X.-l. Chang, and Q.~Zhao, ``Differential electromagnetic analysis on aes cryptographic system,'' in \emph{2009 Second Pacific-Asia Conference on Web Mining and Web-based Application}, 2009, pp. 120--123.

\bibitem[Frieslaar and Irwin(2018)]{Frieslaar2018}
I.~Frieslaar and B.~Irwin, ``Developing an electromagnetic noise generator to protect a raspberry pi from side channel analysis,'' \emph{SAIEE Africa Research Journal}, vol. 109, no.~2, pp. 85--101, 2018.

\bibitem[Alam et~al.(2021)Alam, Yilmaz, Werner, Samwel, Zajic, Genkin, Yarom, and Prvulovic]{Alam2021}
M.~Alam, B.~Yilmaz, F.~Werner, N.~Samwel, A.~Zajic, D.~Genkin, Y.~Yarom, and M.~Prvulovic, ``Nonce@once: A single-trace em side channel attack on several constant-time elliptic curve implementations in mobile platforms,'' in \emph{2021 IEEE European Symposium on Security and Privacy (EuroS\&P)}, 2021, pp. 507--522.

\bibitem[Soares et~al.(2011)Soares, Calazans, Moraes, Maurine, and Torres]{soares2011robust}
R.~Soares, N.~Calazans, F.~Moraes, P.~Maurine, and L.~Torres, ``A robust architectural approach for cryptographic algorithms using gals pipelines,'' \emph{IEEE Design \& Test of Computers}, vol.~28, no.~5, pp. 62--71, 2011.

\bibitem[Mittal et~al.(2021)Mittal, Gupta, and Srivastava]{MITTAL2021102163}
\BIBentryALTinterwordspacing
S.~Mittal, H.~Gupta, and S.~Srivastava, ``A survey on hardware security of dnn models and accelerators,'' \emph{Journal of Systems Architecture}, vol. 117, p. 102163, 2021. [Online]. Available: \url{https://www.sciencedirect.com/science/article/pii/S1383762121001168}
\BIBentrySTDinterwordspacing

\bibitem[Xiao et~al.(2020)Xiao, Xin, and Shen]{Xiao2020}
\BIBentryALTinterwordspacing
Y.~Xiao, J.~Xin, and Y.~Shen, ``Cnn based electromagnetic side channel attacks on soc,'' \emph{IOP Conference Series: Materials Science and Engineering}, vol. 782, no.~3, p. 032055, 3 2020. [Online]. Available: \url{https://dx.doi.org/10.1088/1757-899X/782/3/032055}
\BIBentrySTDinterwordspacing

\bibitem[Wang et~al.(2020)Wang, Wang, and Dubrova]{Wang2020}
\BIBentryALTinterwordspacing
R.~Wang, H.~Wang, and E.~Dubrova, ``Far field em side-channel attack on aes using deep learning,'' in \emph{Proceedings of the 4th ACM Workshop on Attacks and Solutions in Hardware Security}, ser. ASHES'20.\hskip 1em plus 0.5em minus 0.4em\relax New York, NY, USA: Association for Computing Machinery, 2020, p. 35–44. [Online]. Available: \url{https://doi.org/10.1145/3411504.3421214}
\BIBentrySTDinterwordspacing

\bibitem[Camurati et~al.(2020)Camurati, Francillon, and Standaert]{camurati2020understanding}
G.~Camurati, A.~Francillon, and F.-X. Standaert, ``Understanding screaming channels: From a detailed analysis to improved attacks,'' \emph{IACR Transactions on Cryptographic Hardware and Embedded Systems}, pp. 358--401, 2020.

\bibitem[Benadjila et~al.(2020)Benadjila, Prouff, Strullu, Cagli, and Dumas]{benadjila2020deep}
R.~Benadjila, E.~Prouff, R.~Strullu, E.~Cagli, and C.~Dumas, ``Deep learning for side-channel analysis and introduction to ascad database,'' \emph{Journal of Cryptographic Engineering}, vol.~10, no.~2, pp. 163--188, 2020.

\bibitem[Yu et~al.(2021)Yu, Shan, Panoff, and Jin]{Yu2021}
H.~Yu, H.~Shan, M.~Panoff, and Y.~Jin, ``Cross-device profiled side-channel attacks using meta-transfer learning,'' in \emph{2021 58th ACM/IEEE Design Automation Conference (DAC)}, 2021, pp. 703--708.

\bibitem[Jap et~al.(2020)Jap, Yli-M{\"a}yry, Ito, Ueno, Bhasin, and Homma]{Jap2020}
D.~Jap, V.~Yli-M{\"a}yry, A.~Ito, R.~Ueno, S.~Bhasin, and N.~Homma, ``Practical side-channel based model extraction attack on tree-based machine learning algorithm,'' in \emph{Applied Cryptography and Network Security Workshops}.\hskip 1em plus 0.5em minus 0.4em\relax Cham: Springer International Publishing, 2020, pp. 93--105.

\bibitem[Pan et~al.(2021)Pan, Yang, Li, You, Ji, Chen, Hu, and Xue]{Pan2021}
H.~Pan, L.~Yang, H.~Li, C.-W. You, X.~Ji, Y.-C. Chen, Z.~Hu, and G.~Xue, ``Magthief: Stealing private app usage data on mobile devices via built-in magnetometer,'' in \emph{2021 18th Annual IEEE International Conference on Sensing, Communication, and Networking (SECON)}, 2021, pp. 1--9.

\bibitem[Moradi et~al.(2019)Moradi, Van~Acker, Vanherpen, and Denil]{moradi2018model}
M.~Moradi, B.~Van~Acker, K.~Vanherpen, and J.~Denil, ``Model-implemented hybrid fault injection for simulink (tool demonstrations),'' in \emph{Cyber Physical Systems. Model-Based Design}, R.~Chamberlain, W.~Taha, and M.~T{\"o}rngren, Eds.\hskip 1em plus 0.5em minus 0.4em\relax Cham: Springer International Publishing, 2019, pp. 71--90.

\bibitem[Bhunia and Tehranipoor(2019)]{BHUNIA2019245}
\BIBentryALTinterwordspacing
S.~Bhunia and M.~Tehranipoor, ``Chapter 10 - physical attacks and countermeasures,'' in \emph{Hardware Security}.\hskip 1em plus 0.5em minus 0.4em\relax Morgan Kaufmann, 2019, pp. 245--290. [Online]. Available: \url{https://www.sciencedirect.com/science/article/pii/B9780128124772000150}
\BIBentrySTDinterwordspacing

\bibitem[Lim et~al.(2020)Lim, Lee, and Han]{Lim2020}
S.~Lim, J.~Lee, and D.-G. Han, ``Improved differential fault attack on lea by algebraic representation of modular addition,'' \emph{IEEE Access}, vol.~8, pp. 212\,794--212\,802, 2020.

\bibitem[Gunathilake et~al.(2022)Gunathilake, Al-Dubai, Buchanan, and Lo]{Gunathilake2021}
N.~A. Gunathilake, A.~Al-Dubai, W.~J. Buchanan, and O.~Lo, ``Electromagnetic side-channel attack resilience against present lightweight block cipher,'' in \emph{2022 6th International Conference on Cryptography, Security and Privacy (CSP)}, 2022, pp. 51--55.

\bibitem[Hell and Westman(2020)]{Hell2020}
\BIBentryALTinterwordspacing
M.~Hell and O.~Westman, ``Electromagnetic side-channel attack on aes using low-end equipment,'' \emph{ECTI Transactions on Computer and Information Technology (ECTI-CIT)}, vol.~14, no.~2, p. 139–148, Apr. 2020. [Online]. Available: \url{https://ph01.tci-thaijo.org/index.php/ecticit/article/view/239925}
\BIBentrySTDinterwordspacing

\bibitem[Chusseau et~al.(2014)Chusseau, Omarouayache, Raoult, Jarrix, Maurine, Tobich, Bover, Vrignon, Shepherd, Le, Berthier, Rivière, Robisson, and Ribotta]{Chusseau2014}
L.~Chusseau, R.~Omarouayache, J.~Raoult, S.~Jarrix, P.~Maurine, K.~Tobich, A.~Bover, B.~Vrignon, J.~Shepherd, T.-H. Le, M.~Berthier, L.~Rivière, B.~Robisson, and A.-L. Ribotta, ``Electromagnetic analysis, deciphering and reverse engineering of integrated circuits (e-mata hari),'' pp. 1--6, 2014.

\bibitem[Nakamura et~al.(2015)Nakamura, Hayashi, Homma, Mizuki, Aoki, and Sone]{Nakamura2015}
K.~Nakamura, Y.-i. Hayashi, N.~Homma, T.~Mizuki, T.~Aoki, and H.~Sone, ``Method for estimating fault injection time on cryptographic devices from em leakage,'' in \emph{2015 IEEE International Symposium on Electromagnetic Compatibility (EMC)}, 2015, pp. 235--240.

\bibitem[Dehbaoui et~al.(2013)Dehbaoui, Mirbaha, Moro, Dutertre, and Tria]{dehbaoui2013electromagnetic}
A.~Dehbaoui, A.-P. Mirbaha, N.~Moro, J.-M. Dutertre, and A.~Tria, ``Electromagnetic glitch on the aes round counter,'' in \emph{International Workshop on Constructive Side-Channel Analysis and Secure Design}.\hskip 1em plus 0.5em minus 0.4em\relax Springer, 2013, pp. 17--31.

\bibitem[Qiu et~al.(2019)Qiu, Wang, Lyu, and Qu]{10.1145/3319535.3354201}
\BIBentryALTinterwordspacing
P.~Qiu, D.~Wang, Y.~Lyu, and G.~Qu, ``Voltjockey: Breaching trustzone by software-controlled voltage manipulation over multi-core frequencies,'' in \emph{Proceedings of the 2019 ACM SIGSAC Conference on Computer and Communications Security}, ser. CCS '19.\hskip 1em plus 0.5em minus 0.4em\relax New York, NY, USA: Association for Computing Machinery, 2019, p. 195–209. [Online]. Available: \url{https://doi-org.ucd.idm.oclc.org/10.1145/3319535.3354201}
\BIBentrySTDinterwordspacing

\bibitem[Ravi et~al.(2020)Ravi, Bhasin, Roy, and Chattopadhyay]{Ravi2020}
\BIBentryALTinterwordspacing
P.~Ravi, S.~Bhasin, S.~S. Roy, and A.~Chattopadhyay, ``{Drop by Drop you break the rock-Exploiting generic vulnerabilities in Lattice-based PKE/KEMs using EM-based Physical Attacks},'' \emph{Cryptology ePrint Archive}, 2020. [Online]. Available: \url{https://eprint.iacr.org/2020/549}
\BIBentrySTDinterwordspacing

\bibitem[Wang et~al.(2021)Wang, Iyer, Xie, Li, Mathew, Kumar, Orshansky, Yilmaz, and Kulkarni]{Wang2021}
M.~Wang, V.~V. Iyer, S.~Xie, G.~Li, S.~K. Mathew, R.~Kumar, M.~Orshansky, A.~E. Yilmaz, and J.~P. Kulkarni, ``Physical design strategies for mitigating fine-grained electromagnetic side-channel attacks,'' in \emph{2021 IEEE Custom Integrated Circuits Conference (CICC)}, 2021, pp. 1--2.

\bibitem[Sayakkara and Le-Khac(2021)]{Sayakkara2021}
A.~P. Sayakkara and N.-A. Le-Khac, ``Electromagnetic side-channel analysis for iot forensics: Challenges, framework, and datasets,'' \emph{IEEE Access}, vol.~9, pp. 113\,585--113\,598, 2021.

\bibitem[Chatterjee et~al.(2019)Chatterjee, Cao, Raychowdhury, and Sen]{Chatterjee2019}
B.~Chatterjee, N.~Cao, A.~Raychowdhury, and S.~Sen, ``Context-aware intelligence in resource-constrained iot nodes: Opportunities and challenges,'' \emph{IEEE Design \& Test}, vol.~36, no.~2, pp. 7--40, 2019.

\bibitem[Yoshida et~al.(2019)Yoshida, Kubota, Shiozaki, and Fujino]{Yoshida2019}
K.~Yoshida, T.~Kubota, M.~Shiozaki, and T.~Fujino, ``Model-extraction attack against fpga-dnn accelerator utilizing correlation electromagnetic analysis,'' in \emph{2019 IEEE 27th Annual International Symposium on Field-Programmable Custom Computing Machines (FCCM)}, 2019, pp. 318--318.

\bibitem[Sosa et~al.(2021)Sosa, Dyka, Kabin, and Langendörfer]{Sosa2021}
O.~A. Sosa, Z.~Dyka, I.~Kabin, and P.~Langendörfer, ``{Simulation of Electromagnetic Emanation of Cryptographic ICs: Tools, Methods, Problems},'' in \emph{2021 IEEE East-West Design \& Test Symposium (EWDTS)}, 2021, pp. 1--5.

\bibitem[Bela{\"i}d et~al.(2020)Bela{\"i}d, Dagand, Mercadier, Rivain, and Wintersdorff]{Belaid2020}
S.~Bela{\"i}d, P.-{\'E}. Dagand, D.~Mercadier, M.~Rivain, and R.~Wintersdorff, ``Tornado: Automatic generation of probing-secure masked bitsliced implementations,'' in \emph{Advances in Cryptology -- EUROCRYPT 2020}.\hskip 1em plus 0.5em minus 0.4em\relax Cham: Springer International Publishing, 2020, pp. 311--341.

\bibitem[Vosoughi et~al.(2020)Vosoughi, Wang, and Köse]{Vosoughi2020}
A.~Vosoughi, L.~Wang, and S.~Köse, ``Voltage regulator assisted lightweight countermeasure against fault injection attacks,'' \emph{arXiv preprint arXiv:2001.03230}, 2020.

\bibitem[Dey et~al.(2022)Dey, Park, Pundir, Saha, Shuvo, Mehta, Asadi, Rahman, Farahmandi, and Tehranipoor]{Dey2022}
\BIBentryALTinterwordspacing
S.~Dey, J.~Park, N.~Pundir, D.~Saha, A.~M. Shuvo, D.~Mehta, N.~Asadi, F.~Rahman, F.~Farahmandi, and M.~Tehranipoor, ``Secure physical design,'' \emph{Cryptology ePrint Archive}, 2022. [Online]. Available: \url{https://eprint.iacr.org/2022/891}
\BIBentrySTDinterwordspacing

\bibitem[Abdellatif(2019)]{Abdellatif2019}
K.~M. Abdellatif, ``Towards efficient alignment for electromagnetic side channel attacks,'' in \emph{2019 31st International Conference on Microelectronics (ICM)}, 2019, pp. 118--121.

\bibitem[Kwon et~al.(2021)Kwon, Kim, and Hong]{9400816}
D.~Kwon, H.~Kim, and S.~Hong, ``Non-profiled deep learning-based side-channel preprocessing with autoencoders,'' \emph{IEEE Access}, vol.~9, pp. 57\,692--57\,703, 2021.

\bibitem[Iyer et~al.(2021)Iyer, Wang, Kulkarni, and Yilmaz]{Iyer2021}
V.~V. Iyer, M.~Wang, J.~Kulkarni, and A.~E. Yilmaz, ``A systematic evaluation of em and power side-channel analysis attacks on aes implementations,'' in \emph{2021 IEEE International Conference on Intelligence and Security Informatics (ISI)}, 2021, pp. 1--6.

\bibitem[Kumar et~al.(2020)Kumar, Suresh, Kar, Satpathy, Anders, Kaul, Agarwal, Hsu, Chen, Krishnamurthy, De, and Mathew]{Kumar2020}
R.~Kumar, V.~Suresh, M.~Kar, S.~Satpathy, M.~A. Anders, H.~Kaul, A.~Agarwal, S.~Hsu, G.~K. Chen, R.~K. Krishnamurthy, V.~De, and S.~K. Mathew, ``{A 4900-mu m(2) 839-Mb/s Side-Channel Attack-Resistant AES-128 in 14-nm CMOS With Heterogeneous Sboxes, Linear Masked MixColumns, and Dual-Rail Key Addition},'' \emph{IEEE Journal of Solid-State Circuits}, vol.~55, no.~4, pp. 945--955, 2020.

\bibitem[Smith(1997)]{smith1997scientist}
S.~W. Smith, \emph{The Scientist and Engineer's Guide to Digital Signal Processing}.\hskip 1em plus 0.5em minus 0.4em\relax California Technical Pub. San Diego, 1997.

\bibitem[Souissi et~al.(2011)Souissi, Elaabid, Debande, Guilley, and Danger]{souissi2011novel}
Y.~Souissi, M.~A. Elaabid, N.~Debande, S.~Guilley, and J.-L. Danger, ``Novel applications of wavelet transforms based side-channel analysis,'' in \emph{Non-Invasive Attack Testing Workshop}, 2011.

\bibitem[Zhou and Standaert(2019)]{Zhou2019}
\BIBentryALTinterwordspacing
Y.~Zhou and F.-X. Standaert, ``Simplified single-trace side-channel attacks on elliptic curve scalar multiplication using fully convolutional networks,'' in \emph{Proceedings of the 40th WIC Symposium on Information Theory in the Benelux}, 2019. [Online]. Available: \url{http://hdl.handle.net/2078.1/226275}
\BIBentrySTDinterwordspacing

\bibitem[Bu et~al.(2018)Bu, Dai, Lu, Cai, and Shan]{Bu2018}
A.~Bu, W.~Dai, M.~Lu, H.~Cai, and W.~Shan, ``Correlation-based electromagnetic analysis attack using haar wavelet reconstruction with low-pass filtering on an fpga implementaion of aes,'' in \emph{2018 17th IEEE International Conference On Trust, Security And Privacy In Computing And Communications/ 12th IEEE International Conference On Big Data Science And Engineering (TrustCom/BigDataSE)}, 2018, pp. 1897--1900.

\bibitem[Plos et~al.(2009)Plos, Hutter, and Feldhofer]{10.1007/978-3-642-10838-9_13}
T.~Plos, M.~Hutter, and M.~Feldhofer, \emph{On Comparing Side-Channel Preprocessing Techniques for Attacking RFID Devices}.\hskip 1em plus 0.5em minus 0.4em\relax Berlin, Heidelberg: Springer-Verlag, 2009, p. 163–177.

\bibitem[Longo et~al.(2015{\natexlab{b}})Longo, De~Mulder, Page, and Tunstall]{longo2015soc}
J.~Longo, E.~De~Mulder, D.~Page, and M.~Tunstall, ``Soc it to em: Electromagnetic side-channel attacks on a complex system-on-chip,'' in \emph{Cryptographic Hardware and Embedded Systems--CHES 2015: 17th International Workshop, Saint-Malo, France, September 13-16, 2015, Proceedings 17}.\hskip 1em plus 0.5em minus 0.4em\relax Springer, 2015, pp. 620--640.

\bibitem[Arafat et~al.(2021)Arafat, Guo, and Awad]{9417659}
A.~A. Arafat, Z.~Guo, and A.~Awad, ``Vr-spy: A side-channel attack on virtual key-logging in vr headsets,'' in \emph{2021 IEEE Virtual Reality and 3D User Interfaces (VR)}, 2021, pp. 564--572.

\bibitem[Standaert and Archambeau(2008)]{10.1007/978-3-540-85053-3_26}
\BIBentryALTinterwordspacing
F.-X. Standaert and C.~Archambeau, ``Using subspace-based template attacks to compare and combine power and electromagnetic information leakages,'' in \emph{Proceeding Sof the 10th International Workshop on Cryptographic Hardware and Embedded Systems}, ser. CHES '08.\hskip 1em plus 0.5em minus 0.4em\relax Berlin, Heidelberg: Springer-Verlag, 2008, p. 411–425. [Online]. Available: \url{https://doi.org/10.1007/978-3-540-85053-3_26}
\BIBentrySTDinterwordspacing

\bibitem[Gao et~al.(2018)Gao, Zhang, Cheng, Zhou, and Cao]{10.1145/3195970.3196042}
\BIBentryALTinterwordspacing
Y.~Gao, H.~Zhang, W.~Cheng, Y.~Zhou, and Y.~Cao, ``Electro-magnetic analysis of gpu-based aes implementation,'' in \emph{Proceedings of the 55th Annual Design Automation Conference}, ser. DAC '18.\hskip 1em plus 0.5em minus 0.4em\relax New York, NY, USA: Association for Computing Machinery, 2018. [Online]. Available: \url{https://doi.org/10.1145/3195970.3196042}
\BIBentrySTDinterwordspacing

\bibitem[Eisenbarth et~al.(2007)Eisenbarth, Kumar, Paar, Poschmann, and Uhsadel]{4397176}
T.~Eisenbarth, S.~Kumar, C.~Paar, A.~Poschmann, and L.~Uhsadel, ``A survey of lightweight-cryptography implementations,'' \emph{IEEE Design \& Test of Computers}, vol.~24, no.~6, pp. 522--533, 2007.

\bibitem[Park and Kim(2014)]{park2014differential}
M.~Park and J.~Kim, ``Differential fault analysis of the block cipher lea,'' \emph{Journal of the Korea Institute of Information Security \& Cryptology}, vol.~24, no.~6, pp. 1117--1127, 2014.

\bibitem[Jap and Breier(2015)]{10.1007/978-3-319-24315-3_27}
D.~Jap and J.~Breier, ``Differential fault attack on lea,'' in \emph{Information and Communication Technology}, I.~Khalil, E.~Neuhold, A.~M. Tjoa, L.~D. Xu, and I.~You, Eds.\hskip 1em plus 0.5em minus 0.4em\relax Cham: Springer International Publishing, 2015, pp. 265--274.

\bibitem[Nozaki et~al.(2017)Nozaki, Iwase, Ikezaki, and Yoshikawa]{nozaki2017differential}
Y.~Nozaki, T.~Iwase, Y.~Ikezaki, and M.~Yoshikawa, ``Differential electromagnetic analysis for present and its evaluation with several selection functions,'' \emph{Journal of International Council on Electrical Engineering}, vol.~7, no.~1, pp. 137--141, 2017.

\bibitem[{Le Bouder} et~al.(2016){Le Bouder}, Barry, Couroussé, Lanet, and Lashermes]{Bouder2016}
H.~{Le Bouder}, T.~Barry, D.~Couroussé, J.~Lanet, and R.~Lashermes, ``A template attack against verify pin algorithms,'' in \emph{Proceedings of the 13th International Joint Conference on e-Business and Telecommunications - SECRYPT, (ICETE 2016)}, INSTICC.\hskip 1em plus 0.5em minus 0.4em\relax SciTePress, 2016, pp. 231--238.

\bibitem[Simon and Patel(2015)]{Simon2015}
P.~Simon and P.~Patel, ``Side channel analysis of sim cards using combined higher order statistical techniques,'' in \emph{International Conference on Cyber Warfare and Security}.\hskip 1em plus 0.5em minus 0.4em\relax Academic Conferences International Limited, 2015, p. 525.

\bibitem[Li(2015)]{li2015ultra}
W.~Li, ``{An Ultra-Lightweight Side-Channel Resistant Crypto for Pervasive Devices},'' \emph{{International Journal of Multimedia and Ubiquitous Engineering}}, vol.~{10}, pp. 173--186, 2015.

\bibitem[Patranabis et~al.(2019)Patranabis, Roy, Chakraborty, Nagar, Singh, Mukhopadhyay, and Ghosh]{Patranabis2019}
\BIBentryALTinterwordspacing
S.~Patranabis, D.~B. Roy, A.~Chakraborty, N.~Nagar, A.~Singh, D.~Mukhopadhyay, and S.~Ghosh, ``{Lightweight Design-for-security Strategies for Combined Countermeasures Against Side Channel and Fault Analysis in IoT Applications},'' \emph{Journal of Hardware and Systems Security}, 2019. [Online]. Available: \url{https://link.springer.com/article/10.1007/s41635-018-0049-y}
\BIBentrySTDinterwordspacing

\bibitem[Naija et~al.(2017)Naija, Beroulle, and Machhout]{Naija2017}
Y.~Naija, V.~Beroulle, and M.~Machhout, ``Low cost countermeasure at authentication protocol level against electromagnetic side channel attacks on rfid tags,'' \emph{International Journal of Advanced Computer Science and Applications}, vol.~8, no.~11, 2017.

\bibitem[Yoshikawa et~al.(2016)Yoshikawa, Nozaki, and Asahi]{yoshikawa2016twine}
M.~Yoshikawa, Y.~Nozaki, and K.~Asahi, ``Electromagnetic analysis attack for a lightweight block cipher twine,'' in \emph{2016 IEEE/ACES International Conference on Wireless Information Technology and Systems (ICWITS) and Applied Computational Electromagnetics (ACES)}, 2016, pp. 1--2.

\bibitem[Yoshikawa and Nozaki(2016)]{yoshikawa2016prince}
M.~Yoshikawa and Y.~Nozaki, ``Electromagnetic analysis attack for a lightweight cipher prince,'' in \emph{2016 IEEE International Conference on Cybercrime and Computer Forensic (ICCCF)}, 2016, pp. 1--6.

\bibitem[Kasper et~al.(2012)Kasper, Oswald, and Paar]{kasper2011side}
T.~Kasper, D.~Oswald, and C.~Paar, ``Side-channel analysis of cryptographic rfids with analog demodulation,'' in \emph{RFID. Security and Privacy}, A.~Juels and C.~Paar, Eds.\hskip 1em plus 0.5em minus 0.4em\relax Springer Berlin Heidelberg, 2012, pp. 61--77.

\bibitem[Plos et~al.(2012)Plos, Aigner, Baier, Feldhofer, Hutter, Korak, and Wenger]{Plos2012}
T.~Plos, M.~Aigner, T.~Baier, M.~Feldhofer, M.~Hutter, T.~Korak, and E.~Wenger, ``Semi-passive rfid development platform for implementing and attacking security tags,'' \emph{International Journal of RFID Security and Cryptography}, vol.~1, no.~1, pp. 16--24, 2012.

\bibitem[Naija et~al.(2018{\natexlab{a}})Naija, Beroulle, and Machhout]{Naija2018a}
\BIBentryALTinterwordspacing
Y.~Naija, V.~Beroulle, and M.~Machhout, ``{Security Enhancements of a Mutual Authentication Protocol used in a HF Full-fledged RFID Tag},'' \emph{Journal of Electronic Testing}, vol.~34, pp. 291--304, 2018. [Online]. Available: \url{https://link.springer.com/article/10.1007/s10836-018-5725-x}
\BIBentrySTDinterwordspacing

\bibitem[Carluccio et~al.(2005)Carluccio, Lemke, and Paar]{carluccio2005electromagnetic}
D.~Carluccio, K.~Lemke, and C.~Paar, ``Electromagnetic side channel analysis of a contactless smart card: First results,'' in \emph{ECrypt Workshop on RFID and Lightweight Crypto}, 2005.

\bibitem[Naija et~al.(2018{\natexlab{b}})Naija, Beroulle, and Machhout]{Naija2018}
Y.~Naija, V.~Beroulle, and M.~Machhout, ``Electromagnetic attack test platform for validating rfid tag architectures,'' in \emph{2018 6th International EURASIP Workshop on RFID Technology (EURFID)}, 2018, pp. 1--7.

\bibitem[Kasper et~al.(2009)Kasper, Oswald, and Paar]{kasper2009side}
T.~Kasper, D.~Oswald, and C.~Paar, ``Em side-channel attacks on commercial contactless smartcards using low-cost equipment,'' in \emph{International Workshop on Information Security Applications}.\hskip 1em plus 0.5em minus 0.4em\relax Springer, 2009, pp. 79--93.

\bibitem[Xu et~al.(2018)Xu, Zhu, Wang, Du, Choo, Zhang, and Gai]{xu2018side}
R.~Xu, L.~Zhu, A.~Wang, X.~Du, K.-K.~R. Choo, G.~Zhang, and K.~Gai, ``Side-channel attack on a protected rfid card,'' \emph{IEEE Access}, vol.~6, pp. 58\,395--58\,404, 2018.

\bibitem[Oren and Shamir(2007)]{oren2007remote}
Y.~Oren and A.~Shamir, ``Remote password extraction from rfid tags,'' \emph{IEEE Transactions on Computers}, vol.~56, no.~9, pp. 1292--1296, 2007.

\bibitem[Du et~al.(2007)Du, Xiao, Guizani, and Chen]{du2007effective}
\BIBentryALTinterwordspacing
X.~Du, Y.~Xiao, M.~Guizani, and H.-H. Chen, ``An effective key management scheme for heterogeneous sensor networks,'' \emph{Ad Hoc Networks}, vol.~5, no.~1, pp. 24--34, 2007, security Issues in Sensor and Ad Hoc Networks. [Online]. Available: \url{https://www.sciencedirect.com/science/article/pii/S1570870506000412}
\BIBentrySTDinterwordspacing

\bibitem[Du et~al.(2009)Du, Guizani, Xiao, and Chen]{du2009transactions}
X.~Du, M.~Guizani, Y.~Xiao, and H.-H. Chen, ``Transactions papers a routing-driven elliptic curve cryptography based key management scheme for heterogeneous sensor networks,'' \emph{IEEE Transactions on Wireless Communications}, vol.~8, no.~3, pp. 1223--1229, 2009.

\bibitem[Hei et~al.(2013)Hei, Du, Lin, and Lee]{hei2013pipac}
X.~Hei, X.~Du, S.~Lin, and I.~Lee, ``Pipac: Patient infusion pattern based access control scheme for wireless insulin pump system,'' in \emph{2013 Proceedings IEEE Annual Joint Conference: INFOCOM, IEEE Computer and Communications Societies}, 2013, pp. 3030--3038.

\bibitem[Wu et~al.(2016)Wu, Du, and Wu]{wu2015effective}
L.~Wu, X.~Du, and J.~Wu, ``Effective defense schemes for phishing attacks on mobile computing platforms,'' \emph{IEEE Transactions on Vehicular Technology}, vol.~65, no.~8, pp. 6678--6691, 2016.

\bibitem[Cheng et~al.(2017)Cheng, Fu, Du, Luo, and Guizani]{cheng2017lightweight}
Y.~Cheng, X.~Fu, X.~Du, B.~Luo, and M.~Guizani, ``A lightweight live memory forensic approach based on hardware virtualization,'' \emph{Information Sciences}, vol. 379, pp. 23--41, 2017.

\bibitem[Kim et~al.(2012)Kim, Kim, and Park]{KIM20122899}
\BIBentryALTinterwordspacing
T.~H. Kim, C.~Kim, and I.~Park, ``Side channel analysis attacks using am demodulation on commercial smart cards with seed,'' \emph{Journal of Systems and Software}, vol.~85, no.~12, pp. 2899--2908, 2012, self-Adaptive Systems. [Online]. Available: \url{https://www.sciencedirect.com/science/article/pii/S016412121200194X}
\BIBentrySTDinterwordspacing

\bibitem[Lu et~al.(2010)Lu, Boey, Hodgers, and O'Neill]{Lu2010}
Y.~Lu, K.~H. Boey, P.~Hodgers, and M.~O'Neill, ``Seed masking implementations against power analysis attacks,'' in \emph{2010 IEEE Asia Pacific Conference on Circuits and Systems}, 2010, pp. 1199--1202.

\bibitem[Hutter et~al.(2007)Hutter, Mangard, and Feldhofer]{hutter2007power}
M.~Hutter, S.~Mangard, and M.~Feldhofer, ``Power and em attacks on passive 13.56 mhz rfid devices,'' in \emph{Cryptographic Hardware and Embedded Systems - CHES 2007}, P.~Paillier and I.~Verbauwhede, Eds.\hskip 1em plus 0.5em minus 0.4em\relax Springer Berlin Heidelberg, 2007, pp. 320--333.

\bibitem[Kasper et~al.(2010)Kasper, von Maurich, Oswald, and Paar]{kasper2010cloning}
T.~Kasper, I.~von Maurich, D.~Oswald, and C.~Paar, ``Cloning cryptographic rfid cards for 25\$,'' in \emph{5th Benelux Workshop on Information and System Security. Nijmegen, Netherlands}, 2010.

\bibitem[Behzad(2007)]{behzad2007wireless}
A.~Behzad, \emph{Wireless LAN radios: System Definition to Transistor Design}.\hskip 1em plus 0.5em minus 0.4em\relax John Wiley \& Sons, 2007.

\bibitem[Chaman et~al.(2018)Chaman, Wang, Sun, Hassanieh, and Roy~Choudhury]{chaman2018ghostbuster}
\BIBentryALTinterwordspacing
A.~Chaman, J.~Wang, J.~Sun, H.~Hassanieh, and R.~Roy~Choudhury, ``Ghostbuster: Detecting the presence of hidden eavesdroppers,'' in \emph{Proceedings of the 24th Annual International Conference on Mobile Computing and Networking}, ser. MobiCom '18.\hskip 1em plus 0.5em minus 0.4em\relax New York, NY, USA: Association for Computing Machinery, 2018, p. 337–351. [Online]. Available: \url{https://doi.org/10.1145/3241539.3241580}
\BIBentrySTDinterwordspacing

\bibitem[Camurati et~al.(2018)Camurati, Poeplau, Muench, Hayes, and Francillon]{camurati2018screaming}
G.~Camurati, S.~Poeplau, M.~Muench, T.~Hayes, and A.~Francillon, ``Screaming channels: When electromagnetic side channels meet radio transceivers,'' in \emph{Proceedings of the 2018 ACM SIGSAC Conference on Computer and Communications Security}, 2018, pp. 163--177.

\bibitem[Duan et~al.(2015)Duan, Hongxin, Qiang, Xinjie, and Pengfei]{Duan2015}
\BIBentryALTinterwordspacing
L.~Duan, Z.~Hongxin, L.~Qiang, Z.~Xinjie, and H.~Pengfei, ``{Electromagnetic Side-channel Attack Based on PSO Directed Acyclic Graph SVM},'' \emph{The Journal of China Universities of Posts and Telecommunications}, vol.~22, no.~5, pp. 10--15, 2015. [Online]. Available: \url{https://www.sciencedirect.com/science/article/pii/S1005888515606744}
\BIBentrySTDinterwordspacing

\bibitem[Hori et~al.(2012)Hori, Katashita, Sasaki, and Satoh]{hori2012electromagnetic}
Y.~Hori, T.~Katashita, A.~Sasaki, and A.~Satoh, ``Electromagnetic side-channel attack against 28-nm fpga device,'' \emph{Pre-proceedings of the 13th International Workshop on Information Security Applications (WISA 2012)}, p.~84, 2012.

\bibitem[Das and Sen(2020)]{Das2020a}
\BIBentryALTinterwordspacing
D.~Das and S.~Sen, ``Electromagnetic and power side-channel analysis: Advanced attacks and low-overhead generic countermeasures through white-box approach,'' \emph{Cryptography}, vol.~4, no.~4, 2020. [Online]. Available: \url{https://www.mdpi.com/2410-387X/4/4/30}
\BIBentrySTDinterwordspacing

\bibitem[Yilmaz et~al.(2020)Yilmaz, Prvulovic, and Zajić]{Yilmaz2020a}
B.~B. Yilmaz, M.~Prvulovic, and A.~Zajić, ``Electromagnetic side channel information leakage created by execution of series of instructions in a computer processor,'' \emph{IEEE Transactions on Information Forensics and Security}, vol.~15, pp. 776--789, 2020.

\bibitem[Yu et~al.(2020)Yu, Ma, Yang, Zhao, and Jin]{Yu2020}
H.~Yu, H.~Ma, K.~Yang, Y.~Zhao, and Y.~Jin, ``Deepem: Deep neural networks model recovery through em side-channel information leakage,'' in \emph{2020 IEEE International Symposium on Hardware Oriented Security and Trust (HOST)}, 2020, pp. 209--218.

\bibitem[Iyer and Yilmaz(2019)]{Iyer2019}
V.~V. Iyer and A.~E. Yilmaz, ``An adaptive acquisition approach to localize electromagnetic information leakage from cryptographic modules,'' in \emph{2019 IEEE Texas Symposium on Wireless and Microwave Circuits and Systems (WMCS)}, 2019, pp. 1--6.

\bibitem[Montminy et~al.(2013)Montminy, Baldwin, Temple, and Oxley]{Montminy2013}
D.~P. Montminy, R.~O. Baldwin, M.~A. Temple, and M.~E. Oxley, ``Differential electromagnetic attacks on a 32-bit microprocessor using software defined radios,'' \emph{IEEE Transactions on Information Forensics and Security}, vol.~8, no.~12, pp. 2101--2114, 2013.

\bibitem[Guri et~al.(2016)Guri, Monitz, and Elovici]{guri2016usbee}
M.~Guri, M.~Monitz, and Y.~Elovici, ``Usbee: Air-gap covert-channel via electromagnetic emission from usb,'' in \emph{2016 14th Annual Conference on Privacy, Security and Trust (PST)}, 2016, pp. 264--268.

\bibitem[Kocher et~al.(1999)Kocher, Jaffe, and Jun]{kocher1999differential}
P.~Kocher, J.~Jaffe, and B.~Jun, ``Differential power analysis,'' in \emph{Advances in Cryptology --- CRYPTO' 99}, M.~Wiener, Ed.\hskip 1em plus 0.5em minus 0.4em\relax Springer Berlin Heidelberg, 1999, pp. 388--397.

\bibitem[Agrawal et~al.(2003)Agrawal, Rao, and Rohatgi]{agrawal2003multi}
D.~Agrawal, J.~R. Rao, and P.~Rohatgi, ``Multi-channel attacks,'' in \emph{Cryptographic Hardware and Embedded Systems - CHES 2003}, C.~D. Walter, {\c{C}}.~K. Ko{\c{c}}, and C.~Paar, Eds.\hskip 1em plus 0.5em minus 0.4em\relax Springer Berlin Heidelberg, 2003, pp. 2--16.

\bibitem[Cheddad et~al.(2010)Cheddad, Condell, Curran, and Mc~Kevitt]{cheddad2010digital}
A.~Cheddad, J.~Condell, K.~Curran, and P.~Mc~Kevitt, ``Digital image steganography: Survey and analysis of current methods,'' \emph{Signal Processing}, vol.~90, no.~3, pp. 727--752, 2010.

\bibitem[Yang and Sample(2017)]{yang2017comm}
\BIBentryALTinterwordspacing
C.~J. Yang and A.~P. Sample, ``Em-comm: Touch-based communication via modulated electromagnetic emissions,'' \emph{Proc. ACM Interact. Mob. Wearable Ubiquitous Technol.}, vol.~1, no.~3, Sep. 2017. [Online]. Available: \url{https://doi.org/10.1145/3130984}
\BIBentrySTDinterwordspacing

\bibitem[Ahmad et~al.(2013)Ahmad, Musa, Nadarajah, Hassan, and Othman]{ahmad2013comparison}
M.~S. Ahmad, N.~E. Musa, R.~Nadarajah, R.~Hassan, and N.~E. Othman, ``Comparison between android and ios operating system in terms of security,'' in \emph{2013 8th International Conference on Information Technology in Asia (CITA)}, 2013, pp. 1--4.

\bibitem[Aboulkassimi et~al.(2011)Aboulkassimi, Agoyan, Freund, Fournier, Robisson, and Tria]{aboulkassimi2011electromagnetic}
D.~Aboulkassimi, M.~Agoyan, L.~Freund, J.~Fournier, B.~Robisson, and A.~Tria, ``Electromagnetic analysis (ema) of software aes on java mobile phones,'' in \emph{2011 IEEE International Workshop on Information Forensics and Security}, 2011, pp. 1--6.

\bibitem[Balasch et~al.(2015)Balasch, Gierlichs, Reparaz, and Verbauwhede]{balasch2015dpa}
J.~Balasch, B.~Gierlichs, O.~Reparaz, and I.~Verbauwhede, ``Dpa, bitslicing and masking at 1 ghz,'' in \emph{International Workshop on Cryptographic Hardware and Embedded Systems}.\hskip 1em plus 0.5em minus 0.4em\relax Springer, 2015, pp. 599--619.

\bibitem[K{\"o}nighofer(2008)]{konighofer2008fast}
R.~K{\"o}nighofer, ``A fast and cache-timing resistant implementation of the aes,'' in \emph{Topics in Cryptology -- CT-RSA 2008}, T.~Malkin, Ed.\hskip 1em plus 0.5em minus 0.4em\relax Springer Berlin Heidelberg, 2008, pp. 187--202.

\bibitem[Belgarric et~al.(2016)Belgarric, Fouque, Macario-Rat, and Tibouchi]{belgarric2016side}
P.~Belgarric, P.-A. Fouque, G.~Macario-Rat, and M.~Tibouchi, ``Side-channel analysis of weierstrass and koblitz curve ecdsa on android smartphones,'' in \emph{Cryptographers’ Track at the RSA Conference}.\hskip 1em plus 0.5em minus 0.4em\relax Springer, 2016, pp. 236--252.

\bibitem[Bukasa et~al.(2018)Bukasa, Lashermes, Le~Bouder, Lanet, and Legay]{Bukasa2017}
S.~K. Bukasa, R.~Lashermes, H.~Le~Bouder, J.-L. Lanet, and A.~Legay, ``How trustzone could be bypassed: Side-channel attacks on a modern system-on-chip,'' in \emph{Information Security Theory and Practice}.\hskip 1em plus 0.5em minus 0.4em\relax Cham: Springer International Publishing, 2018, pp. 93--109.

\bibitem[{Hu, Xiao-yang} et~al.(2018){Hu, Xiao-yang}, {Chen, Kai-yan}, {Zhang, Yang}, {Guo, Dong-xin}, and {Wei, Yan-hai}]{Hu2018}
\BIBentryALTinterwordspacing
{Hu, Xiao-yang}, {Chen, Kai-yan}, {Zhang, Yang}, {Guo, Dong-xin}, and {Wei, Yan-hai}, ``{Research on Electromagnetic Side-channel Signal Extraction for Mobile Device PCM-9589F Multi-COM},'' \emph{MATEC Web Conference}, vol. 232, p. 04022, 2018. [Online]. Available: \url{https://www.matec-conferences.org/articles/matecconf/abs/2018/91/matecconf_eitce2018_04022/matecconf_eitce2018_04022.html}
\BIBentrySTDinterwordspacing

\bibitem[Lin et~al.(2017)Lin, Yu, Zhang, Yang, Zhang, and Zhao]{lin2017survey}
J.~Lin, W.~Yu, N.~Zhang, X.~Yang, H.~Zhang, and W.~Zhao, ``{A Survey on Internet of Things: Architecture, Enabling Technologies, Security and Privacy, and Applications},'' \emph{IEEE Internet of Things Journal}, vol.~4, no.~5, pp. 1125--1142, 2017.

\bibitem[Boozer et~al.(2021)Boozer, John, and Mukherjee]{Boozer2021}
\BIBentryALTinterwordspacing
A.~A. Boozer, A.~John, and T.~Mukherjee, ``{Internet of Things Software and Hardware Architectures and Their Impacts on Forensic Investigations: Current Approaches and Challenges},'' \emph{Journal of Digital Forensics, Security and Law}, vol.~16, no.~4, 2021. [Online]. Available: \url{https://commons.erau.edu/jdfsl/vol16/iss2/4/}
\BIBentrySTDinterwordspacing

\bibitem[Balogh et~al.(2021)Balogh, Gallo, Ploszek, Špaček, and Zajac]{Balogh2021}
\BIBentryALTinterwordspacing
S.~Balogh, O.~Gallo, R.~Ploszek, P.~Špaček, and P.~Zajac, ``Iot security challenges: Cloud and blockchain, postquantum cryptography, and evolutionary techniques,'' \emph{Electronics}, vol.~10, no.~21, 2021. [Online]. Available: \url{https://www.mdpi.com/2079-9292/10/21/2647}
\BIBentrySTDinterwordspacing

\bibitem[Tirumaladass et~al.(2020)Tirumaladass, Axelsson, Dougherty, Rasool, and Eldefrawy]{Tirumaladass2020}
V.~Tirumaladass, S.~Axelsson, M.~Dougherty, M.~A. Rasool, and M.~H. Eldefrawy, ``Deep learning-based electromagnetic side-channel analysis for the investigation of iot devices,'' in \emph{2020 Second International Conference on Inventive Research in Computing Applications (ICIRCA)}, 2020, pp. 150--156.

\bibitem[Sayakkara et~al.(2020{\natexlab{b}})Sayakkara, Le-Khac, and Scanlon]{Sayakkara2020a}
\BIBentryALTinterwordspacing
A.~Sayakkara, N.-A. Le-Khac, and M.~Scanlon, ``{EMvidence: A Framework for Digital Evidence Acquisition from IoT Devices through Electromagnetic Side-Channel Analysis},'' \emph{Forensic Science International: Digital Investigation}, vol.~32, p. 300907, 2020. [Online]. Available: \url{https://www.sciencedirect.com/science/article/pii/S2666281720300020}
\BIBentrySTDinterwordspacing

\bibitem[Levina et~al.(2021)Levina, Varyukhin, Kaplun, Zamansky, and van~der Linden]{Levina2021}
A.~Levina, V.~Varyukhin, D.~Kaplun, A.~Zamansky, and D.~van~der Linden, ``A case study exploring side-channel attacks on pet wearables.'' \emph{IAENG International Journal of Computer Science}, vol.~48, no.~4, 2021.

\bibitem[Lewis(2021)]{jaggerlewis}
J.~Lewis, ``Jagger lewis,'' \url{https://www.jagger-lewis.com/en-en/home}, 2021, accessed: November, 2022.

\bibitem[Dinu and Kizhvatov(2018)]{Dinu2018}
\BIBentryALTinterwordspacing
D.~Dinu and I.~Kizhvatov, ``Em analysis in the iot context: Lessons learned from an attack on thread,'' \emph{IACR Transactions on Cryptographic Hardware and Embedded Systems}, vol. 2018, no.~1, p. 73–97, Feb. 2018. [Online]. Available: \url{https://tches.iacr.org/index.php/TCHES/article/view/833}
\BIBentrySTDinterwordspacing

\bibitem[Durvaux and Durvaux(2020)]{Durvaux2020}
\BIBentryALTinterwordspacing
F.~Durvaux and M.~Durvaux, ``Sca-pitaya: A practical and affordable side-channel attack setup for power leakage--based evaluations,'' \emph{Digital Threats}, vol.~1, no.~1, Mar. 2020. [Online]. Available: \url{https://doi.org/10.1145/3371393}
\BIBentrySTDinterwordspacing

\bibitem[Agosta et~al.(2018)Agosta, Barenghi, Pelosi, and Scandale]{Agosta2018}
\BIBentryALTinterwordspacing
G.~Agosta, A.~Barenghi, G.~Pelosi, and M.~Scandale, ``Reactive side-channel countermeasures: Applicability and quantitative security evaluation,'' \emph{Microprocessors and Microsystems}, vol.~62, pp. 50--60, 2018. [Online]. Available: \url{https://www.sciencedirect.com/science/article/pii/S0141933118301790}
\BIBentrySTDinterwordspacing

\bibitem[Jevtic et~al.(2021)Jevtic, Ylitolva, Calonge, Ojanen, Santti, and Koskinen]{Jevtic2021}
R.~Jevtic, M.~Ylitolva, C.~Calonge, M.~Ojanen, T.~Santti, and L.~Koskinen, ``Em side-channel countermeasure for switched-capacitor dc–dc converters based on amplitude modulation,'' \emph{IEEE Transactions on Very Large Scale Integration (VLSI) Systems}, vol.~29, no.~6, pp. 1061--1072, 2021.

\bibitem[Singh et~al.(2019{\natexlab{a}})Singh, Kar, Mathew, Rajan, De, and Mukhopadhyay]{Singh2019}
A.~Singh, M.~Kar, S.~Mathew, A.~Rajan, V.~De, and S.~Mukhopadhyay, ``25.3 a 128b aes engine with higher resistance to power and electromagnetic side-channel attacks enabled by a security-aware integrated all-digital low-dropout regulator,'' in \emph{2019 IEEE International Solid-State Circuits Conference - (ISSCC)}, 2019, pp. 404--406.

\bibitem[Das et~al.(2021{\natexlab{a}})Das, Ghosh, Raychowdhury, and Sen]{Das2021a}
D.~Das, S.~Ghosh, A.~Raychowdhury, and S.~Sen, ``Em/power side-channel attack: White-box modeling and signature attenuation countermeasures,'' \emph{IEEE Design \& Test}, vol.~38, no.~3, pp. 67--75, 2021.

\bibitem[Nassar et~al.(2012)Nassar, Souissi, Guilley, and Danger]{nassar2012rsm}
M.~Nassar, Y.~Souissi, S.~Guilley, and J.-L. Danger, ``Rsm: A small and fast countermeasure for aes, secure against 1st and 2nd-order zero-offset scas,'' in \emph{2012 Design, Automation \& Test in Europe Conference \& Exhibition (DATE)}, 2012, pp. 1173--1178.

\bibitem[Chong et~al.(2021{\natexlab{a}})Chong, Ng, Chen, Lwin, Kyaw, Ho, Chang, and Gwee]{Chong2021}
K.-S. Chong, J.-S. Ng, J.~Chen, N.~K.~Z. Lwin, N.~A. Kyaw, W.-G. Ho, J.~Chang, and B.-H. Gwee, ``Dual-hiding side-channel-attack resistant fpga-based asynchronous-logic aes: Design, countermeasures and evaluation,'' \emph{IEEE Journal on Emerging and Selected Topics in Circuits and Systems}, vol.~11, no.~2, pp. 343--356, 2021.

\bibitem[Shan et~al.(2014)Shan, Shi, Fu, Zhang, Tian, Xu, Yang, and Li]{Shan2014}
\BIBentryALTinterwordspacing
W.~Shan, L.~Shi, X.~Fu, X.~Zhang, C.~Tian, Z.~Xu, J.~Yang, and J.~Li, ``A side-channel analysis resistant reconfigurable cryptographic coprocessor supporting multiple block cipher algorithms,'' in \emph{Proceedings of the 51st Annual Design Automation Conference}, ser. DAC '14.\hskip 1em plus 0.5em minus 0.4em\relax New York, NY, USA: Association for Computing Machinery, 2014, p. 1–6. [Online]. Available: \url{https://doi.org/10.1145/2593069.2593077}
\BIBentrySTDinterwordspacing

\bibitem[Khan et~al.(2017)Khan, Wanchoo, Mumcu, and Köse]{khan2017implications}
A.~W. Khan, T.~Wanchoo, G.~Mumcu, and S.~Köse, ``Implications of distributed on-chip power delivery on em side-channel attacks,'' in \emph{2017 IEEE International Conference on Computer Design (ICCD)}, 2017, pp. 329--336.

\bibitem[Judy et~al.(2022)Judy, Smith, Wallace, and Chen]{judy2022electromagnetic}
R.~Judy, A.~Smith, L.~Wallace, and X.~Chen, ``Electromagnetic waveform characterization for side-channel attacks on aes encryption,'' in \emph{2022 IEEE Physical Assurance and Inspection of Electronics (PAINE)}, 2022, pp. 1--7.

\bibitem[Gao et~al.(2023)Gao, Zhang, Ma, He, and Zhao]{gao2023eo}
\BIBentryALTinterwordspacing
Y.~Gao, Q.~Zhang, H.~Ma, J.~He, and Y.~Zhao, ``Eo-shield: A multi-function protection scheme against side channel and focused ion beam attacks,'' in \emph{Proceedings of the 28th Asia and South Pacific Design Automation Conference}, ser. ASPDAC '23.\hskip 1em plus 0.5em minus 0.4em\relax New York, NY, USA: Association for Computing Machinery, 2023, p. 670–675. [Online]. Available: \url{https://doi.org/10.1145/3566097.3567924}
\BIBentrySTDinterwordspacing

\bibitem[He et~al.(2011)He, Pizarro, de~la Torre, Portilla, and Riesgo]{He2011a}
\BIBentryALTinterwordspacing
W.~He, C.~Pizarro, E.~de~la Torre, J.~Portilla, and T.~Riesgo, ``{SCA Security Verification on Wireless Sensor Network Node},'' \emph{VLSI Circuits and Systems V}, vol. 8067, pp. 298--312, 2011. [Online]. Available: \url{https://www.spiedigitallibrary.org/conference-proceedings-of-spie/8067/80670W/SCA-security-verification-on-wireless-sensor-network-node/10.1117/12.887537.short}
\BIBentrySTDinterwordspacing

\bibitem[Frieslaar and Irwin(2017)]{Frieslaar2017a}
I.~Frieslaar and B.~Irwin, ``Investigating the electromagnetic side channel leakage from a raspberry pi,'' in \emph{2017 Information Security for South Africa (ISSA)}, 2017, pp. 24--31.

\bibitem[Pammu et~al.(2016{\natexlab{b}})Pammu, Chong, and Gwee]{Pammu2016a}
A.~A. Pammu, K.-S. Chong, and B.-H. Gwee, ``Highly secured arithmetic hiding based s-box on aes-128 implementation,'' in \emph{2016 International Symposium on Integrated Circuits (ISIC)}, 2016, pp. 1--4.

\bibitem[Zhou and Kong(2019)]{Kong2019}
W.-h. Zhou and F.-t. Kong, ``Electromagnetic side channel attack against embedded encryption chips,'' in \emph{2019 IEEE 19th International Conference on Communication Technology (ICCT)}, 2019, pp. 140--144.

\bibitem[Carlier et~al.(2005)Carlier, Chabanne, Dottax, and Pelletier]{carlier2005generalizing}
V.~Carlier, H.~Chabanne, E.~Dottax, and H.~Pelletier, ``Generalizing square attack using side-channels of an aes implementation on an fpga,'' in \emph{International Conference on Field Programmable Logic and Applications, 2005.}\hskip 1em plus 0.5em minus 0.4em\relax IEEE, 2005, pp. 433--437.

\bibitem[Singh et~al.(2019{\natexlab{b}})Singh, Kar, Mathew, Rajan, De, and Mukhopadhyay]{singh2018improved}
A.~Singh, M.~Kar, S.~K. Mathew, A.~Rajan, V.~De, and S.~Mukhopadhyay, ``Improved power/em side-channel attack resistance of 128-bit aes engines with random fast voltage dithering,'' \emph{IEEE Journal of Solid-State Circuits}, vol.~54, no.~2, pp. 569--583, 2019.

\bibitem[Chong et~al.(2021{\natexlab{b}})Chong, Ng, Chen, Lwin, Kyaw, Ho, Chang, and Gwee]{9424617}
K.-S. Chong, J.-S. Ng, J.~Chen, N.~K.~Z. Lwin, N.~A. Kyaw, W.-G. Ho, J.~Chang, and B.-H. Gwee, ``Dual-hiding side-channel-attack resistant fpga-based asynchronous-logic aes: Design, countermeasures and evaluation,'' \emph{IEEE Journal on Emerging and Selected Topics in Circuits and Systems}, vol.~11, no.~2, pp. 343--356, 2021.

\bibitem[Chawla et~al.(2020)Chawla, Singh, Kumar, Kar, and Mukhopadhyay]{chawla2020securing}
N.~Chawla, A.~Singh, H.~Kumar, M.~Kar, and S.~Mukhopadhyay, ``{Securing IoT Devices Using Dynamic Power Management: Machine Learning Approach},'' \emph{IEEE Internet of Things Journal}, vol.~8, no.~22, pp. 16\,379--16\,394, 2020.

\bibitem[Buchanan and Woodward(2017)]{buchanan2017will}
W.~Buchanan and A.~Woodward, ``Will quantum computers be the end of public key encryption?'' \emph{Journal of Cyber Security Technology}, vol.~1, no.~1, pp. 1--22, 2017.

\bibitem[Mavroeidis et~al.(2018)Mavroeidis, Vishi, Zych, and J{\o}sang]{mavroeidis2018impact}
V.~Mavroeidis, K.~Vishi, M.~D. Zych, and A.~J{\o}sang, ``\BIBforeignlanguage{English}{The impact of quantum computing on present cryptography},'' \emph{\BIBforeignlanguage{English}{International Journal of Advanced Computer Science and Applications}}, vol.~9, no.~3, 2018.

\bibitem[Fernández-Caramès and Fraga-Lamas(2020)]{fernandez2020towards}
T.~M. Fernández-Caramès and P.~Fraga-Lamas, ``Towards post-quantum blockchain: A review on blockchain cryptography resistant to quantum computing attacks,'' \emph{IEEE Access}, vol.~8, pp. 21\,091--21\,116, 2020.

\bibitem[Sendrier(2017)]{sendrier2017code}
N.~Sendrier, ``Code-based cryptography: State of the art and perspectives,'' \emph{IEEE Security \& Privacy}, vol.~15, no.~4, pp. 44--50, 2017.

\bibitem[D'Anvers et~al.(2019)D'Anvers, Tiepelt, Vercauteren, and Verbauwhede]{10.1145/3338467.3358948}
\BIBentryALTinterwordspacing
J.-P. D'Anvers, M.~Tiepelt, F.~Vercauteren, and I.~Verbauwhede, ``Timing attacks on error correcting codes in post-quantum schemes,'' in \emph{Proceedings of ACM Workshop on Theory of Implementation Security Workshop}, ser. TIS'19.\hskip 1em plus 0.5em minus 0.4em\relax New York, NY, USA: Association for Computing Machinery, 2019, p. 2–9. [Online]. Available: \url{https://doi.org/10.1145/3338467.3358948}
\BIBentrySTDinterwordspacing

\bibitem[Das et~al.(2021{\natexlab{b}})Das, Danial, Golder, Modak, Maity, Chatterjee, Seo, Chang, Varna, Krishnamurthy, Mathew, Ghosh, Raychowdhury, and Sen]{Das2020b}
D.~Das, J.~Danial, A.~Golder, N.~Modak, S.~Maity, B.~Chatterjee, D.-H. Seo, M.~Chang, A.~L. Varna, H.~K. Krishnamurthy, S.~Mathew, S.~Ghosh, A.~Raychowdhury, and S.~Sen, ``Em and power sca-resilient aes-256 through >350× current-domain signature attenuation and local lower metal routing,'' \emph{IEEE Journal of Solid-State Circuits}, vol.~56, no.~1, pp. 136--150, 2021.

\bibitem[Iyer and Yılmaz(2022)]{Iyer2022}
V.~V. Iyer and A.~E. Yılmaz, ``An anova method to rapidly assess information leakage near cryptographic modules,'' \emph{IEEE Transactions on Electromagnetic Compatibility}, vol.~64, no.~4, pp. 915--929, 2022.

\bibitem[Danial et~al.(2021)Danial, Das, Golder, Ghosh, Raychowdhury, and Sen]{Danial2021}
\BIBentryALTinterwordspacing
J.~Danial, D.~Das, A.~Golder, S.~Ghosh, A.~Raychowdhury, and S.~Sen, ``Em-x-dl: Efficient cross-device deep learning side-channel attack with noisy em signatures,'' \emph{Journal on Emerging Technologies in Computing Systems}, vol.~18, no.~1, Sep. 2021. [Online]. Available: \url{https://doi.org/10.1145/3465380}
\BIBentrySTDinterwordspacing

\bibitem[Mukhtar and Kong(2018)]{Mukhtar2018a}
N.~Mukhtar and Y.~Kong, ``Hyper-parameter optimization for machine-learning based electromagnetic side-channel analysis,'' in \emph{2018 26th International Conference on Systems Engineering (ICSEng)}, 2018, pp. 1--7.

\bibitem[Ghosh et~al.(2022)Ghosh, Nath, Das, Ghosh, and Sen]{Ghosh2022}
A.~Ghosh, M.~Nath, D.~Das, S.~Ghosh, and S.~Sen, ``Electromagnetic analysis of integrated on-chip sensing loop for side-channel and fault-injection attack detection,'' \emph{IEEE Microwave and Wireless Components Letters}, vol.~32, no.~6, pp. 784--787, 2022.

\bibitem[Veyrat-Charvillon and Standaert(2010)]{Veyrat-Charvillon2010}
N.~Veyrat-Charvillon and F.-X. Standaert, ``Adaptive chosen-message side-channel attacks,'' in \emph{Proceedings of the 8th International Conference on Applied Cryptography and Network Security}, ser. ACNS'10.\hskip 1em plus 0.5em minus 0.4em\relax Berlin, Heidelberg: Springer-Verlag, 2010, p. 186–199.

\bibitem[Choudary and Kuhn(2018)]{Choudary2018}
M.~O. Choudary and M.~G. Kuhn, ``Efficient, portable template attacks,'' \emph{IEEE Transactions on Information Forensics and Security}, vol.~13, no.~2, pp. 490--501, 2018.

\bibitem[Tunstall(2017)]{Tunstall2017}
\BIBentryALTinterwordspacing
M.~Tunstall, \emph{Smart Card Security}.\hskip 1em plus 0.5em minus 0.4em\relax Cham: Springer International Publishing, 2017, pp. 217--251. [Online]. Available: \url{https://doi.org/10.1007/978-3-319-50500-8_9}
\BIBentrySTDinterwordspacing

\bibitem[Nomata et~al.(2016)Nomata, Matsubayashi, Sawada, and Satoh]{Nomata2016}
Y.~Nomata, M.~Matsubayashi, K.~Sawada, and A.~Satoh, ``Comparison of side-channel attack on cryptographic cirucits between old and new technology fpgas,'' in \emph{2016 IEEE 5th Global Conference on Consumer Electronics}, 2016, pp. 1--4.

\bibitem[He et~al.(2020)He, Ma, Guo, Zhao, and Jin]{He2020a}
J.~He, H.~Ma, X.~Guo, Y.~Zhao, and Y.~Jin, ``Design for em side-channel security through quantitative assessment of rtl implementations,'' in \emph{2020 25th Asia and South Pacific Design Automation Conference (ASP-DAC)}, 2020, pp. 62--67.

\bibitem[Kar et~al.(2018{\natexlab{a}})Kar, Singh, Mathew, Rajan, De, and Mukhopadhyay]{Kar2018a}
M.~Kar, A.~Singh, S.~K. Mathew, A.~Rajan, V.~De, and S.~Mukhopadhyay, ``Reducing power side-channel information leakage of aes engines using fully integrated inductive voltage regulator,'' \emph{IEEE Journal of Solid-State Circuits}, vol.~53, no.~8, pp. 2399--2414, 2018.

\bibitem[Poggi et~al.(2022)Poggi, Maurine, Ordas, Sarafianos, and Raoult]{Poggi2022}
D.~Poggi, P.~Maurine, T.~Ordas, A.~Sarafianos, and J.~Raoult, ``{EM Emission Modeling for Secure IC Design},'' in \emph{2021 13th International Workshop on the Electromagnetic Compatibility of Integrated Circuits (EMC Compo)}, 2022, pp. 39--44.

\bibitem[De~Mulder et~al.(2006)De~Mulder, Ors, Preneel, and Verbauwhede]{de2006differential}
E.~De~Mulder, S.~B. Ors, B.~Preneel, and I.~Verbauwhede, ``Differential electromagnetic attack on an fpga implementation of elliptic curve cryptosystems,'' in \emph{2006 World Automation Congress}, 2006, pp. 1--6.

\bibitem[Soares et~al.(2010)Soares, Calazans, Lomn\'{e}, Dehbaoui, Maurine, and Torres]{10.1145/1854153.1854183}
\BIBentryALTinterwordspacing
R.~I. Soares, N.~L.~V. Calazans, V.~Lomn\'{e}, A.~Dehbaoui, P.~Maurine, and L.~Torres, ``A gals pipeline des architecture to increase robustness against dpa and dema attacks,'' ser. SBCCI '10.\hskip 1em plus 0.5em minus 0.4em\relax New York, NY, USA: Association for Computing Machinery, 2010, p. 115–120. [Online]. Available: \url{https://doi.org/10.1145/1854153.1854183}
\BIBentrySTDinterwordspacing

\bibitem[Shan et~al.(2015)Shan, Fu, and Xu]{Shan2015}
W.~Shan, X.~Fu, and Z.~Xu, ``A secure reconfigurable crypto ic with countermeasures against spa, dpa, and ema,'' \emph{IEEE Transactions on Computer-Aided Design of Integrated Circuits and Systems}, vol.~34, no.~7, pp. 1201--1205, 2015.

\bibitem[Maistri et~al.(2013)Maistri, Tiran, Maurine, Koren, and Leveugle]{Maistri2013}
P.~Maistri, S.~Tiran, P.~Maurine, I.~Koren, and R.~Leveugle, ``Countermeasures against em analysis for a secured fpga-based aes implementation,'' in \emph{2013 International Conference on Reconfigurable Computing and FPGAs (ReConFig)}, 2013, pp. 1--6.

\bibitem[Das et~al.(2019)Das, Nath, Chatterjee, Ghosh, and Sen]{Das2019a}
D.~Das, M.~Nath, B.~Chatterjee, S.~Ghosh, and S.~Sen, ``Stellar: A generic em side-channel attack protection through ground-up root-cause analysis,'' in \emph{2019 IEEE International Symposium on Hardware Oriented Security and Trust (HOST)}, 2019, pp. 11--20.

\bibitem[Dyrkolbotn and Snekkenes(2009)]{dyroaksa}
G.~O. Dyrkolbotn and E.~Snekkenes, ``Modified template attack: Detecting address bus signals of equal hamming weight,'' in \emph{The Norwegian Information Security Conference (NISK)}, 2009, pp. 43--56.

\bibitem[Picek et~al.(2018)Picek, Heuser, Wu, Alippi, and Regazzoni]{Picek2018}
\BIBentryALTinterwordspacing
S.~Picek, A.~Heuser, L.~Wu, C.~Alippi, and F.~Regazzoni, ``{When Theory meets Practice: A Framework for Robust Profiled Side-channel Analysis},'' \emph{Cryptology ePrint Archive}, 2018. [Online]. Available: \url{https://eprint.iacr.org/2018/1123}
\BIBentrySTDinterwordspacing

\bibitem[Callan et~al.(2016)Callan, Behrang, Zajic, Prvulovic, and Orso]{Callan2016}
\BIBentryALTinterwordspacing
R.~Callan, F.~Behrang, A.~Zajic, M.~Prvulovic, and A.~Orso, ``Zero-overhead profiling via em emanations,'' in \emph{Proceedings of the 25th International Symposium on Software Testing and Analysis}, ser. ISSTA 2016.\hskip 1em plus 0.5em minus 0.4em\relax New York, NY, USA: Association for Computing Machinery, 2016, p. 401–412. [Online]. Available: \url{https://doi.org/10.1145/2931037.2931065}
\BIBentrySTDinterwordspacing

\bibitem[Jiang and Pavlidis(2021)]{Jiang2021}
M.~Jiang and V.~F. Pavlidis, ``A probe placement method for efficient electromagnetic attacks,'' in \emph{SMACD / PRIME 2021; International Conference on SMACD and 16th Conference on PRIME}, 2021, pp. 1--4.

\bibitem[Pammu et~al.(2019)Pammu, Ho, Lwin, Chong, and Gwee]{Pammu2018}
A.~A. Pammu, W.-G. Ho, N.~K.~Z. Lwin, K.-S. Chong, and B.-H. Gwee, ``A high throughput and secure authentication-encryption aes-ccm algorithm on asynchronous multicore processor,'' \emph{IEEE Transactions on Information Forensics and Security}, vol.~14, no.~4, pp. 1023--1036, 2019.

\bibitem[Kar et~al.(2018{\natexlab{b}})Kar, Singh, Ghosh, Mathew, Rajan, De, Beyah, and Mukhopadhyay]{Kar2018}
\BIBentryALTinterwordspacing
M.~Kar, A.~Singh, S.~Ghosh, S.~Mathew, A.~Rajan, V.~De, R.~Beyah, and S.~Mukhopadhyay, ``{Blindsight: Blinding EM side-channel leakage using built-in fully integrated inductive voltage regulator},'' \emph{arXiv Preprint}, 2018. [Online]. Available: \url{https://arxiv.org/abs/1802.09096}
\BIBentrySTDinterwordspacing

\bibitem[Liu et~al.(2012)Liu, Feng, Yuan, and Gao]{Liu2012}
B.~Liu, H.~Feng, Z.~Yuan, and Y.~Gao, ``Learning to attack from electromagnetic emanation,'' in \emph{2012 6th Asia-Pacific Conference on Environmental Electromagnetics (CEEM)}, 2012, pp. 202--205.

\bibitem[Le et~al.(2007)Le, Clediere, Serviere, and Lacoume]{Le2007}
T.-H. Le, J.~Clediere, C.~Serviere, and J.-L. Lacoume, ``Noise reduction in side channel attack using fourth-order cumulant,'' \emph{IEEE Transactions on Information Forensics and Security}, vol.~2, no.~4, pp. 710--720, 2007.

\bibitem[Schlumberger et~al.(2021)Schlumberger, Wildermann, and Teich]{Schlumberger2021}
J.~Schlumberger, S.~Wildermann, and J.~Teich, ``Corsica: A framework for conducting real-world side-channel analysis,'' in \emph{2021 11th IFIP International Conference on New Technologies, Mobility and Security (NTMS)}, 2021, pp. 1--5.

\bibitem[Ravi et~al.(2019)Ravi, Jungk, and Bhasin]{Ravi2019}
P.~Ravi, B.~Jungk, and S.~Bhasin, ``Single trace electromagnetic side-channel attacks on fpga implementation of elliptic curve cryptography,'' in \emph{2019 Joint International Symposium on Electromagnetic Compatibility, Sapporo and Asia-Pacific International Symposium on Electromagnetic Compatibility (EMC Sapporo/APEMC)}, 2019, pp. 1--4.

\bibitem[Nomikos et~al.(2020)Nomikos, Papadimitriou, Stergiopoulos, Koutras, Psarakis, and Kotzanikolaou]{Nomikos2020}
K.~Nomikos, A.~Papadimitriou, G.~Stergiopoulos, D.~Koutras, M.~Psarakis, and P.~Kotzanikolaou, ``On a security-oriented design framework for medical iot devices: The hardware security perspective,'' in \emph{2020 23rd Euromicro Conference on Digital System Design (DSD)}, 2020, pp. 301--308.

\bibitem[Cammarota et~al.(2018)Cammarota, Banerjee, and Rosenberg]{Cammarota2018}
\BIBentryALTinterwordspacing
R.~Cammarota, I.~Banerjee, and O.~Rosenberg, ``{Machine learning IP protection},'' \emph{2018 IEEE/ACM International}, 2018. [Online]. Available: \url{https://dl.acm.org/doi/abs/10.1145/3240765.3270589}
\BIBentrySTDinterwordspacing

\bibitem[Sayakkara et~al.(2020{\natexlab{c}})Sayakkara, Le-Khac, and Scanlon]{Sayakkara2020c}
\BIBentryALTinterwordspacing
A.~Sayakkara, N.-A. Le-Khac, and M.~Scanlon, ``{Facilitating Electromagnetic Side-Channel Analysis for IoT Investigation: Evaluating the EMvidence Framework},'' \emph{Forensic Science International: Digital Investigation}, vol.~33, p. 301003, 2020. [Online]. Available: \url{https://www.sciencedirect.com/science/article/pii/S2666281720302523}
\BIBentrySTDinterwordspacing

\bibitem[Seçkiner and Köse(2021)]{Seckiner2021a}
S.~Seçkiner and S.~Köse, ``Preprocessing of the physical leakage information to combine side-channel distinguishers,'' \emph{IEEE Transactions on Very Large Scale Integration (VLSI) Systems}, vol.~29, no.~12, pp. 2052--2063, 2021.

\bibitem[Kang et~al.(2009)Kang, Choi, Chung, Cho, and Han]{Kang2009}
Y.~S. Kang, D.~H. Choi, B.~H. Chung, H.~S. Cho, and D.-G. Han, ``Efficient key detection method in the correlation electromagnetic analysis using peak selection algorithm,'' \emph{Journal of Communications and Networks}, vol.~11, no.~6, pp. 556--563, 2009.

\bibitem[Gunathilake et~al.(2021)Gunathilake, Al-Dubai, Buchanan, and Lo]{Gunathilake2021a}
N.~A. Gunathilake, A.~Al-Dubai, W.~J. Buchanan, and O.~Lo, ``Electromagnetic analysis of an ultra-lightweight cipher: Present,'' in \emph{10th International Conference on Cryptography and Information Security (CRYPIS 2021)}, vol.~11, 2021, pp. 185--205.

\bibitem[Chawla et~al.(2019)Chawla, Singh, Kar, and Mukhopadhyay]{Chawla2019}
N.~Chawla, A.~Singh, M.~Kar, and S.~Mukhopadhyay, ``Application inference using machine learning based side channel analysis,'' pp. 1--8, 2019.

\bibitem[Chawla et~al.(2021)Chawla, Kumar, and Mukhopadhyay]{Chawla2021}
N.~Chawla, H.~Kumar, and S.~Mukhopadhyay, ``Machine learning in wavelet domain for electromagnetic emission based malware analysis,'' \emph{IEEE Transactions on Information Forensics and Security}, vol.~16, pp. 3426--3441, 2021.

\bibitem[Lorenzo et~al.(2019)Lorenzo, McDonald, Andel, Glisson, and Russ]{Lorenzo2019}
F.~Lorenzo, J.~T. McDonald, T.~R. Andel, W.~B. Glisson, and S.~Russ, ``Evaluating side channel resilience in iphone 5c unlock scenarios,'' in \emph{2019 SoutheastCon}, 2019, pp. 1--7.

\bibitem[Leignac et~al.(2019)Leignac, Potin, Rigaud, Dutertre, and Ponti\'{e}]{Leignac2019}
\BIBentryALTinterwordspacing
P.~Leignac, O.~Potin, J.-B. Rigaud, J.-M. Dutertre, and S.~Ponti\'{e}, ``Comparison of side-channel leakage on rich and trusted execution environments,'' in \emph{Proceedings of the Sixth Workshop on Cryptography and Security in Computing Systems}, ser. CS2 '19.\hskip 1em plus 0.5em minus 0.4em\relax New York, NY, USA: Association for Computing Machinery, 2019, p. 19–22. [Online]. Available: \url{https://doi.org/10.1145/3304080.3304084}
\BIBentrySTDinterwordspacing

\bibitem[Tuveri et~al.(2018)Tuveri, Hassan, Garcia, and Brumley]{Tuveri2018}
\BIBentryALTinterwordspacing
N.~Tuveri, S.~u. Hassan, C.~P. Garcia, and B.~B. Brumley, ``Side-channel analysis of sm2: A late-stage featurization case study,'' in \emph{Proceedings of the 34th Annual Computer Security Applications Conference}, ser. ACSAC '18.\hskip 1em plus 0.5em minus 0.4em\relax New York, NY, USA: Association for Computing Machinery, 2018, p. 147–160. [Online]. Available: \url{https://doi.org/10.1145/3274694.3274725}
\BIBentrySTDinterwordspacing

\bibitem[Khan et~al.(2018)Khan, Alam, Zajic, and Prvulovic]{Khan2018}
\BIBentryALTinterwordspacing
H.~A. Khan, M.~Alam, A.~Zajic, and M.~Prvulovic, ``{Detailed Tracking of Program Control Flow Using Analog Side-channel Signals: A Promise for IoT Malware Detection and a Threat for Many Cryptographic Implementations},'' in \emph{Cyber Sensing 2018}, I.~V. Ternovskiy and P.~Chin, Eds., vol. 10630, International Society for Optics and Photonics.\hskip 1em plus 0.5em minus 0.4em\relax SPIE, 2018, p. 1063005. [Online]. Available: \url{https://doi.org/10.1117/12.2304382}
\BIBentrySTDinterwordspacing

\bibitem[Wakabayashi et~al.(2018)Wakabayashi, Maruyama, Mori, Goto, Kinugawa, and Hayashi]{Wakabayashi2018}
S.~Wakabayashi, S.~Maruyama, T.~Mori, S.~Goto, M.~Kinugawa, and Y.-I. Hayashi, ``A feasibility study of radio-frequency retroreflector attack,'' in \emph{Proceedings of the 12th USENIX Conference on Offensive Technologies}, ser. WOOT'18.\hskip 1em plus 0.5em minus 0.4em\relax USA: USENIX Association, 2018, p.~4.

\bibitem[Garc\'{\i}a et~al.(2020)Garc\'{\i}a, Ul~Hassan, Tuveri, Gridin, Aldaya, and Brumley]{Garcia2020}
C.~P. Garc\'{\i}a, S.~Ul~Hassan, N.~Tuveri, I.~Gridin, A.~C. Aldaya, and B.~B. Brumley, ``Certified side channels,'' in \emph{Proceedings of the 29th USENIX Conference on Security Symposium}, ser. SEC'20.\hskip 1em plus 0.5em minus 0.4em\relax USA: USENIX Association, 2020.

\end{thebibliography}

\newpage 
\clearpage
\onecolumn
\appendix

\section{Classification of the Systematically Reviewed Literature}

\begin{table}[!h]
\centering
\caption{Listing and Classification of Included Literature (Grouped by Targeted Device Type)}
\label{tab:legend}
\begin{tabular}{|>{\scriptsize}c|>{\scriptsize}l|}

\hline
\multicolumn{2}{|c|}{\footnotesize\textbf{Legend}} \\ \hline
\faCircle & High Relevancy/Applicability  \\ \hline
\rotatebox[origin=c]{90}{\faAdjust} & Medium Relevancy/Applicability \\ \hline
\faCircle[regular] & Low Relevancy/Applicability \\ \hline
- & No Relevancy/Applicability \\ \hline
\faLaptop  & Computer \\ \hline
\faTablet*  & Mobile Device \\ \hline
\faNetworkWired & Network Device \\ \hline
\cincludegraphics[scale=0.035,trim=0 0.9cm 0 -1cm, clip]{Images/internet-of-things.png}  & Internet of Things Devices \\ \hline
\cincludegraphics[scale=0.015]{Images/microcontroller.png} & Integrated Circuit/FPGA/Microcontroller/Microprocessor \\ \hline
\includegraphics[scale=0.09, trim=0 0 0 -1cm, clip]{Images/multiple.png} & Multiple Devices \\ \hline
\includegraphics[scale=0.04,trim=0 0 0 -2cm, clip]{Images/oscilloscope.png} & Oscilloscope \\ \hline
\faWaveSquare & Software Defined Radio\\ \hline

\cmark & Yes \\ \hline
\xmark & No \\ \hline
\end{tabular}
\end{table}


\small
\begin{longtable*}{|>{\footnotesize}p{5.7cm}|*{12}{>{\centering\arraybackslash\scriptsize}p{0.3cm}|}>{\centering\arraybackslash\scriptsize}p{0.6cm}|}
\hline
\textbf{Reference} & \rotatebox{90}{\textbf{AES Encryption}} & \rotatebox{90}{\textbf{DES Encryption}} & \rotatebox{90}{\textbf{ECC Encryption}} & \rotatebox{90}{\textbf{EEA Encryption}} & \rotatebox{90}{\textbf{RSA Encryption}} & \rotatebox{90}{\textbf{Other Encryption}} &\rotatebox{90}{\textbf{Instrumented?}} & \rotatebox{90}{\textbf{Device Type}} & \rotatebox{90}{\textbf{EM Capture}} & \rotatebox{90}{\textbf{Machine Learning  }} & \rotatebox{90}{\textbf{Deep Learning}} & \rotatebox{90}{\textbf{Key Recovery}} \\ 
\hline
\endfirsthead

\multicolumn{13}{l}{\small \textsf{Table \ref{tab:legend}: Listing and Classification of Included Literature (Grouped by Targeted Device Type), continued}}\\[1ex]
\hline
\textbf{Reference} & \rotatebox{90}{\textbf{AES Encryption}} & \rotatebox{90}{\textbf{DES Encryption}} & \rotatebox{90}{\textbf{ECC Encryption}} & \rotatebox{90}{\textbf{EEA Encryption}} & \rotatebox{90}{\textbf{RSA Encryption}} & \rotatebox{90}{\textbf{Other Encryption}} & \rotatebox{90}{\textbf{Instrumented?}} & \rotatebox{90}{\textbf{Device Type}} & \rotatebox{90}{\textbf{EM Capture}} & \rotatebox{90}{\textbf{Machine Learning  }} & \rotatebox{90}{\textbf{Deep Learning}} & \rotatebox{90}{\textbf{Key Recovery}} \\ 
\hline
\endhead
\hline
\endfoot




\multicolumn{12}{c}{Targeted Device: IC/FPGA/Microcontroller/Microprocessor} \\ \hline

\citet{Das2020b}          &       \faCircle       &       -       &       -       &      -        &       -       &    -     & \cmark               &            \cincludegraphics[scale=0.015]{Images/microcontroller.png}          &         \cincludegraphics[scale=0.04]{Images/oscilloscope.png}        &             -              &             -           &            \rotatebox[origin=c]{90}{\faAdjust}           \\ \hline
\citet{Iyer2021}          &       \faCircle       &       -       &       -       &      -        &       -       &    -     & \cmark               &            \cincludegraphics[scale=0.015]{Images/microcontroller.png}           &       \cincludegraphics[scale=0.04]{Images/oscilloscope.png}          &          -               &           -             &           \faCircle            \\
\hline

\end{longtable*}

\end{document}